\documentclass{aa}
\usepackage{txfonts}
\usepackage{graphicx}
\usepackage{natbib}
\bibpunct{(}{)}{;}{a}{}{,} 
\usepackage{color}
\newcommand{\emm}[1]{\ensuremath{#1}}   
\newcommand{\emr}[1]{\emm{\mathrm{#1}}} 

\newcommand{\unit}[1]{\emm{\, \emr{#1}}}
\newcommand{\ala}{\mbox{\rlap{\hbox{\lower4pt\hbox{$\sim$}}}\hbox{$<$}}} 

\newcommand{\Lsun}{\unit{L_\odot}}
\newcommand{\Msun}{\unit{M_\odot}}

\newcommand{\nht}{\ifmmode {{\rm NH}_3} \else {NH{\bas 3}} \fi}
\newcommand{\as}{\ifmmode {^{\scriptscriptstyle\prime\prime}}
        \else $^{\scriptscriptstyle\prime\prime}$\fi}

\begin{document}

\title{A dual-frequency sub-arcsecond study of proto-planetary disks at mm wavelengths:
First evidence for radial variations of the dust properties.\\
  \thanks{PdBI is operated by IRAM, which is supported by INSU/CNRS (France), MPG
  (Germany), and IGN (Spain).}}

\author{
 S.~Guilloteau \inst{1,2},
 A.~Dutrey \inst{1,2},
 V.~Pi\'etu \inst{3}
 and Y.~Boehler \inst{1,2}
 }
\institute{Universit\'e de Bordeaux, Observatoire Aquitain des Sciences de l'Univers,  BP 89, F-33271 Floirac, France
  \and{CNRS/INSU - UMR5804, Laboratoire d'Astrophysique de Bordeaux;  BP 89, F-33271 Floirac, France}\\
  \email{Anne.Dutrey@obs.u-bordeaux1.fr, Stephane.Guilloteau@obs.u-bordeaux1.fr}\\ \email{Yann.Boehler@obs.u-bordeaux1.fr}
 \and{IRAM, 300 rue de la Piscine, 38400 Saint Martin d'H\`eres, France.}
 \email{pietu@iram.fr}}

\offprints{S.Guilloteau, \email{Stephane.Guilloteau@obs.u-bordeaux1.fr}}

\date{Received 14-Jun-2010, Accepted 02-Mar-2011} %

\authorrunning{Guilloteau et al.} %
\titlerunning{Dual frequency mm imaging of proto-planetary disks}
\abstract
{Proto-planetary disks are thought to provide the initial environment for planetary system formation.
The dust and gas distribution and its evolution with time is one of the key elements in the process.}
{We attempt to characterize the radial distribution of dust in disks
around a sample of young stars from an observational point of view, and, when possible, in
a model-independent way, by using parametric laws.}
{We used the IRAM PdBI interferometer to provide very high angular resolution (down to 0.4$''$ in some sources) observations of the continuum at 1.3~mm and 3~mm around a sample of T~Tauri stars in the Taurus-Auriga region. The sample includes single and multiple systems, with a total of 23 individual disks. We used track-sharing observing mode to minimize the biases. We fitted these data with two kinds of models: a "truncated power law" model and a model presenting an exponential decay at the disk edge ("viscous" model).}
{
Direct evidence for tidal truncation is found in the multiple systems.
The temperature of the mm-emitting dust is constrained in a few systems.
Unambiguous evidence for large grains is obtained by resolving out disks with very low values of the dust
emissivity index $\beta$. In most disks that are sufficiently resolved 
at two different wavelengths,
we find a radial dependence of $\beta$, which appears to increase from low values (as low as 0) at the center to about 1.7 -- 2 at the disk edge. The same behavior could apply to all studied disks. It introduces further ambiguities in interpreting the brightness profile, because the regions with apparent $\beta \approx 0$ can also be interpreted as being optically thick when their brightness temperature is high enough. Despite the added uncertainty on the dust absorption coefficient,  the characteristic size of the disk appears to increase with a higher estimated star age.}
{These results provide the first direct evidence of the radial dependence of the grain size in proto-planetary disks. Constraints of the surface density distributions and their evolution remain ambiguous because of a degeneracy with the $\beta(r)$ law.}


\maketitle %

\section{Introduction}

The gas and dust surface densities of proto-planetary disks appear as one of the key parameters in the formation of  planetary systems.
For example, the formation mechanism of giant planets remains a debated problem. Competing models are the core-accretion mechanism
\citep[e.g.][]{Hubickyj+etal_2005}, which faces apparent timescale difficulties, and the gravitational instability \citep[e.g.][]{Boss_1997,Rice+etal_2005},
which requires massive disks. Determining the dust and gas densities as a function of age of the proto-planetary disks would be a major step
to decide the relative importance of the various processes that potentially lead to planet formation.

However, there is no ideal way to measure these densities. H$_2$ remains the more abundant molecule in proto-planetary disks but is difficult to observe because it only possesses quadrupolar rotation lines in the mid-IR. The gas column density is thus usually estimated
from molecular tracers such as CO or less abundant molecules \citep{Pietu+etal_2007}. Uncertainties linked to a poor accuracy on the molecular abundance and its variation across the disk owing to the chemical behavior of the observed molecule usually affect the results \citep{dutrey+etal_2007}.
The dust surface density can, in theory, be directly derived from the dust brightness temperature. However, the dust emissivity is still poorly known and the accuracy on the surface density depends on the knowledge of the dust properties (composition, size, etc\ldots ) and its radial and vertical variations through the disk. Finally, the dust-to-gas ratio may also vary with radius.

In all cases, high angular resolution is required to derive the surface density profile because the typical size of disks range from 100 AU to 1000 AU.
Attempts have also been made in the optical, using scattered light images \citep{Burrows+etal_1996}, but they are hampered by the need to extrapolate the density structure from the upper layers to the disk mid-plane. Other methods include silhouette disks against the bright background of HII regions: \citet{McCaughrean+ODell_1996} showed that steep edges (power law exponent $\sim -4.5$, or exponential taper) were needed to reproduce the ``proplyds'' in Orion, but this cannot be extrapolated inward because of the high opacities.

The mm domain is better suited to sample the bulk of the disk. However, the high angular resolution required, at least better than $1''$, implies the use of large mm/submm interferometers. For the dust emission, early attempts include the 3 mm study of \cite{Dutrey+etal_1996} with the IRAM array, the 2 mm survey of \citet{Kitamura+etal_2002} using NRO, and more recently the 1.3 -- 0.8 mm study performed by \citet{Andrews+Williams_2007} with the SMA.
These studies were interpreted in a simplified framework of truncated power laws for the surface densities.

High-resolution studies for the gas are even more difficult. Using the same simplified model,
the CO outer radius is in general found to be much larger than the dust-derived outer radius
\citep[e.g.][]{Dutrey+etal_1998,Simon+etal_2000,Isella+etal_2006}. This is confirmed through CO isotopologue studies
in several sources, such as AB\,Aur \citep{Pietu+etal_2005},  DM\,Tau, LkCa\,15, and MWC\,480 \citep{Pietu+etal_2007}. Although this may be interpreted as changing dust properties with radius, \citet{Hughes+etal_2008} suggested this could also be caused by a different surface density distribution, with an exponentially tapered fall-off of the density with radius. At the resolution of their observations, $\simeq 1''$, the truncated power law and the softened-edge version are indistinguishable.

A similar approach has been used by \citet{Isella+etal_2009} to interpret a $\simeq 0.7''$ resolution 1.3 mm survey with CARMA, and by \citet{Andrews+etal_2009} for SMA observations at 0.8 mm.
%

All these analysis were based on single frequency imaging, although the overall SED is often used to provide additional constraints on the disk parameters. For thermal emission, the only observable is the brightness distribution of the dust at frequency $\nu$
\begin{eqnarray}
 T_b(\nu,r) & = & (1-e^{-\tau(\nu,r)}) J_\nu(\nu,T_d(r)) \\
  &   = &  (1-e^{-\kappa(\nu,r) \Sigma(r)}) J_\nu(\nu,T_d(r)) ,
  \label{eq:tb}
  \end{eqnarray}
where $J_\nu(\nu,T)$ is the Planck function. At least, measurements at three different frequencies are required to  independently constrain $\Sigma(r)$, $T_d(r)$ and  $\kappa(\nu,r)$. In the mm domain, the dust is mostly optically thin and the Rayleigh-Jeans approximation
valid in many cases,
\begin{equation}
  T_b(\nu,r) \approx \kappa(\nu,r) \Sigma(r) T_d(r) .
  \label{eq:tbrj}
\end{equation}
To first order, this allows the separation of the evolution of
$\kappa(\nu,r)$ from that of $\Sigma(r) T(r)$ with measurements at two frequencies, only.
Resolved images are needed at both wavelengths to remove the degeneracy between an optically thick core and
possible radial variations of the spectral index $\beta$. Recently, \citet{Isella+etal_2010} reported a first resolved dual-frequency study of RY\,Tau and DG\,Tau, while \citet{Banzatti+etal_2010} published a resolved multi-frequency study of CQ\,Tau.

In this paper we report on
a high angular resolution ($0.4$ to $1''$), dual-frequency survey of $\sim 20$ of circumstellar disks located in the Taurus-Auriga complex, 8 of which have sub-arcsecond angular resolution at both 2.7 and 1.3 mm.
Observations are described in Section \ref{sec:obs}. Section \ref{sec:model} presents the disk models that we used and the analysis we performed using our specifically developed method. In Section \ref{sec:results} describes
the results of this analysis. The consequences and interpretations are presented in  Section \ref{sec:discussion}. We then conclude in Section \ref{sec:conc}.

\section{Sample, observations and data reduction}
\label{sec:obs}

\begin{table*}
\caption{Stellar properties of the sources in the sample}
\label{tab:stars}\small
\begin{tabular}{lccccccccc}
 \hline \hline
 \noalign{\smallskip} Star & Sp. type & $T_{\rm eff}$ & $A_{\rm V}$ & $L_{\rm star}/L_\odot$ & $M_{\rm star}/M_\odot$& $\log t \pm \sigma_{\log t}$ ($t$ in yr) & Ref.$^{3}$ & \\
 \noalign{\smallskip}\hline
 \noalign{\smallskip}
 BP Tau & K7 & 4060 & 0.49 & $\rm 0.65^{+0.13}_{-0.1}$ & 0.78 $\pm$ 0.08 & 6.51 $\pm$ 0.12 & 1 \\

 CI Tau & K7 & 4060 & 1.77 & $\rm 0.96^{+1.36}_{-0.34}$ & 0.76 $\pm$ 0.08 & 6.27 $\pm$ 0.28 & 1 \\
 CQ Tau & A8/F2  & 7200  & ..     &    12$^{+4}_{-4}$  & 1.7 & 6.7 & 2      \\

 CY Tau & M1 & 3720 & 0.1 & 0.4 $^{+0.09}_{-0.07}$ &  0.48 $\pm$ 0.05 & 6.37 $\pm$ 0.11 & 1 \\

 \textbf{DG Tau} & K7-M0 & 4000 & ... &  6.36 &  0.7 & $5.45 \pm 0.15$ & 3 \\

 DL Tau & K7 & 4060 & ... &  1.12 & 0.7 & $6.23 \pm 0.15$ & 3 \\

 DM Tau & M1 & 3720 & 0 & $\rm 0.16^{+0.24}_{-0.07}$ & 0.47 $\pm$ 0.06 & 6.87 $\pm$ 0.34 & 1 \\

 \textit{DQ Tau} & M0 & 3850 & 0.97 & 0.91 & 0.55 & $6.00 \pm 0.15$ & 3 \\

 FT Tau & C  &  $< 5000$    &     & 0.38 & $[0.7,1.0]$ & $> 6.0 $ & 3, this work  \\

 GM Aur & K3 & 4730 & 0.14 & $\rm 1.23^{+1.07}_{-0.47}$ & 1.37 $\pm$ 0.17 & 6.87 $\pm$ 0.23 & 1 \\
 LkCa 15 & K5 & 4350 & 0.62 & $\rm 0.85^{+0.3}_{-0.2}$ & 1.12 $\pm$ 0.08 & 6.7 $\pm$ 0.16 & 1 \\
 MWC 480 & A4 & 8460 &  & 11.5 &  1.8 & 6.7 & 4 \\

 MWC 758 & A3 &      &  & 11 & 1.8 & 6.7 & 2 \\

\textit{UZ Tau E} & M1 & 3720 & 1.49 & $> 0.88$ &  $0.5$ & $6.20 \pm 0.15$ & 3 \\

\textit{UZ Tau W} & M3 & 3470 & 0.83 & $> 0.38$ &  $0.35$ & $6.20 \pm 0.15$ & 3 \\

 \textbf{HL Tau} & K7 & 4060 & ... & 6.60 & 0.7 & $5.45 \pm 0.15$ & 3 \\

 \textbf{\textit{HH 30}} &   M2-M3 & 3500 & ... & $[0.2,0.9]$ &  0.25 & $ 6.2 \pm 0.2$ & 5\\

 \textbf{DG Tau b}  &  &     &   &  $>0.02$ & 3 \\

 \textit{T Tau N}  & K0 & 5250 & 1.39 & 15.5 &  1.9 & $ 6.70 \pm 0.15$ & 3 \\

 \textit{Haro 6-10 a/b} & K3 & 4800 &  & $[1.8,3.3]$ &  1.5-1.8 & $ 6.3 \pm 0.2$ & 6 \\
 Haro 6-13 & M0 & 3850 & .. & 2.1 &  0.55 & $5.70 \pm 0.15$ & 7 \\
 Haro 6-33 & M0.5 & 3850 & .. & 0.76 &  0.55 & $6.17 \pm 0.15$ & 7 \\

 \hline
 \end{tabular}
 \vspace{0.1cm}\\
{References for observational properties: (1) \cite{Bertout+etal_2007};
  (2) \cite{Chapillon+etal_2008}; (3) \cite{Kenyon+Hartmann_1995}; (4)
   \cite{Pietu+etal_2007}; (5) \cite{Guilloteau+etal_2008,Pety+etal_2006}; (6) \cite{Prato+etal_2009}; (7)   \cite{Schaefer+etal_2009}.
   Note that ages have been derived homogeneously, using the \cite{Siess+etal_2000} tracks, but do not necessarily correspond to values cited in the other papers.}
\end{table*}

Table \ref{tab:stars} indicates the properties of the sources in the sample.
The sample contains classical T Tauri stars or late-type HAe stars, single or multiples (in italics), and a few embedded sources with optical jets and molecular outflows like DG Tau, DG Tau-b, HL Tau, and HH 30 (in boldface). Properties were obtained from the quoted literature. For homogeneity, all ages
were derived using the \citet{Siess+etal_2000} evolutionary tracks, directly from the work of  \citet{Bertout+etal_2007} when available, or re-derived using the cited estimates of luminosity and spectral types. These stellar ages tend to be somewhat higher (factor 1.5) than derived from other evolutionary tracks \citep{Dantona+Mazzitelli_1997,Palla+Stahler_1999}, although even higher ages can be obtained using  the \citet{Baraffe+etal_1998} tracks.  Note that the evolutionary tracks remain ill constrained, and no available set reproduces the constraints derived from the kinematic masses, see \citet{Simon+etal_2000} and the small corrections brought by more accurate measurements of \citet{Pietu+etal_2007} and \citet{Dutrey+etal_2008}. However, all evolutionary tracks produce a similar ordering of the ages, at least in the 0.5 - 1.5 \Msun\ range of masses, which dominate our sample. Because the DG\,Tau-b luminosity is unknown, its age is completely uncertain. Since it still displays an active molecular outflow, we have tentatively assumed it to be 1 Myr old, but with large uncertainties. For \object{GM Aur}, the mass derived by \citet{Bertout+etal_2007} is somewhat larger than the kinematically derived value $1.00 \pm 0.05\Msun$ from \citet{Dutrey+etal_2008}. Accordingly, its age may be overestimated by about 50\%.

Part of the survey was made by simultaneously  observing at 2.7 or 3.4 mm and 1.3 mm in the winter seasons between Nov 1995 and Oct 1998 using the dual frequency receivers on Plateau de Bure (see \citet{Simon+etal_2000} for a description of these observations). Sources were observed in track-sharing mode, typically six to eight at a time. In all cases, the intensity scale was calibrated by using MWC\,349 as flux calibrator. This method ensures an homogeneous calibration across the sample, specially for the spectral index determination as MWC\,349 has a precisely characterized spectral index of 0.6.

Additional high angular resolution with 750 m baselines data was collected from \citet{Pietu+etal_2006} for MWC 480 and LkCa 15, simultaneously at 110 and 220 GHz. For HH\,30 we used the data from  \citet{Guilloteau+etal_2008}.

Higher angular observations (baselines up to 760 m) were also obtained in Feb 2007 at 1.3 mm, and Feb 2008 at 2.7 mm, again in track-sharing mode among six to eight sources, with the new dual-polarization, single frequency receivers. MWC\,349 served as flux calibrator, but in addition MWC\,480 was used as an internal flux-scale consistency check, because it is compact, bright enough and independently measured.

The main survey reaches angular resolution of $0.5\times 0.3''$ at 1.3 mm and a factor of 2 lower at 2.7 mm. Phase stability was good
during the main survey observations: most observations are noise-limited, rather than dynamic-range limited. Dynamic range only limits
the brightest sources HL\,Tau, T\,Tau (which were observed only during the first period) and, to a lesser extent DG\,Tau and MWC\,480, for which the effective noise is twice the thermal
noise.

Some sources also have 2.7\,mm data from previous studies \citep{Dutrey+etal_1996}.
In addition, more limited angular resolution data from
\citet{Schaefer+etal_2009} for \object{Haro 6-13} and \object{Haro 6-33} ($1.2\times 0.7''$ resolution) and \citet{Chapillon+etal_2008} for \object{MWC 758} and \object{CQ Tau} (about $1.3''$ resolution) are also included for completeness.

\begin{table*}
\caption{Derived positions, beam sizes, and proper motions}
\label{tab:position}
\begin{tabular}{l|cccc|cc|cc|cc}
\hline
(1) & (2) & (3) & (4) & (5) & (6) & (7) & (8) & (9) & (10) & (11) \\
Source & R.A. & Dec. & Beam Size & PA & $\mu_a$ &  $\mu_b$ &  $\mu_a$ &  $\mu_b$  \\
     & \multicolumn{2}{c}{J2000.0} & (arcsec) & ($^\circ$) & \multicolumn{2}{c|}{measured, mas/yr}  & \multicolumn{2}{c}{adopted, mas/yr} & \multicolumn{2}{c}{Ducourant et al, mas/yr}  \\
\hline
BP\,Tau& 04:19:15.834&  29:06:26.98 & $(  0.71\times  0.49 )$ &  16. & $9.4 \pm 1.0$ & $-31.9 \pm 1.0$ & 8 & -30  & $6 \pm 2$ & $-29 \pm 2$ \\
CI\,Tau& 04:33:52.014&  22:50:30.06 & $(  0.53\times  0.30 )$ &  25. & $13.4 \pm 2.0$ & $-14 .0\pm 2.0 $ & 12 & -14 & $10 \pm 6$ & $-16 \pm 6$\\
CQ\,Tau& 05:35:58.481&  24:44:54.14 & $(  1.60\times  1.58 )$ &  38. & &  & 0 & -24  & $0 \pm 2$ & $-24 \pm 2$\\
CY\,Tau& 04:17:33.729&  28:20:46.86 & $(  0.56\times  0.30 )$ &  15. & $14.1 \pm 1.0$ & $-25.7 \pm 1.0$ & 12 & -25 & $12 \pm 2$ & $-24 \pm 2$ \\
DG\,Tau& 04:27:04.694&  26:06:16.10 & $(  0.66\times  0.36 )$ &  16. & $8.7 \pm 0.5$ & $-16.7 \pm 0.5$ &  10 & -15 & $3\pm2$ & $-24 \pm 2$ \\
DL\,Tau& 04:33:39.077&  25:20:38.10 & $(  0.61\times  0.36 )$ &  25. & $13.7 \pm 1.0$ & $-14.7 \pm 1.0$ & 14 & -14 & 7 & -22 (a) \\ 
DM\,Tau& 04:33:48.736&  18:10:09.99 & $(  0.65\times  0.30 )$ &  18. & $16.7 \pm 1.0$ & $-14.2 \pm 1.5$ & 14 & -16 & $11\pm7$ & $-19 \pm 7$\\
DQ\,Tau& 04:46:53.064&  17:00:00.09 & $(  0.57\times  0.29 )$ &  24. & $1 \pm 3$  & $-5 \pm 3$  & 0 & -6 & $0 \pm 7$ & $-6 \pm 7$\\
FT\,Tau& 04:23:39.188&  24:56:14.28 & $(  0.52\times  0.29 )$ &  25. & $12.8 \pm 1.5 $ & $-19.1 \pm 1.5$ & 16 & -21 & 11  & -19 (a)\\
GM\,Aur& 04:55:10.985&  30:21:59.43 & $(  1.15\times  1.02 )$ &  59. & $12.4 \pm 1.3 $ & $-4.7 \pm 1.3$ & 11 & -6 & $3\pm 6$ & $-26 \pm 6$ \\
LkCa\,15& 04:39:17.794&  22:21:03.43 & $(  0.70\times  0.46 )$ &  26. & $24 \pm 2$ & $-18 \pm 2$ &  8 & -15 &
$8 \pm2$ & $-15 \pm 2$ \\
MWC\,480& 04:58:46.265&  29:50:36.98 & $(  1.01\times  0.57 )$ &  10. & $5.4 \pm 0.6$ & $-23.6 \pm 0.8$  & 6 & -23 & $6.2 \pm 1.3$ & $-23.8 \pm 0.8$ (b) \\
MWC\,758& 05:30:27.538&  25:19:57.26 & $(  1.52\times  1.31 )$ &  13. & & & 6 & -26 & $5.2 \pm 1.4$ & $-26.0 \pm 0.6$ (b) \\
UZ\,Tau\,E& 04:32:43.071&  25:52:31.07 & $(  1.15\times  0.70 )$ &  14. & $1 \pm 6$ & $-38 \pm 6$ &  2 & -26
& $2 \pm6$ & $-26 \pm 6$ \\
UZ\,Tau\,W& 04:32:42.808&  25:52:31.39 & $(  1.16\times  0.71 )$ &  13. & & & & \\
HL\,Tau& 04:31:38.413&  18:13:57.55 & $(  0.94\times  0.54 )$ &  17. & & & 14 & -20 & $8 \pm 6$ & $-22 \pm 6$ (c) \\
HH\,30& 04:31:37.468&  18:12:24.21 & $(  0.60\times  0.32 )$ &  22. & $ 9 \pm 4$ & $-8 \pm 4$ & \textit{8} & \textit{-12} \\
DG\,Tau\,b& 04:27:02.562&  26:05:30.50 & $(  0.54\times  0.26 )$ &  16.& & & 10 & -15 \\
T\,Tau & 04:21:59.435&  19:32:06.36 & $(  1.13\times  0.86 )$ &  41. &  & & 14 & -12 & $14 \pm 2$ & $-12 \pm 2$\\
Haro\,6-10\,N& 04:29:23.729&  24:33:01.52 & $(  0.89\times  0.56 )$ &  19. & & & 10 & -20 \\
Haro\,6-10\,S& 04:29:23.736&  24:33:00.26 & $(  0.89\times  0.56 )$ &  19. & & & & \\
Haro\,6-13& 04:32:15.419&  24:28:59.49 & $(  1.21\times  0.77 )$ &  39. & & & [0] & [0] & $5 \pm 7$ & $-21 \pm 7$ \\
Haro\,6-33& 04:41:38.825&  25:56:26.77 & $(  1.16\times  0.77 )$ &  41. & & & 10 & -20 \\
\hline
\end{tabular}\vspace{0.1cm} \\
{All positions refer to Epoch J2000. Col 6-7 indicate the proper motions derived from our data
($\mu_a = \mu_\alpha cos(\delta)$ and $\mu_b = \mu_\delta$). Col 8-9 indicate the values adopted in the analysis,
in general a weighted average of our measurements and those of \citet{Ducourant+etal_2005}, except for
(a) data from \citet{Itoh+etal_2008}, (b) from the Hipparcos catalog \citet{Perryman+etal_1997}, (c) from \citet{Zacharias+etal_2003}. For HH\,30, the adopted value was discussed in \citet{Guilloteau+etal_2008}.}
\end{table*}

\begin{figure*} 
   \includegraphics[angle=0, width=16.0cm]{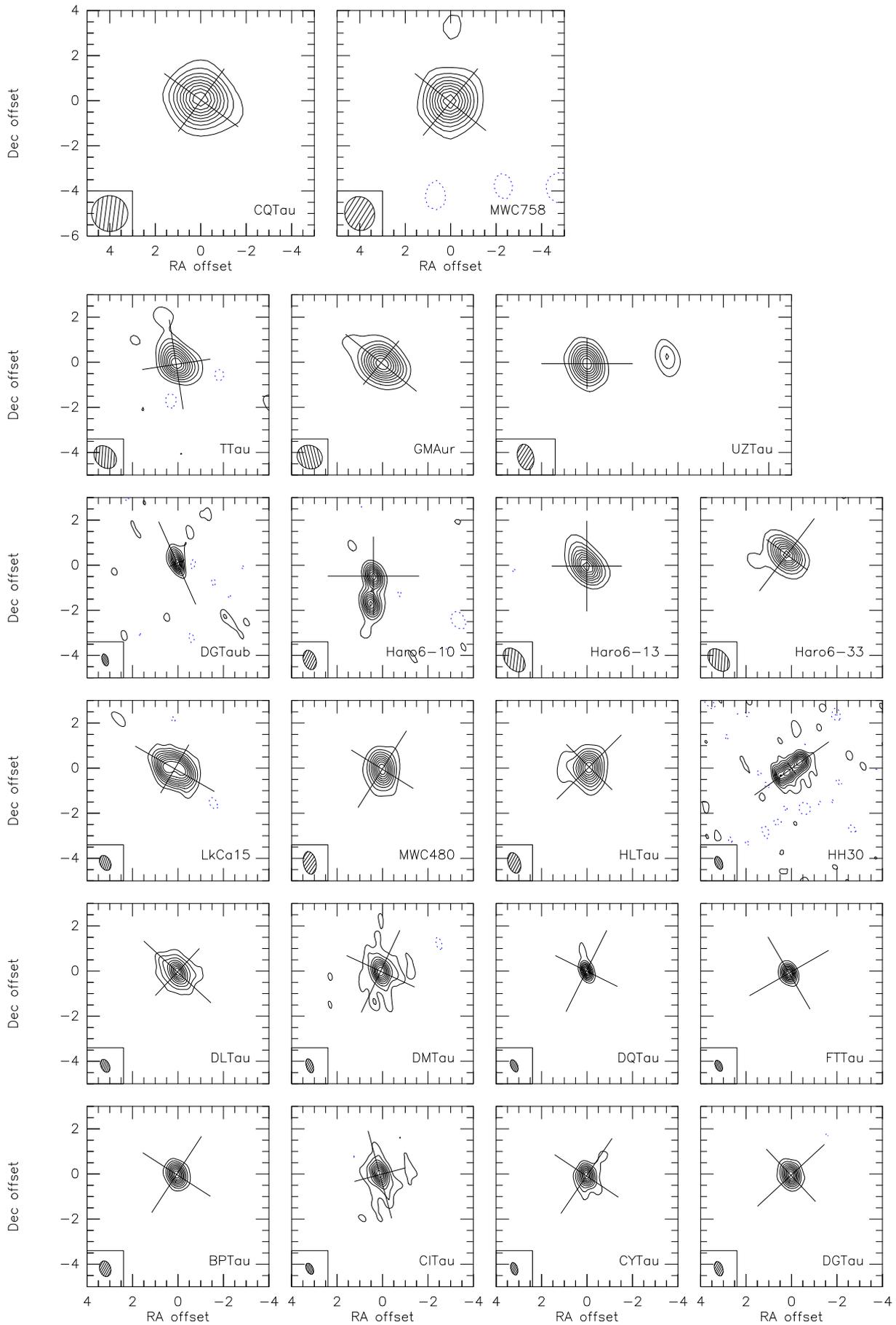}
 \caption{High angular resolution image of the continuum emission
 from the sources observed in the survey at 1.3 mm (230 GHz).
 The contours are relative to the peak intensity, in steps of 10 \%.
 Coordinates are offsets in arcseconds from the reference positions given in Table \ref{tab:position}
 }
 \label{fig:relative1}
 \end{figure*}

\begin{figure*} 
   \includegraphics[angle=0, width=16.0cm]{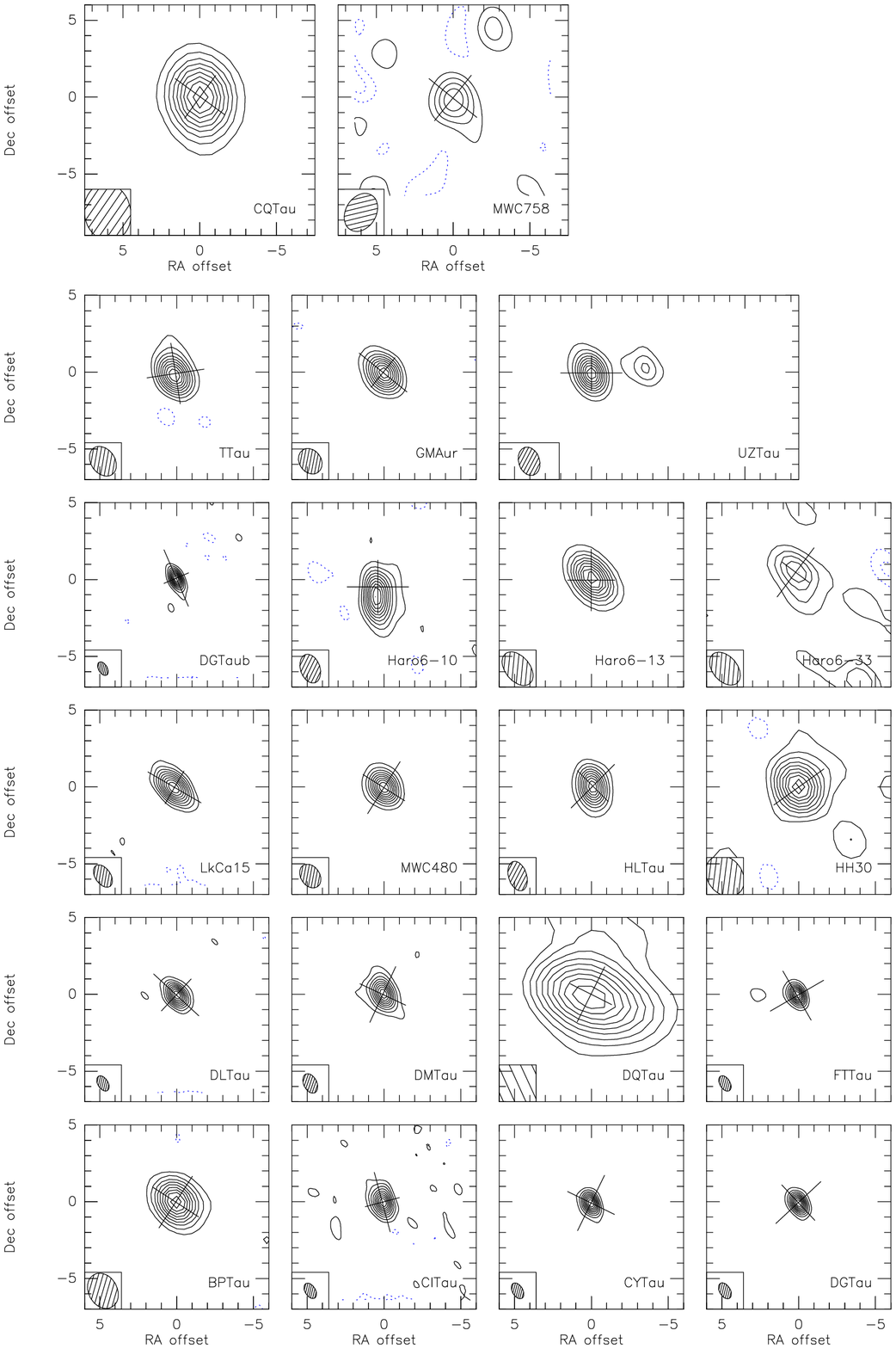}
 \caption{High angular resolution image of the continuum emission
 from the sources observed in the survey at 2.7 or 3.4 mm.
 The contours are relative to the peak intensity, in steps of 10\%, except for the
 weakest sources (Haro 6-33 \& MWC 758) for which the step is 20 \%.
 }
 \label{fig:relative3}
 \end{figure*}

Table \ref{tab:position} indicates the resulting beam sizes for each source at 230 GHz.
The positions indicated in Table \ref{tab:position} are those determined from this study, and are the reference positions for
Fig.\ref{fig:relative1}, \ref{fig:relative3}, \ref{fig:alldmtau},  \ref{fig:uztau}, \ref{fig:fttau} and \ref{fig:allbptau}-\ref{fig:alluztauw}.
 Because the data span more than 12 years of time, correction for the star proper motions is important. The proper motions were taken from the \citet{Ducourant+etal_2005} catalog when available, or determined from our own measurement, as the astrometric accuracy of the Plateau de Bure is high enough to allow measurements to about 2 mas/yr in each direction over a 10 year span when sufficient signal-to-noise is available. The positions are in the J2000.0 system and referred to epoch 2000.0 after correction for proper motion. The positional accuracy is better than 0.05$''$.

\begin{table*}
\begin{tabular}{l|ccc|cccc}
\hline
 (1) & (2) & (3) & (4) & (5) & (6) & (7) & (8) \\
Source &  Major & Minor & PA & 1.3 mm Flux & 2.7 mm Flux & $\alpha$ & $\alpha_{100}$ \\
     &  ($''$)  & ($''$)  & ($^\circ$) & mJy & mJy &  & \\
\hline
BP\,Tau (*) &      0.50$\pm$      0.01&      0.34$\pm$      0.01&       10.$\pm$        2.&      58.2$\pm$       1.3&       4.2$\pm$       0.2&      2.73$\pm$      0.07&      2.39$\pm$      0.06\\
CI\,Tau&      0.74$\pm$      0.01&      0.47$\pm$      0.01&       14.$\pm$        1.&     125.3$\pm$       6.2&      19.0$\pm$       0.8&      2.58$\pm$      0.13&      1.72$\pm$      0.06\\
CQ\,Tau (*) &      0.86$\pm$      0.04&      0.63$\pm$      0.04&       31.$\pm$        8.&     162.4$\pm$       2.6&      13.3$\pm$       0.5&      2.60$\pm$      0.06&      2.60$\pm$      0.05\\
CY\,Tau&      0.55$\pm$      0.01&      0.47$\pm$      0.01&      -15.$\pm$        4.&     111.1$\pm$       2.9&      23.4$\pm$       0.7&      2.13$\pm$      0.08&      1.86$\pm$      0.05\\
DG\,Tau&      0.56$\pm$      0.01&      0.46$\pm$      0.01&       -1.$\pm$        3.&     389.9$\pm$       4.6&      59.5$\pm$       0.9&      2.57$\pm$      0.04&      2.48$\pm$      0.03\\
DL\,Tau&      0.71$\pm$      0.01&      0.56$\pm$      0.01&       29.$\pm$        2.&     204.4$\pm$       1.9&      27.3$\pm$       1.0&      2.75$\pm$      0.06&      1.86$\pm$      0.04\\
DM\,Tau&      0.50$\pm$      0.01&      0.45$\pm$      0.01&      -36.$\pm$        9.&     108.5$\pm$       2.4&      15.6$\pm$       0.4&      2.65$\pm$      0.07&      1.78$\pm$      0.05\\
DQ\,Tau (*) &      0.24$\pm$      0.01&      0.17$\pm$      0.01&      -24.$\pm$        6.&      83.1$\pm$       2.8&       9.6$\pm$       0.8&      2.24$\pm$      0.12&      1.69$\pm$      0.10\\
FT\,Tau&      0.43$\pm$      0.01&      0.40$\pm$      0.01&      -59.$\pm$        8.&      72.5$\pm$       3.9&      18.8$\pm$       0.8&      1.85$\pm$      0.13&      1.65$\pm$      0.04\\
GM\,Aur&      1.05$\pm$      0.05&      0.57$\pm$      0.05&       57.$\pm$        4.&     175.8$\pm$       5.3&      23.7$\pm$       0.8&      2.74$\pm$      0.09&      2.74$\pm$      0.06\\
LkCa\,15&      1.20$\pm$      0.04&      0.91$\pm$      0.04&       65.$\pm$        6.&     109.6$\pm$       2.0&      17.4$\pm$       0.6&      2.52$\pm$      0.07&      2.49$\pm$      0.05\\
MWC\,480&      0.67$\pm$      0.01&      0.55$\pm$      0.01&       22.$\pm$        3.&     289.3$\pm$       2.5&      35.8$\pm$       0.5&      2.86$\pm$      0.03&      2.76$\pm$      0.02\\
MWC\,758&      1.00$\pm$      0.09&      0.82$\pm$      0.10&      -12.$\pm$       22.&      54.8$\pm$       2.0&       7.3$\pm$       1.4&      2.76$\pm$      0.31&      2.77$\pm$      0.30\\
UZ\,Tau\,E&      0.75$\pm$      0.01&      0.45$\pm$      0.01&      -89.$\pm$        2.&     149.9$\pm$       1.4&      22.9$\pm$       0.6&      2.57$\pm$      0.05&      2.58$\pm$      0.04\\
UZ\,Tau\,W&      0.40$\pm$      0.04&      0.33$\pm$      0.03&      -35.$\pm$       24.&      34.3$\pm$       1.3&       6.4$\pm$       0.6&      2.30$\pm$      0.18&      2.29$\pm$      0.14\\
HL\,Tau&      0.87$\pm$      0.01&      0.64$\pm$      0.01&      -45.$\pm$        2.&     818.8$\pm$      10.8&      94.1$\pm$       0.9&      2.96$\pm$      0.03&      2.90$\pm$      0.02\\
HH\,30&      1.43$\pm$      0.02&      0.22$\pm$      0.03&      -55.$\pm$        0.&      19.8$\pm$       0.8&       3.8$\pm$       0.2&      2.26$\pm$      0.13&      2.31$\pm$      0.12\\
DG\,Tau\,b&      0.69$\pm$      0.03&      0.34$\pm$      0.02&       26.$\pm$        2.&     531.4$\pm$       0.0&      83.6$\pm$      12.4&      2.53$\pm$      0.20&      2.02$\pm$      0.09\\
T\,Tau&      0.48$\pm$      0.05&      0.34$\pm$      0.06&        4.$\pm$       17.&     199.7$\pm$       6.2&      48.8$\pm$       1.0&      1.93$\pm$      0.07&      1.97$\pm$      0.05\\
Haro\,6-10\,N&      0.24$\pm$      0.11&      0.09$\pm$      0.06&       53.$\pm$       18.&      43.8$\pm$       3.1&      10.5$\pm$       0.7&      1.95$\pm$      0.19&      1.96$\pm$      0.14\\
Haro\,6-10\,S&      0.37$\pm$      0.05&      0.11$\pm$      0.07&       -2.$\pm$        8.&      46.7$\pm$       3.2&       9.1$\pm$       0.7&      2.24$\pm$      0.20&      2.12$\pm$      0.14\\
Haro\,6-13&      0.52$\pm$      0.03&      0.36$\pm$      0.04&       -1.$\pm$       10.&     113.5$\pm$       4.0&      31.3$\pm$       1.0&      1.76$\pm$      0.09&      1.76$\pm$      0.07\\
Haro\,6-33&      0.57$\pm$      0.11&      0.45$\pm$      0.07&       31.$\pm$       28.&      34.2$\pm$       3.1&       8.0$\pm$       1.0&      1.99$\pm$      0.30&      1.65$\pm$      0.24\\
\hline
\end{tabular}
\label{tab:flux}
\caption{Apparent sizes and orientations  derived from a Gaussian fit (Col 2-4)
to the 1.3 mm data in the $uv$ plane for baselines longer than 100 m.
Total flux at 1.3 and 2.7 mm  (or 3.4 mm for stars with (*) in Col 1) (Col 5-6), and apparent spectral index $\alpha$ (Col 7) are derived from Gaussian fit to all visibilities.  $\alpha_{100}$ (Col 8) is the apparent
spectral index for baselines longer than 100 m.}
\end{table*}

Figure \ref{fig:relative1} is a montage of the 1.3\,mm images of the survey sources, presented in terms of fraction of the peak flux. Figure \ref{fig:relative3} is as Fig.\ref{fig:relative1}, but for 2.7 or 3.4\,mm, depending on the sources. Robust weighting was used to produce these images. Despite the fairly wide range of angular resolutions (from $0.5\times0.3''$ to about $1.5''$), clearly some objects are much more centrally condensed than others. In particular, the most compact sources are the two circumstellar disks in the Haro 6-10 binary.

\section{Modeling}

\label{sec:model}

\subsection{Simple analysis}
\label{sec:model:simple}

The measured flux densities at 1.3 mm and around 3 mm are given in Table \ref{tab:flux} (considering only baselines shorter than 100 m). They result from a simple elliptical Gaussian fit to the $uv$ data. For the orientations and apparent sizes, all baselines were included. Short baseline data, although adequate to measure the overall flux densities and apparent spectral index $\alpha$,
are not suitable to derive characteristic sizes and even position angles. This is because, to first order, disks have power law distributions of the surface density and temperature
and are optically thin at such wavelengths. Thus, when seen at low inclination, ($< 45^\circ$ or so), the surface brightness is a power law of the radius and has no characteristic size. This can bias the apparent position angle, since the apparent half-power size only depends on the angular resolution and the exponent of the power law.
For nearly edge-on disks ($i > 75^\circ$), the disk thickness introduces a characteristic size, because the brightness falls off like a Gaussian in this direction, so the position angle is properly recovered.  Thus, in general, reliable position angles can only be derived with sufficient angular resolution, i.e. from long baseline fits. These properties can explain the different position angles found by previous authors using lower resolution data \cite[e.g.][]{Dutrey+etal_1996,Kitamura+etal_2002}.
Note that these biases on the position angles can also affect analysis made with more elaborate disk models: only sufficiently high angular resolution can provide an unbiased determination of this parameter.

On the other hand, for sources with an apparent core-halo structure, such as \object{DM Tau} or \object{CI Tau}, the long baseline fit tends to represent only the central part and misses substantial flux. The spectral index $\alpha_{100}$ derived from long baseline data (Table~\ref{tab:flux}, Col.8) is systematically smaller than that from the short baseline fit only (Col.7). This indicates either a contribution of an optically thick core and/or dust grain evolution.

\begin{figure}[!ht] 
   \includegraphics[width=7.0cm]{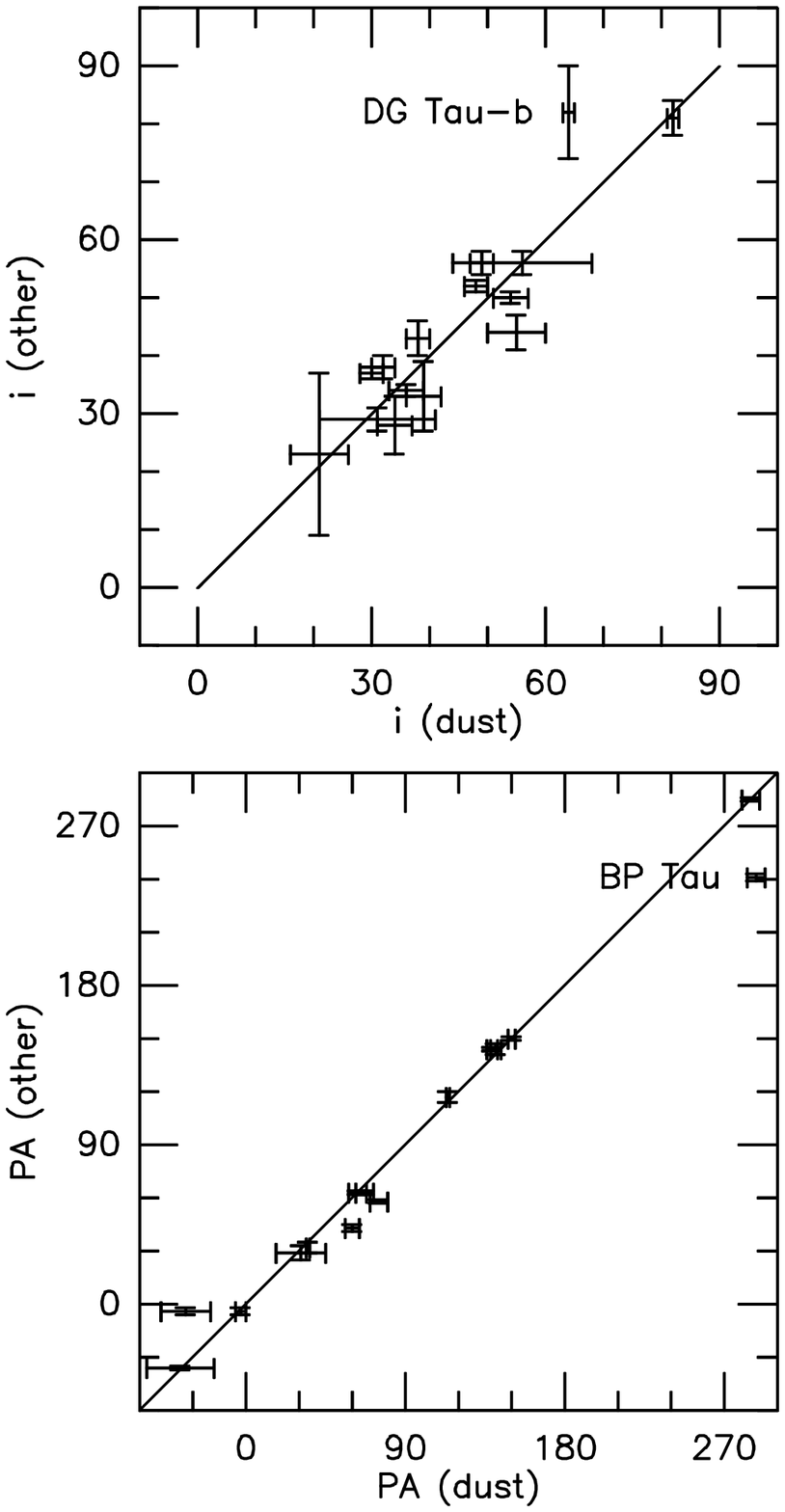}
\caption{Top: disk inclinations measured from dust and other methods (CO or jets).
Bottom: Position angle of the disk rotation axis derived from dust and other methods.
The only discrepant points are for BP\,Tau (orientation) and DG\,Tau-b (inclination).}
   \label{fig:geometry}
\end{figure}

\subsection{Model description}
\label{sec:model:descr}

 Because the apparent size, orientation, and spectral index may depend on the $uv$ coverage when using a simple Gaussian model, we must analyze the data with more realistic brightness distributions.
Because a direct inversion of the brightness profile is impossible, due to the combination of insufficient resolution and the limited signal-to-noise, a global fitting technique using some a-priori model
must be used.
We therefore analyzed the continuum emission in terms of two ``standard'' disk models that differ only in the surface density distribution. Model 1 uses a simple truncated power law, Model 2 an exponentially tapered power law with an arbitrarily large outer radius. The  surface density is characterized by three parameters plus an inner radius in each model. Our approach is to keep the model parametric and simple to avoid as much as possible biases towards a specific physical model for disks.

In Model 1, the surface density is a simple power law with a sharp inner and outer radius:
\begin{equation}\label{eq:power}
    \Sigma(r) = \Sigma_{0} \left(\frac{r}{r_0}\right)^{-p} ,
\end{equation}
for $R_\mathrm{int} < r < R_\mathrm{out}$.

In Model 2, the density is tapered by an exponential edge:
\begin{equation}\label{eq:edge}
\Sigma(r) = \Sigma_0 \left(\frac{r}{R_0}\right)^{-\gamma}  \exp\left(-(r/R_c)^{2-\gamma}\right) .
\end{equation}
Note that Model 1 derives from Model 2 by simply setting $R_c \rightarrow \infty$ and $p=\gamma$ in the above parametrization.
Model 2 is a solution of the self-similar evolution of a viscous disk in which the viscosity is a power law of the radius (with constant exponent in time $\gamma$).

With the inner ($R_\mathrm{int}$) and outer ($R_\mathrm{out}$) radii, the disk mass is given by
\begin{equation}
\label{eq:mass}
M_d = \frac{2 \pi R_0^2 \Sigma_0}{2-\gamma} \left( \frac{R_c}{R_0}\right)^{2-\gamma} \left( \exp \left(-\left(\frac{R_\mathrm{int}}{R_c}\right)^{2-\gamma}\right) - \exp\left(-\left(\frac{R_\mathrm{out}}{R_c}\right)^{2-\gamma}\right) \right) ,
\end{equation}
which for small $R_\mathrm{int}$ and large $R_\mathrm{out}$ yields
\begin{equation}
\label{eq:mass2}
M_d = \frac{2 \pi R_0^2 \Sigma_0}{2-\gamma} \left( \frac{R_c}{R_0}\right)^{2-\gamma} .
\end{equation}
The simple power law case is recovered for $R_c \rightarrow \infty$, by developing to first order
in $(r/R_c)^{2-\gamma}$,
\begin{equation}
M_d = \frac{2 \pi R_0^2 \Sigma_0}{2-\gamma}  \left( \left(\frac{R_\mathrm{out}}{R_0}\right)^{2-\gamma} - \left(\frac{R_\mathrm{int}}{R_0}\right)^{2-\gamma} \right) .
\end{equation}
One can also  used $M_d$ as a free parameter instead of $\Sigma_0$, like in \citet{Andrews+etal_2009}.
Eq.\ref{eq:mass} can also be used to show that $R_c$ is the radius which contains 63 \% of the disk mass,
because $M(r<R_c) = M_d (1-1/e) = 0.63 M_d$ provided $R_\mathrm{out}$ is large enough.

An equivalent parametrization is that described by \citet{Isella+etal_2009}
\begin{equation}\label{eq:isella}
\Sigma(r) = \Sigma_t \left(\frac{R_t}{r}\right)^{\gamma} \exp\left(\frac{1-(r/R_t)^{2-\gamma})}{2(2-\gamma)}\right) .
\end{equation}
The parameterizations using $R_t$ or $M_d$ become ill defined for $\gamma = 2$, which makes them less suited for use in a minimization scheme than the simple parametric expression of Eq.\ref{eq:edge} (for which only $R_c$ is unconstrained, as the surface density
becomes a power law). $R_t$ is related to $R_c$ by
\begin{equation}\label{eq:rtrc}
    R_t = R_c \left( \frac{1}{2(2-\gamma)}\right)^\frac{1}{2-\gamma} .
\end{equation}
$R_t/R_c$ is close to 0.5 for all values of $\gamma$ below 1, reaches 1 for $\gamma=1.5$, then diverges for $\gamma \rightarrow 2$.
In the framework of self-similar viscous evolution \citep{LyndenBell+Pringle_1974,Hartmann+etal_1998}, it can be shown that $R_t$ is the radius at which the net mass flux changes sign.

In both models, the temperature is assumed to be a simple power law of the radius
\begin{equation}\label{eq:temperature}
    T(r) = T_0 (r/R_0)^{-q} .
\end{equation}
The disks are thus vertically isothermal.
To allow a homogeneous comparison, we used $T_{100} = T(100~\mathrm{AU}) = 15$ K and $q = 0.4$, except when those parameters can be constrained by the observations. The validity and impact of this assumption will be discussed in Sec.\ref{sec:temperature}.

Similar analyses have been used by \citet{Kitamura+etal_2002} and \citet{Andrews+Williams_2007} for their 2\,mm and 0.8\,mm data respectively. Most previous studies \citep{Kitamura+etal_2002,Andrews+Williams_2007,Isella+etal_2009} used the thin disk approximation to compute
visibilities. Here, because our sample includes highly inclined objects, we assume that the disks are flared, with a scale height varying as a power law of the radius $h(r) = H_{100} (r/100 \mathrm{AU})^{-h}$. For all but the two highly inclined objects (HH\,30 and DG\,Tau-b), we
used  $H_{100} = 16$ AU and $h=-1.25$. These values agree with those derived using the gas temperature  determined from CO observations whenever available, and the stellar mass, either from kinematic determination \citep{Simon+etal_2000} or standard evolutionary tracks. The results are, however, completely independent of the assumed scale height, which justifies \textit{a posteriori} the thin disk approximation used by previous authors. However, for the two highly inclined objects, $H_{100}$ and the exponent $h$ had to be used as adjustable parameters.

The inner radius $R_\mathrm{int}$ is also  not significant in general, except for a few special sources that display inner cavities, such as GM\,Aur, HH\,30 and LkCa\,15 (see Sec.\ref{sec:holes}). We fixed it to 1 AU, but in general, any value lower than about 3-4 AU would not change the results. For Model 2, we used for $R_\mathrm{out}$ the outer radius derived from CO observations when available. If not, we set it to 500 AU. These outer radii are large enough to have negligible influence on the results.

Each model has thus a priori five free intrinsic parameters: two for temperature $T_0$ and $q$,  three for the surface density $\Sigma_0$, $p$ or $\gamma$, and $R_\mathrm{out}$ or $R_c$, plus the inclination, orientation and position.

The dust opacity as a function of wavelength and radius completes the description. In a first step, we assume it to be independent of radius and described by the following  prescription
\begin{equation}\label{eq:kappa}
    \kappa(\nu) = \kappa_{230} (\nu/230 \mathrm{GHz})^{\beta_m} ,
\end{equation}
with $\kappa_{230} = 2~\mathrm{cm}^2{\rm g}^{-1}$ (per gram of dust).
This introduces one additional parameter, the mean dust emissivity index $\beta_m$. This is similar to the \citet{Beckwith+etal_1990} results, but using a different pivot frequency to avoid further dependence of the derived disk mass on $\beta$. The dust model used by  \citet{Andrews+Williams_2007} and \citet{Andrews+etal_2009} also results in $\beta_m=1$, but with a slightly different absorption coefficient $\kappa_{230} = 2.2 ~\mathrm{cm}^2{\rm g}^{-1}$.
Finally, we also assume that the gas-to-dust ratio is constant and equal to 100. In a second step, we shall relax the assumption of constant $\kappa(\nu)$ as a function of radius $r$, see Sec.\ref{sub:radial:dust}.

Appendix \ref{app:fitting} (available on-line only) illustrates some of the possible degeneracy between the various models, in particular between
constant dust properties with an optically thick inner region, and variable dust properties.

\subsection{Fitting method}
\label{sec:model:method}

For the inclination and orientation, we used the accurate determination from the CO kinematics when possible. Values derived from optical observations (scattered light images, or optical jets) or molecular jets were used for some sources for which the disk kinematics is not known. Independent fits of these parameters from the dust emission were also performed to check the consistency of the results: see Table \ref{tab:geometry} and references therein. We stress, however, that the uncertainties on the disk inclination and orientation do not significantly affect the derived radial structure.

At each observed frequency, the radiative transfer equation is solved by a simple ray-tracing algorithm, and model images are generated. Great care has been taken to avoid numerical precision problems caused by finite grid effects.  The numerical integration is typically performed on a 128 x 128 grid, with 512 points along the line of sight. Two oversampling techniques are used to enhance the accuracy while keeping computational costs reasonable. First, the overall image is interpolated (by bilinear interpolation) by a factor 2 before computing the model visibilities. Second, the inner 64 x 64 pixels are re-computed on this finer grid with a smaller step along the line of sight (64 x 64 x 1024). Larger numbers were used for the largest disk. This results in effective pixel sizes of 2 to 7 AU in (x,y), depending on the outer disk radius used in the model, and steps 4 to 8 times smaller along the line of sight.

A modified Levenberg-Marquardt method  was used to derive the disk parameters by a non-linear least squares fit of the modeled visibilities directly to the observed $uv$ data, as detailed by \citet{Pietu+etal_2007}. Like all methods, L-M minimization can be trapped in local minima
when the starting point is too far away from the solution. We alleviate this problem by using multiple re-starts when needed, and also
by adapting the step size used to compute the gradient. We found empirically that using steps equal to half a sigma on each parameter provided stable results.
Error bars were derived from the covariance matrix, except when the parameter coupling was too strong (e.g. between $R_c$ and $\gamma$ in Model 2 for $\gamma$ larger than about 1.5).
In that case, the multi-parameter fit was reduced to a one parameter problem by finding the best fit for several values of this parameter, and determining the error bars from the resulting $\chi^2$ distribution.
Data at several available wavelengths are fitted simultaneously by the same model, which allows us to constrain $\beta_m$. However, whenever data at very nearby frequencies (220 and 230 GHz, for example) exist, only one was considered in this process, because even small absolute calibration error could result in a strong bias on the value of $\beta_m$. In the dual frequency fit, the long wavelength (2.7 or 3.4\,mm) data do not in general influence  the derived surface density law, because of their lower angular resolution, but only serve to determine $\beta_m$.
Because the geometric parameters are largely decoupled from the disk intrinsic parameters, the simultaneous fit of dual-frequency
data sets used (in general) four parameters: $\Sigma_0$, $p$ ($\gamma$ in Model 2), $R_\mathrm{out}$ (or $\gamma$ and $R_c$ in Model 2), and $\beta_m$.  Additional parameters ($T_0,q$ or $H_0,h$) were also fitted simultaneously when needed.
Separate fits were also made at 1.3 and 2.7 mm for the few sources were the angular resolution at 2.7 mm is sufficient, or when data sets at 1.4 mm also existed: in these cases, $\beta_m$ was set at the value found from the dual frequency analysis, and only the three remaining parameters were fitted together.

The choice of the pivot radius $R_0$ in Equations \ref{eq:power}-\ref{eq:edge} is important. There is always an optimal value that  minimizes the error on $\Sigma_0$, which depends on the angular resolution and source surface density profile \citep[see discussion in][]{Pietu+etal_2007}. Using a non-optimal value results in a coupling of $\Sigma_0$ with $p$ for Model 1, and $\gamma,R_c$ in Model 2. Another different pivot radius is also required for $T_0$ when the source is sufficiently optically thick and resolved to constrain $T_0,q$.

Two stars required a specific treatment: the binary Haro 6-10 and the quadruple UZ\,Tau, which have two disks in the field of view. For Haro 6-10, a simple Gaussian model of the emission from the other disk was subtracted before the analysis of each disk. For UZ\,Tau, the procedure was more elaborate. First a Gaussian model of the emission from the companion (UZ\,Tau W) was subtracted, and the remaining emission from UZ\,Tau E was analyzed. Then, the best-fit model of UZ\,Tau E was subtracted from the original data, and the emission from UZ\,Tau W analyzed separately.

\begin{figure}[!ht] 
   \includegraphics[width=7.0cm]{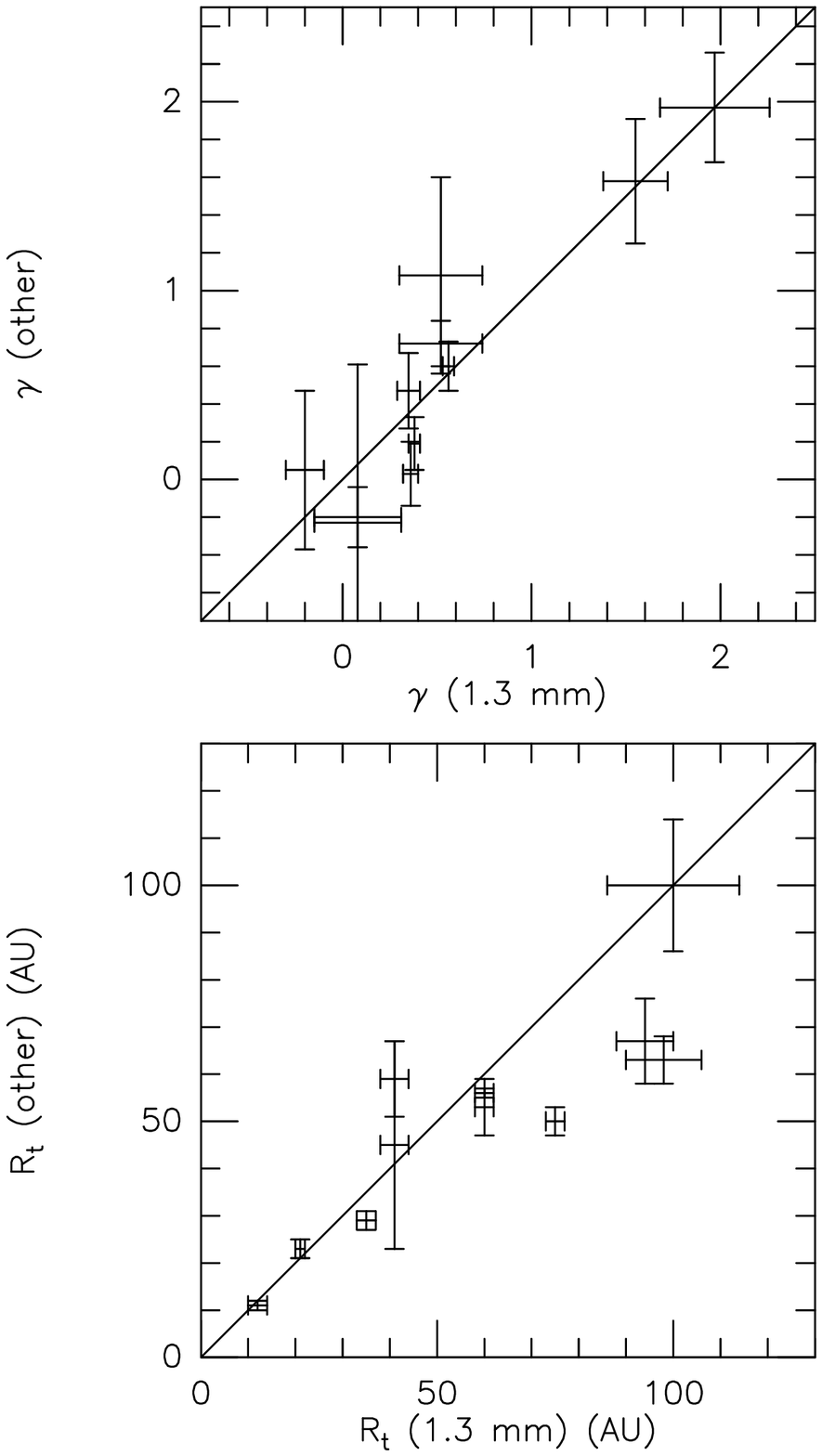}
\caption{Transition radius (bottom) and characteristic exponent $\gamma$ derived from independent data sets.}
   \label{fig:compare}
\end{figure}

\begin{table}[!ht]
\caption{Derived inclinations}
\label{tab:geometry}
\begin{tabular}{l|cc|cc}
\hline \hline
Source &  $PA$ & $i$ &  $PA_\mathrm{CO}$ & $i_\mathrm{CO}$ \\
     & $^\circ$  & $^\circ$  & $^\circ$  & $^\circ$  \\
\hline
BP\,Tau&     107$\pm$     5&      39$\pm$     3&    -119$\pm$     2&      33$\pm$     6\\
CI\,Tau&     285$\pm$     5&      55$\pm$     5&     285$\pm$     1&      44$\pm$     3\\
CQ\,Tau&     -37$\pm$    19&     -31$\pm$    10&     -36$\pm$     1&     -29$\pm$     2\\
CY\,Tau&      63$\pm$     5&      34$\pm$     3&      63$\pm$     1&      28$\pm$     5\\
DG\,Tau&      60$\pm$     4&      32$\pm$     2&      43$\pm$     2&      38$\pm$     2\\
DG\,Tau b&   114$\pm$     1&      64$\pm$     2&     117$\pm$     3&      $> 75$ \\
DL\,Tau&     141$\pm$     3&      38$\pm$     2&     144$\pm$     3&      43$\pm$     3\\
DM\,Tau&      67$\pm$     5&     -36$\pm$     3&      63$\pm$     1&     -34$\pm$     1\\
FT\,Tau&      31$\pm$    14&      21$\pm$     5&      29$\pm$     4&      23$\pm$    14\\
GM\,Aur&     139$\pm$     3&      54$\pm$     3&     144$\pm$     1&      50$\pm$     1\\
HL\,Tau&       42$\pm$    2&      45$\pm$     1&                45 &                 45 \\
LkCa\,15&     150$\pm$     2&      48$\pm$     2&     150$\pm$     1&      52$\pm$     1\\
MWC\,480&      75$\pm$     5&      30$\pm$     2&      58$\pm$     1&      37$\pm$     1\\
MWC\,758&     147$\pm$   292&     -11$\pm$   249&     141$\pm$     1&      18$\pm$    36\\
UZ\,Tau E&      -3$\pm$     3&     131$\pm$     2&      -4$\pm$     2&     124$\pm$     2\\
UZ\,Tau W&     -34$\pm$    14&     124$\pm$    12&      -4$\pm$     2&     124$\pm$     2\\
HH\,30&      35$\pm$     1&      98$\pm$     1&      32$\pm$     3&      99$\pm$     3\\
\hline
\end{tabular}\\
\vspace{0.2cm}\\
{Position angles are those of the disk \textit{rotation} axis.
The inclinations $i_\mathrm{CO}$ have been derived from CO observations except for UZ\,Tau W
(assumed to be equal to that of UZ\,Tau E), DG\,Tau, and DG\,Tau-b, which come from \citet{Eisloffel+Mundt_1998}. For HL\,Tau,
the CO outflow defines $i_\mathrm{CO}$ and $PA_\mathrm{CO}$. Conventions for PA and $i$
use the rotation axis orientation as described by \citet{Pietu+etal_2007}.}
\end{table}

\begin{table*}
\caption{Derived parameters for the viscous and power law models}
\label{tab:main}
\begin{tabular}{l|cccc|c|ccc|c}
\hline
\hline
 (1) & (2) & (3) & (4) & (5) & (6) & (7) & (8) & (9) & (10) \\
Source & Mass & $R_c$ & $R_t$ & $\gamma$ & $\Delta\chi^2$ & $\Sigma_{100}$ & $R_\mathrm{out}$ & p & $\beta_m$ \\
     & $0.001 \Msun$  & AU & AU  &  &  & g.cm$^{-2}$ & AU &  &  \\
\hline
BP Tau&       5.4$\pm$       0.2&       43$\pm$     2&       22$\pm$     1&     -0.04$\pm$      0.12&      14&    3.88$\pm$    0.11&      57$\pm$     1&      0.40$\pm$      0.07&      0.65$\pm$      0.04\\
CI Tau&      37.1$\pm$       2.7&      166$\pm$    10&       81$\pm$     4&      0.30$\pm$      0.04&      33&    2.59$\pm$    0.06&     201$\pm$     4&      0.59$\pm$      0.03&      0.68$\pm$      0.05\\
CQ Tau&       6.3$\pm$       0.4&       86$\pm$     8&       41$\pm$     3&      0.61$\pm$      0.25&       0&    0.49$\pm$    0.03&     188$\pm$    30&      1.35$\pm$      0.15&      0.75$\pm$      0.05\\
CY Tau&      16.5$\pm$       0.6&       67$\pm$     2&       32$\pm$     1&      0.28$\pm$      0.06&      35&    3.55$\pm$    0.07&      92$\pm$     2&      0.82$\pm$      0.04&      0.16$\pm$      0.03\\
DG Tau&      36.0$\pm$       2.0&        9$\pm$     2&       12$\pm$     8&      1.56$\pm$      0.11&     -33&    3.51$\pm$    0.06&     198$\pm$    27&      2.74$\pm$      0.08&      1.31$\pm$      0.05\\
DL Tau&      49.0$\pm$       1.0&      148$\pm$     4&       72$\pm$     2&      0.37$\pm$      0.03&      63&    4.48$\pm$    0.04&     179$\pm$     2&      0.67$\pm$      0.02&      0.70$\pm$      0.03\\
DM Tau&      31.1$\pm$       1.6&      180$\pm$    10&       86$\pm$     5&      0.54$\pm$      0.03&     -17&    2.65$\pm$    0.03&     274$\pm$    16&      0.56$\pm$      0.02&      0.78$\pm$      0.04\\
DQ Tau (*) &      12.1$\pm$       4.2&       11$\pm$    20&       25$\pm$    50&      1.63$\pm$      0.13&    -278&    0.43$\pm$    0.01&     439$\pm$   534&      2.03$\pm$      0.06&      0.35$\pm$      0.15\\
FT Tau&       7.7$\pm$       0.3&       43$\pm$     1&       21$\pm$     1&     -0.17$\pm$      0.09&      13&    5.31$\pm$    0.12&      57$\pm$     0&      0.40$\pm$      0.06&     -0.13$\pm$      0.04\\
GM Aur (*)&      27.0$\pm$       3.6&       98$\pm$    24&      \textit{$> 80$}&      1.53$\pm$      0.07&       3&    2.55$\pm$    0.02&     578$\pm$   184&      2.02$\pm$      0.05&      1.02$\pm$      0.06\\
LkCa 15&      28.4$\pm$       1.4&      109$\pm$     3&       55$\pm$     1&     -0.23$\pm$      0.17&      17&    4.90$\pm$    0.10&     178$\pm$     7&      1.66$\pm$      0.12&      1.27$\pm$      0.05\\
MWC 480&     182.3$\pm$      11.2&       81$\pm$     5&       39$\pm$     4&      0.75$\pm$      0.17&      28&    9.08$\pm$    0.15&     155$\pm$     6&      1.86$\pm$      0.07&      1.74$\pm$      0.05\\
MWC 758&      10.6$\pm$       1.5&      102$\pm$    27&       52$\pm$    15&      0.54$\pm$      0.52&       0&    0.95$\pm$    0.07&     187$\pm$    50&      1.09$\pm$      0.30&      1.53$\pm$      0.27\\
UZ Tau E&      24.1$\pm$       0.7&       79$\pm$     2&       39$\pm$     1&      0.12$\pm$      0.08&      27&    4.96$\pm$    0.15&     115$\pm$     5&      0.72$\pm$      0.07&      0.74$\pm$      0.04\\
UZ Tau W&       3.5$\pm$       0.2&       50$\pm$     2&       23$\pm$     8&      1.05$\pm$      0.46&      -1&    0.35$\pm$    0.02&     128$\pm$    43&      1.66$\pm$      0.21&      0.39$\pm$      0.14\\
HL Tau&      90.6$\pm$       4.1&       40$\pm$    15&       22$\pm$     2&      1.32$\pm$      0.08&     -75&   12.73$\pm$    0.35&     280$\pm$    26&      2.62$\pm$      0.11&      1.97$\pm$      0.07\\
HH 30&       8.1$\pm$       0.4&      102$\pm$     2&       62$\pm$     2&     -2.41$\pm$      0.42&       0&    2.50$\pm$    0.11&     123$\pm$     3&     -0.56$\pm$      0.39&      0.47$\pm$      0.08\\
DG Tau b (*) &      67.9$\pm$      29.6&       81$\pm$    15&       \textit{48$\pm$    18} &      1.18$\pm$      0.18&      -7&    5.67$\pm$    1.49&     303$\pm$    23&      1.95$\pm$      0.10&      0.94$\pm$      0.12\\
T Tau&       0.1$\pm$       0.05&        8$\pm$     2&        8$\pm$     2&     -1$\pm$      1 &      -6&    $> 5$ &      67$\pm$    20&      [1] &      0.48$\pm$      0.50\\
Haro 6-10 N&       0.6$\pm$       0.1&       17$\pm$     3&        5$\pm$     3&      [0] &       3&   $>10$ &      14$\pm$     1&      [1] &      [1.0] \\
Haro 6-10 S&       0.5$\pm$       0.1&       10$\pm$     2&        4$\pm$     3&      [0] &       0&   $>10$ &      14$\pm$    1&  [1] &      [1.0] \\
Haro 6-13&      17.3$\pm$       7.7&       19$\pm$    41&       10$\pm$     1&      1.00$\pm$      2.39&       2&    3.56$\pm$    1.26&      90$\pm$    32&      1.03$\pm$      0.94&      0.08$\pm$      0.07\\
Haro 6-33&       6.8$\pm$       1.6&   $> 50$  &   - &      1.48$\pm$      0.15&       0&    0.28$\pm$    0.02&     439$\pm$   616&      1.57$\pm$      0.17&      0.41$\pm$      0.26\\
\hline
\end{tabular}\vspace{0.1cm}\\
{(*) Error bars on $R_t$ to be considered with caution, see text. Negative $\Delta\chi^2$ indicates that the power law model is better. Values in brackets indicate fixed parameters.}
\end{table*}

\begin{table*}
\caption{Comparison of values derived from independent data sets at similar wavelengths}
\label{tab:compare}
\begin{tabular}{l|cc|cc|cc|cc}
\hline
\hline
Source &  $R_{out}$ & $p$ &  $R_{out}$ & $p$ & $R_t$ & $\gamma$ &  $R_t$ & $\gamma$  \\
     & AU  &    & AU &   & AU  &    & AU &   \\
     & \multicolumn{2}{c|}{at 1.3 mm} & \multicolumn{2}{c|}{at 1.4 mm}
     & \multicolumn{2}{c|}{at 1.3 mm} & \multicolumn{2}{c}{at 1.4 mm} \\
\hline
LkCa\,15&    198$\pm$    15&      1.59$\pm$      0.19&      179$\pm$     8&      1.62$\pm$      0.11
  &      60$\pm$     2&      0.08$\pm$      0.23&      56$\pm$     1&     -0.20$\pm$      0.16 \\
MWC480&     153$\pm$     6&      1.77$\pm$      0.09&      188$\pm$     8&      1.75$\pm$      0.06
 &     41$\pm$     3&      0.52$\pm$      0.22&     59$\pm$     8&      0.72$\pm$      0.12\\
\hline
\end{tabular}
\end{table*}


\begin{table*}
\caption{Comparison of values derived from two different wavelengths}
\label{tab:comparewave}
\begin{tabular}{l|cc|cc|cc|cc}
\hline
\hline
Source &  $R_{out}$ & $p$ &  $R_{out}$ & $p$ &  $R_t$ & $\gamma$ &  $R_t$ & $\gamma$  \\
     & AU  &    & AU &    & AU  &    & AU &   \\
     & \multicolumn{2}{c|}{at 1.3 mm} & \multicolumn{2}{c|}{at 2.8 mm}
     & \multicolumn{2}{c|}{at 1.3 mm} & \multicolumn{2}{c}{at 2.8 mm} \\
\hline
CI\,Tau&      215$\pm$     6&      0.58$\pm$      0.03&     186$\pm$    13&      0.61$\pm$      0.10
  &       98$\pm$     8&      0.36$\pm$      0.04&      63$\pm$     5&      0.03$\pm$      0.17 \\
CY\,Tau&      104$\pm$     3&      0.90$\pm$      0.03&     108$\pm$    10&      1.22$\pm$      0.12
  &       35$\pm$     2&      0.35$\pm$      0.06&      29$\pm$     2&      0.47$\pm$      0.20 \\
DG\,Tau&      188$\pm$    30&      2.69$\pm$      0.10&     401$\pm$     7&      2.89$\pm$      0.02
  &       12$\pm$     2&      1.55$\pm$      0.17&      11$\pm$     1&      1.58$\pm$      0.33 \\
DL\,Tau&      181$\pm$     1&      0.65$\pm$      0.02&     146$\pm$     7&      0.72$\pm$      0.09
  &       75$\pm$     2&      0.38$\pm$      0.03&      50$\pm$     3&      0.19$\pm$      0.14 \\
DM\,Tau&      285$\pm$    24&      0.55$\pm$      0.02&     250$\pm$    37&      0.76$\pm$      0.07
  &       94$\pm$     6&      0.56$\pm$      0.03&      67$\pm$     9&      0.60$\pm$      0.13 \\
FT\,Tau&       57$\pm$     1&      0.41$\pm$      0.06&      85$\pm$    12&      1.03$\pm$      0.24
  &       21$\pm$     1&     -0.20$\pm$      0.10&      23$\pm$     2&      0.05$\pm$      0.42 \\
LkCa\,15&     198$\pm$    15&      1.59$\pm$      0.19&     168$\pm$    33&      1.70$\pm$      0.65
  &       60$\pm$     2&      0.08$\pm$      0.23&      53$\pm$     6&     -0.23$\pm$      0.84 \\
MWC\,480&      153$\pm$     6&      1.77$\pm$      0.09&     170$\pm$    51&      2.07$\pm$      0.32
  &       41$\pm$     3&      0.52$\pm$      0.22&      45$\pm$    22&      1.08$\pm$      0.52 \\
\hline
\end{tabular}
\end{table*}

All results are presented in  Table \ref{tab:main}. A comparison of the results obtained from
independent data sets at similar wavelengths is shown in Table \ref{tab:compare}, which shows the excellent agreement of the constrained parameters (see also Fig.\ref{fig:compare}). In addition, the good agreement of geometric parameters with determinations from other studies is a further proof of the data quality (see Table \ref{tab:geometry}, and Fig.\ref{fig:geometry}).

\paragraph {Simple power law}
Results for the surface density parameters, $\Sigma_{100},p$ and $R_\mathrm{out}$, are presented in Col. 7-10 of Table \ref{tab:main}. For most sources, the emission is largely optically thin, so the derived surface density will scales as roughly $1/T_0$, but the outer radius remains essentially unaffected by the choice of the temperature. The only exceptions are the \object{T Tau} and Haro\,6-10, which are essentially optically thick disks. \object{FT Tau} and \object{Haro 6-13} may also be attributed to thick disks.

\paragraph{Exponential edge}
We generally used Eq.\ref{eq:edge} to first locate the minimum. However, because of the direct dependency between the parameters, the errors on $R_t$ and $M_d$ were obtained by re-fitting the data using these parameters
as primary parameters rather than $R_c$ and $\Sigma_0$.  Note that while the error on $\Sigma_t$ may become very large, $\Sigma_0$ is generally  constrained with a very similar accuracy as in Model 1.

Results are presented in Cols 2-5 of Table \ref{tab:main}. It was difficult to adjust this model to a few sources, among which were the apparently optically thick sources T\,Tau and Haro\,6-10, and the single stars DQ\,Tau, DG Tau-b, and GM\,Aur. For the three latter stars, the best-fit power law has an index of $p=2$. In this case, the expression in Eq.\ref{eq:edge} attempts to fit $\gamma = 2$ and diverges.
Finding the best fit requires the determination of the best transition radius $R_c$ and its errorbar for all values of $\gamma$ ranging from 0.6 to 1.9 (by steps of 0.1).
The relative errors on $R_t$ are  generally larger than for $R_c$, because $R_t/R_c$ diverges for $\gamma \rightarrow 2$.
No constraint on $R_t$ is possible for DQ\,Tau. For GM\,Aur, only a lower limit is obtained, while
for DG\,Tau-b, $R_t$ is very marginally constrained: at the $2 \sigma$ level, any value is acceptable. The error bars should be taken with care in those cases. A similar procedure was used for Haro 6-13 and Haro 6-33, for which $R_t$ remains unconstrained at the $2\sigma$ level.


Col 6 of Table \ref{tab:main} indicates the difference in $\chi^2$ between Model 1 and Model 2. A positive value indicates Model 2 (the viscous disk) provides an \textit{apparently} better fit than the truncated power law. The significance of this result will be discussed in Section \ref{sec:discussion:density}.

Deprojected, circularly averaged visibility profiles are displayed in the middle column of Figure \ref{fig:alldmtau} for DM\,Tau and Figures \ref{fig:allbptau}-\ref{fig:alluztauw} for the others sources (in Appendix \ref{app:sources}, available on-line only). These deprojected visibilities only serve as an illustration of the fit results, but not to determine the parameter values and their errors.

\begin{figure*}[ht] 
   \includegraphics[angle=270,width=18.0cm]{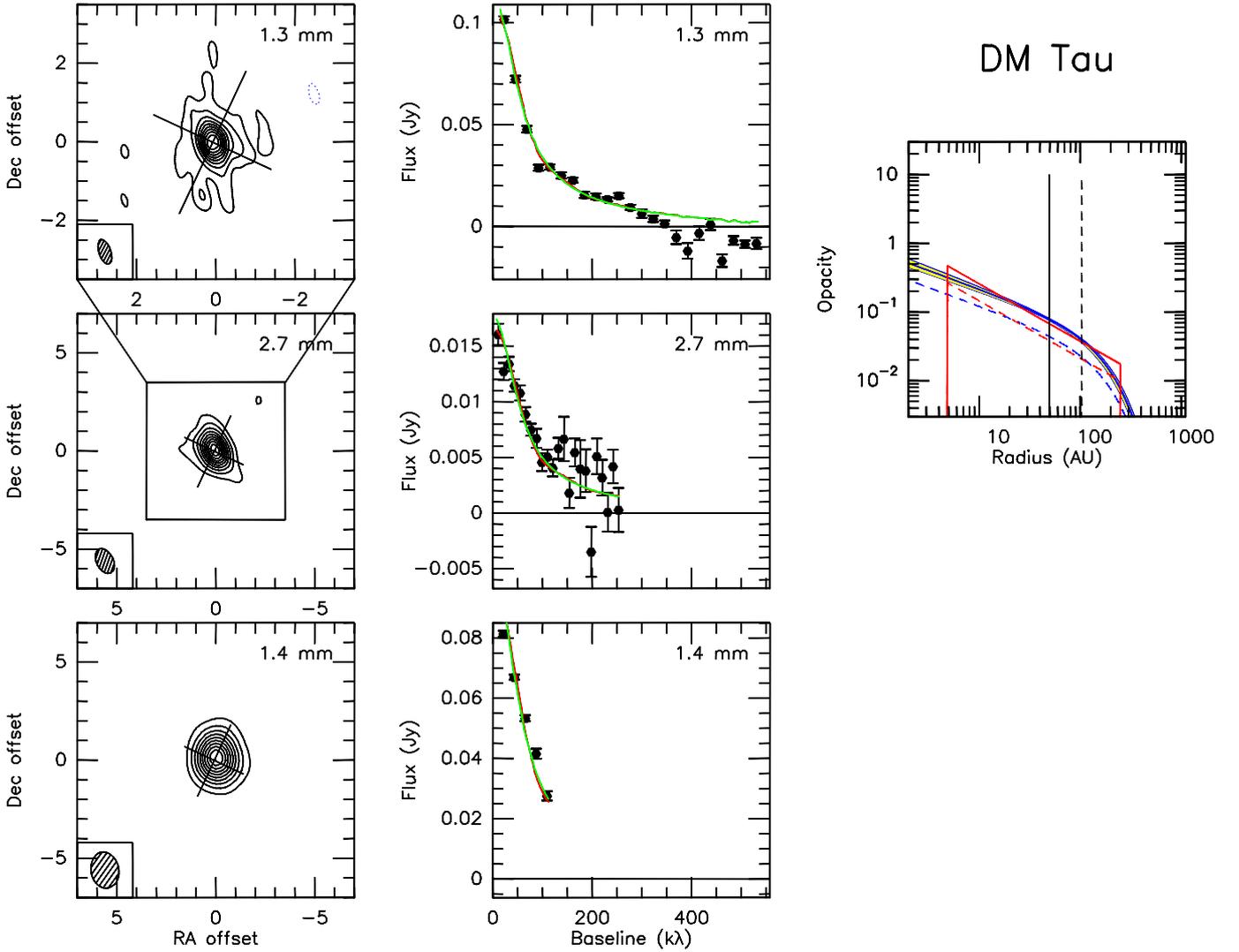}
\caption{Left row: High angular resolution images of DM Tau. On top, high resolution 1.3 mm image, in the middle, the 2.7\,mm (or 3.4\,mm for some sources) image with a box indicating the size of the 1.3 mm view. For sources (like this one) that have independent data sets at other wavelengths (1.4 or 3.4\,mm), a lower panel displays the corresponding image for the same area as above. All contours are 10 \% of the peak value to illustrate consistently the apparent sizes and low level extensions.  Contour level is 2 mJy/beam ($3.5\sigma$) at 1.3 mm, 0.78 mJy/beam ($2 \sigma$) at 2.7\,mm, and 6 mJy/beam ($7 \sigma$) at 1.4\,mm.
Middle row: Deprojected and circularly averaged visibilities and best-fit models for each wavelength.
Red is for power law, green for exponential edge. Right panel: Best-fit opacity profile (perpendicular to the disk plane, i.e.
$\kappa_\nu \Sigma(r)$) for the 1.3 mm and long wavelength models. The continuous line is for the short wavelength, the dashed line for the long wavelength. The vertical lines indicate the effective angular resolution.}
   \label{fig:alldmtau}
\end{figure*}

\section{Results}

\label{sec:results}


\subsection{The dust temperature}
\label{sec:temperature}

For the assumed temperature law, our treatment differs quite significantly from those of \citet{Kitamura+etal_2002} and \citet{Andrews+Williams_2007}, who assumed that the temperature derived from IR-emitting dust by fitting the SED also applies to the mm emitting dust. However, strong vertical temperature gradients are expected in disks \citep[e.g.][]{Dalessio+etal_1999}.

Because the mm emission comes from cold dust around the disk mid-plane, using a power law for the dust temperature distribution is an oversimplification.  The dust temperature is expected to follow three different regimes, depending on whether the disk is optically thick or thin for absorption of the incident radiation and re-emission of its own radiation. The two extreme regimes predict $\approx 1/\sqrt{r}$ temperature dependence, and are connected by a nearly constant temperature (or even slightly rising) region (``plateau''), whose extent depends on the source radial opacity profile \citep{Dalessio+etal_1999,Chiang+Goldreich_1997}. A more self-consistent approach was taken by \citet{Isella+etal_2009}, who derived dust opacities from the Mie theory assuming a specific dust composition and grain size distribution, and solve for the dust temperature in the two-layer approximation of \citet{Chiang+Goldreich_1997}. However, in this case the derived dust temperature depends (by an unknown amount) on the assumed dust composition.  Furthermore, using a single temperature for all grain sizes is an oversimplification. The dust thermal balance is largely dominated by the IR radiation
\citep[see][]{Chiang+Goldreich_1997}. Because the opacities are not gray, the temperature of dust grains is expected to depend on their size. The details will depend on the exact behavior of the dust emissivity as a function of wavelength, but  generally larger grains are expected to be colder \citep{Wolf_2003+Tdust,Chapillon+etal_2008}. Yet, these grains dominate the mm emission that we are observing.

Our approach of keeping the dust temperature as a parametric law allows us to directly measure the effective temperature of the emitting grains whenever the angular resolution is sufficient to resolve the optically thick core of the  disk. Furthermore, we can estimate the impact of the temperature uncertainty on the derived surface density parameters. Such a step-by-step approach allows us to understand and quantify the existing couplings between the dust parameters, the disk temperature and the disk surface density.

Because the flux scales as $T\times\Sigma$, the assumed values for the temperature may affect the derived shapes of $\Sigma(r)$.  In Model 1, the exponent $p$ will be directly affected, because $p+q$ is preserved for pure optically thin emission.
This is confirmed by our analysis for both models (see Appendix \ref{app:temperature}). However, the effects are small because our adopted value for $q=0.4$ is a good first order approximation of most (reasonable) temperature profiles. In Model 2, an inappropriate temperature profile may affect $R_c$, because this parameter is constrained by the steepening of the emission as function of increasing radius. Again, Appendix \ref{app:temperature} (available online only) shows the effect is limited, $R_c$ being affected by at most 20 \%.

\begin{figure}
   \includegraphics[angle=270,width=8.0cm]{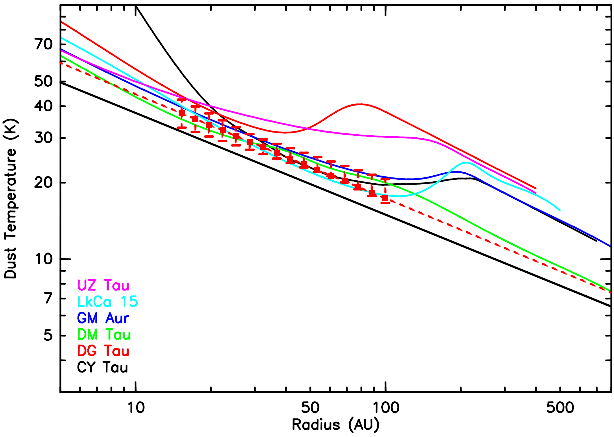}
\caption{Temperature laws derived by \citet{Isella+etal_2009} (color curves, one for each source) compared to our assumed
power law (black continuous line). The dashed red line indicate the best fit power law for DG\,Tau, and the error bars indicate
the $\pm 1 \sigma$ range in the region where this power law is constrained, i.e. about 20 to 100 AU. }
\label{fig:temperature}
\end{figure}

In a few sources, \citet{Isella+etal_2009} derived dust temperature as a function of radius from a joint modeling of the SED and 1.3\,mm images. We used the temperatures displayed in their Fig.7 as an input in our modeling to check the magnitude of the effects in all sources we have in common. The results are presented in Table~\ref{tab:tempisella}. The temperature law has no visible influence on the pivot radius, $R_t$, and affects $\gamma$ by at most 0.1-0.2. Our used temperature laws are displayed on top of those of \cite{Isella+etal_2009} in Fig.\ref{fig:temperature}.
From Table~\ref{tab:tempisella} and Appendix \ref{app:temperature} we conclude that the uncertainties in our assumed dust temperature distribution do not significantly affect the shape of the derived surface density distribution.

\begin{table*}
\caption{Temperature derived from partially optically thick disks}
\label{tab:temperature}
\begin{tabular}{l|cc|c|cc}
\hline \hline
Source &  $T_k$ & $q$ & $R_0$ &  $T_k$ & $q$ \\
     & (K)  &    & AU & (K) &   \\
     & \multicolumn{2}{c|}{Viscous} & & \multicolumn{2}{c}{Power} \\
\hline
DG\,Tau  &   26.0$\pm$      2.2&      0.56$\pm$      0.09&   30&      28.5$\pm$      1.9&      0.41$\pm$      0.09\\
MWC\,480 &   13.2$\pm$      1.9&      0.65$\pm$      0.09&   40&      16.6$\pm$      1.9&      0.42$\pm$      0.09\\
HL\,Tau  &   25.2$\pm$      1.9&      0.39$\pm$      0.09&   55&      24.9$\pm$      1.9&      0.40$\pm$      0.09\\
T\,Tau   &   23.3$\pm$      1.9&     -0.32$\pm$      0.09&   40&      16.0$\pm$      1.9&      0.36$\pm$      0.09\\
DG\,Tau-b &  21.6$\pm$      1.6&     0.29 $\pm$      0.11&   40&      21.1$\pm$      1.2&      0.35$\pm$      0.10\\
\hline
\end{tabular}\vspace{0.1cm}\\
{$R_0$ is the reference radius at which the temperatures are derived.}
\end{table*}

\begin{table*}
\caption{Effect of the temperature laws}
\label{tab:tempisella}
\begin{tabular}{l|ccc|c|ccc}
\hline
\hline
Source &  $R_t$ & $\gamma$ & $\beta_m$ & $\Delta\chi^2$ & $R_t$ & $\gamma$ & $\beta_m$ \\
     & AU  &    & & AU &   \\
     & \multicolumn{3}{c|}{Simple T} & & \multicolumn{3}{c}{T from \citet{Isella+etal_2009}} \\
\hline
CY\,Tau&       32$\pm$     1&      0.28$\pm$      0.06&    0.17$\pm$    0.04&     2.&      32$\pm$     1&      0.13$\pm$      0.06&    0.13$\pm$    0.03\\
DG\,Tau&       12$\pm$     8&      1.56$\pm$      0.11&    1.45$\pm$    0.12&   193.&      13$\pm$     2&      1.23$\pm$      0.11&    0.95$\pm$    0.04\\
DM\,Tau&       86$\pm$     5&      0.54$\pm$      0.03&    0.77$\pm$    0.04&     4.&      87$\pm$     5&      0.64$\pm$      0.03&    0.73$\pm$    0.04\\
GM\,Aur&      112$\pm$    37&      1.53$\pm$      0.07&    1.02$\pm$    0.08&    12.&     135$\pm$    76&      1.79$\pm$      0.06&    0.93$\pm$    0.06\\
LkCa\,15&       55$\pm$     1&     -0.23$\pm$      0.17&    1.26$\pm$    0.06&      0.&      51$\pm$     1&     -0.27$\pm$      0.16&    1.21$\pm$    0.05\\
UZ\,Tau\,E&       39$\pm$     1&      0.12$\pm$      0.08&    0.74$\pm$    0.04&     1.&      35$\pm$     1&      0.22$\pm$      0.08&    0.62$\pm$    0.04\\
\hline
\end{tabular}\vspace{0.1cm}\\
{A positive value of $\Delta \chi^2$ indicates that the simple power law fit provides a better result than the more complex
temperature profile.}
\end{table*}

However, the disk masses are sensitive to the assumed dust temperature, since they scale to first order as $1/T$. Furthermore, the dust emissivity index $\beta_m$ can also be affected, because the contribution of the optically thick core depends on the dust temperature. The differences in the analysis of the MWC\,480 performed by \cite{Hamidouche+etal_2006} and \cite{Pietu+etal_2007} illustrate the importance of the effect. A similar effect can be seen for DG\,Tau in Table \ref{tab:tempisella}: $\beta_m$ changes by 0.5 between the two hypotheses on the temperature.

From the best-fit values, a few sources in our sample display partially resolved cores that  may be interpreted as optically thick cores, and thus allow a direct determination of the temperature. As detailed in Appendix \ref{app:fitting}, these ``thick cores'' satisfy
two conditions: i) they have the same brightness at both wavelengths, and ii) their brightness distribution is relatively flat, because the temperature is expected to decrease as $r^{-0.4 - 0.7}$ at most. The fitting process indicates that this happens for DG\,Tau, DG\,Tau-b, HL\,Tau, \object{T Tau},  and MWC\,480. The derived values are presented in Table \ref{tab:temperature}. Because the Model 1 and 2 have different opacity distributions (see Fig.\ref{fig:allbptau}-\ref{fig:alluztauw}), they predict different optically thick zones, and thus the temperature  slightly depends on the assumed density model. For T\,Tau, the apparent difference is largely an artifact, because the source is basically a completely optically thick disk, for which the ``viscous'' disk model is poorly constrained. The measured values and slopes justify a posteriori our simple hypothesis for the temperature law.
The dependence is small for \object{DG Tau}, \object{DG Tau-b}, and \object{HL Tau}, though. In the power law model, the extrapolated temperature at 100 AU for DG\,Tau is 17\,K, close to our adopted value of 15 K for all other sources. HL\,Tau is slightly warmer, 19\,K. For DG\,Tau-b, the temperature at 100 AU is 15\,K, but the exponent is slightly lower than 0.4.

Formally, FT\,Tau has both a flat enough brightness distribution and a low apparent $\beta_m$ to be consistent with optically thick dust, but would require a very low dust temperature to match the observed flux densities. A dust temperature of 10 K at 40 AU would just provide adequate flux (the brightness can be obtained from the (apparent) opacities displayed in Fig.\ref{fig:allfttau}). Such a low value seems inconsistent with the relatively luminous and massive central star, so the warmer, optically thin solution with $\beta_m \simeq 0$ is to be preferred.

Among the observed sources, MWC\,480 deserves specific comments concerning the temperature.
In this bright source, the ``thick core'' is quite large, $50-80$ AU. However, its brightness is moderate, which means that when this is interpreted as being optically thick, the derived temperatures are very low \citep[see][]{Pietu+etal_2006}. The large size of the ``thick core'' results in substantial opacity corrections for $\beta_m$, which in turn leads to unrealistic values for Model 2.

An alternate explanation for the relatively flat brightness distribution in the inner part is a warmer, optically thin region with $\beta \simeq 0$.  This is not consistent with the value of $\beta_m$ derived from the integrated flux, and can only happen if $\beta$ varies with radius (see Appendix \ref{app:fitting}). This is studied in Sec.\ref{sub:radial:dust} and MWC\,480 will be rediscussed in more detail in Sec.\ref{sec:sub:mwc480}.

\subsection{Surface densities}
\label{sec:sub:surface}

\cite{Isella+etal_2009} have published a high-resolution (0.7$''$) survey at 1.3 mm of the Taurus region, with several sources
in common to our study.  It has been analyzed in terms of the viscous disk model, and Table \ref{tab:valisella} shows a comparison of the results. Note that in this analysis,  we assumed no inner hole for Lk\,Ca\,15 to provide a consistent comparison, and its apparent deficit of emission in the center is purely explained by a negative value for $\gamma$. In general, our data have a higher resolution and are slightly more sensitive, which results in error bars that are lower than in \citet{Isella+etal_2009}, the only exception is GM\,Aur, for which our resolution is moderate.

The agreement between both studies is reasonable, typically within 2 $\sigma$. The most notable exception is DG\,Tau. DG\,Tau was further studied at higher resolution by \citet{Isella+etal_2010}; the agreement on $R_t$ is reasonable, but they find $\gamma = 0.28 \pm 0.05$ instead of $\gamma = 1.6 \pm 0.1$ in our study. The difference between the two results may be due to the widely different $uv$ coverage, linked to a non symmetric source. Our data are dominated by fairly moderate baseline lengths (up to $300 k\lambda$), while \citet{Isella+etal_2010} find a substantial contribution to the imaginary part of the visibilities
at 1.3\,mm up to $200 k\lambda$ (see their Fig.2 and image in Fig.10).

We also note that the agreement is better on $R_c$ (or $R_t$) than on $\gamma$. This is to be expected, as $R_c$ is a first order parameter (the radius which encloses 63 \% of the disk mass), while $\gamma$ is a second-order parameter (the slope of the surface density distribution).


\begin{table}
\caption{Comparison with other results}
\label{tab:valisella}
\begin{tabular}{l|cc|cc}
\hline
\hline
Source &  $R_t$ & $\gamma$ &  $R_t$ & $\gamma$ \\
     & AU  &    & AU &  \\
     & \multicolumn{2}{c|}{This work} & \multicolumn{2}{c}{Isella et al\.} \\
\hline
CY\,Tau&       32$\pm$     1&      0.28$\pm$      0.06&      55$\pm$     5&     -0.30$\pm$      0.30\\
DG\,Tau&       12$\pm$     8&      1.56$\pm$      0.11&      21$\pm$     1&     -0.50$\pm$      0.20\\
DM\,Tau&       86$\pm$     5&      0.54$\pm$      0.03&      86$\pm$    32&      0.80$\pm$      0.10\\
LkCa\,15&       62$\pm$     1&     -1.24$\pm$      0.12&      60$\pm$     4&     -0.80$\pm$      0.40\\
UZ\,Tau\,E&       39$\pm$     1&      0.12$\pm$      0.08&      43$\pm$    10&      0.80$\pm$      0.40\\
GM\,Aur&      58$\pm$    23&      0.30$\pm$      0.10&      56$\pm$     1&      0.40$\pm$      0.10\\
\hline
\end{tabular}\vspace{0.1cm}\\
{Rightmost columns indicate values derived by \citet{Isella+etal_2009}. The leftmost
columns are our results. For LkCa\,15 and GM\,Aur, we assumed no central hole for
this comparison, and thus
obtain different results from those in Table \ref{tab:main}.}
\end{table}

\subsection{Emissivity index}

$\beta_m$ values have been reported for a number of sources in our sample by \cite{Rodmann+etal_2006} and \citet{Ricci+etal_2010}. Their analysis is different from ours, because $\beta_m$ is derived from spatially unresolved multi-wavelength data, from a fit of the SED. \citet{Rodmann+etal_2006} use a simple power law to derive the spectral index $\alpha$ of the  mm SED between 7 and 1 mm. Overall, the agreement with our results is poor, most likely as a result of several effects. First, \cite{Rodmann+etal_2006} apply a uniform correction for opacity, while we have shown that the existence of optically thick cores affect $\beta_m$ very inhomogeneously, with corrections ranging from 0 to 0.5. Second, the different frequency span must also affect $\beta_m$, because using a power law for the dust emissivity is only an approximation; in particular, the emissivity is expected to steepen at long wavelengths \citep[e.g.][]{Draine_2006}. The agreement with the results of \citet{Ricci+etal_2010} is much better, most likely because they use a more elaborate procedure for the SED fit, in which some estimate of the disk size and surface density slope is used to account for the optical depths effects.

\subsection{Individual objects}
\label{sec:objects}
\subsubsection{Multiple stars}
\label{sec:multiple}

\object{Haro 6-10} stands out as exceptional. Although the formal fit gives marginally optically thin disks and $\beta_m \simeq 0$, this is likely to be an artifact caused by seeing limitation. Indeed, any small ``seeing'' effect spreads out a little emission and makes the source slightly more extended than in reality. This mimics an (optically thin) halo. Thus, Haro 6-10 is best represented by (two) optically thick disks of radii around 15 AU (scaling as  $1/\sqrt{T_0}$ since only the total flux is constrained). This result indicates that the amplitude and phase calibrations are sufficiently accurate to determine sizes as small as 30 AU (total), or about $1/5^{th}$ of the synthesized beam in this case. The inclination cannot be derived for Haro 6-10.
The minimum mass of each disk is $10^{-3} \Msun$ (see Appendix \ref{app:compact}, available online only).

\object{T Tau} was already studied by \cite{Hogerheijde+etal_1997} and \citet{Akeson+etal_1998} in the mm domain.
As in these studies, only the northern member of the multiple system is detected.   Like Haro 6-10, the emission can be explained by a nearly optically thick disk.  Because of the larger size, the seeing effect is negligible and only the optically thick solution is found to be viable. Our best-fit inclination of  $40 \pm 4^\circ$ is somewhat larger than the $\sim 20^\circ$ derived by \citet{Ratzka+etal_2009} from IR studies. However, this only influences the apparent opacities by the ratio of the cos($i$), i.e. about 20 \%.
The minimum mass of the disk is $0.007 \Msun$, assuming the disk is optically thick.

\begin{figure}
   \includegraphics[angle=270,width=8.0cm]{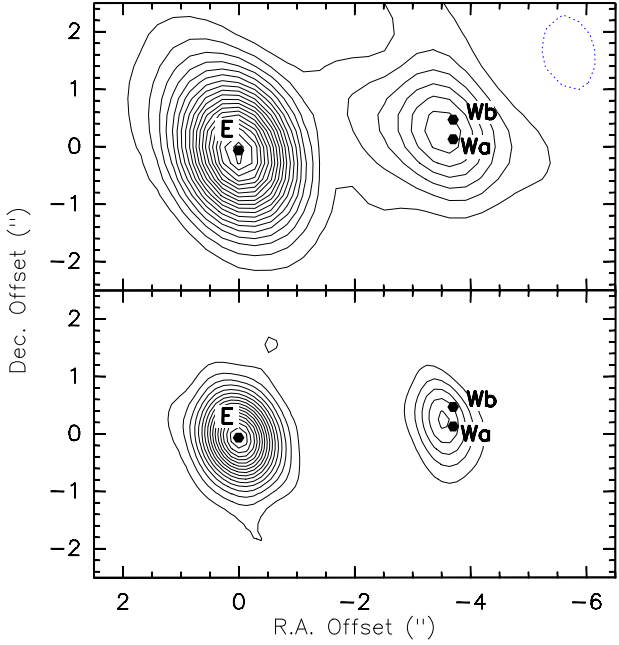}
\caption{Relative positions of the disks and stars in the UZ\,Tau multiple system.
The system geometry is from \citet{Simon+etal_1992}, except that we used a separation
smaller by 1$\sigma$ between UZ\,Tau-E and UZ\,Tau-Wa. Top: 2.7\,mm map, contour step 0.9 mJy/beam (35 mK, 2 $\sigma$). Bottom:
1.3\,mm map, contour step 5 mJy/beam (140 mK, 4 $\sigma$).}
\label{fig:uztau}
\end{figure}

The quadruple system \object{UZ Tau} shows emission from two regions: one around the spectroscopic binary UZ\,Tau East, the other near the optically resolved wider binary UZ\,Tau West
\citep[separation $0.34''$ at $PA \simeq 0$,][]{Simon+etal_1992}. Given the disk inclination of UZ\,Tau East \citep{Simon+etal_2000} which is  confirmed by our new measurements, and assuming disks and orbits are coplanar, the true deprojected separation would be $\sim 100$ AU. Interpreting the emission around UZ\,Tau W as a single disk yields a similar orientation (consistent with coplanar disks) and an outer radius of $120 \pm 45$ AU. This is fairly large compared to the binary separation, and may be difficult to reconcile with tidal truncation.
This result, however, could be an artefact of improper subtraction of the UZ\,Tau East emission because any small (positive) residual emission left around UZ\,Tau East could bias the derivation of the position angle and size. A solution with two circumstellar disks is not totally excluded by our data. From the images, we find that the emission centroid is in between  UZ\,Tau West A and B (see Fig.\ref{fig:uztau}). The displacement observed between 1.3\,mm and 2.7\,mm suggests that the disk around West B disk is more optically thick than that around West A. Under the interpretation of circumstellar disks, their minimum mass is $6\,10^{-4} \Msun$.

The small size of circumstellar disks in known binaries suggests that tidal effects are responsible for their truncation, although a firm conclusion cannot be drawn because the inclination of Haro 6-10 is unknown.

\citet{Mathieu+etal_1997} found \object{DQ Tau} to be a non-eclipsing, double-lined spectroscopic
binary, comprised of two relatively equal-mass stars $M \approx 0.65 \Msun$
with spectral types in the range of K7 to M1 and an orbital period of 15.804 days.
The orbit is eccentric, but with an apastron around 0.28 AU, the tidal cavity should be
much smaller than  1 AU.
DQ\,Tau has been recognized as variable in the mm domain by \citet{Salter+etal_2008}. The variability is caused by interactions between the magnetospheres  when the two stars are near periastron, so that flares happen periodically. The observation dates and derived total flux for each date are given in
Table \ref{tab:dqtau}. No evidence for variability is found in our data as expected, since none of our
observations happened close to periastron.  The measured emission is thus coming purely from the
dusty (circumbinary) disk.

\begin{table} 
\caption{Observed flux densities for DQ\,Tau.}
\label{tab:dqtau}
\begin{tabular}{rccc}
\hline \hline
Date &  Frequency &  Flux Density & Nearest Periastron \\
     &  (GHz)     &   (mJy) &  (days)\\
\hline
1997-12-05  &   90 &       $9.6 \pm 0.7$  & 4\\
1997-12-30  &  90  &      $8.5 \pm 1.1$  & 2\\
\hline
1997-12-05 &    230 &      $72 \pm 2$  & 4 \\
1997-30-12  &  230    &   $84 \pm 5$ & 2\\
2008-02-11  &  230    &   $83 \pm 2^*$ & 5\\
\hline
\end{tabular}\\
(*) Long baseline data only, total flux extrapolated using the
apparent size of {$0.5''$} derived from the Dec 1997 data.
\end{table}

\begin{figure}
   \includegraphics[angle=270,width=7.0cm]{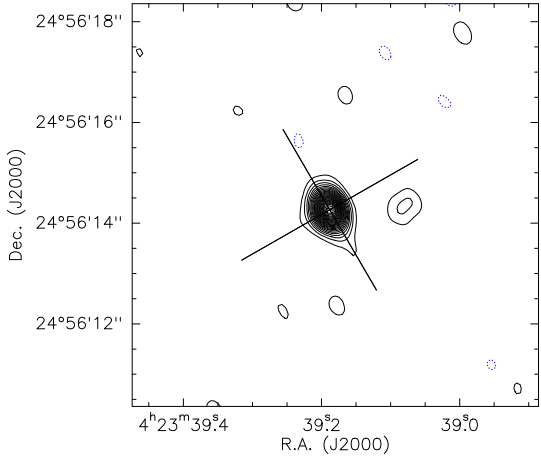}
\caption{1.3 mm emission from the FT\,Tau system showing the $\simeq 6 \sigma$ detection
west of the main object.  Contour steps are 1 mJy/beam, or $2\sigma$.}
\label{fig:fttau}
\end{figure}

Another star is possibly affected by binarity: \object{FT Tau}, which displays a weak, but significant (6 $\sigma$) emission $1.3''$ west of the main star (see Fig.\ref{fig:fttau}), and a very small ($\simeq 60$~AU radius) disk with $\beta_m \simeq 0$ (see Table \ref{tab:main} and Fig.\ref{fig:allfttau}). The position of the secondary peak of mm emission is, however, different from that of the near IR source
found by \citet{Itoh+etal_2008}.

The case of \object{HH 30} is somehow unusual. \citet{Anglada+etal_2007} suggested that HH\,30 is a binary based on the precession of its optical jet, but could not decide between a close binary and a $\simeq 15$ AU separation. \citet{Guilloteau+etal_2008} showed that the deficit of mm emission could be interpreted as a central hole consistent with the tidal truncation in the wide binary model. Here, in Model 2, the inner radius becomes unsignificant: any value below about 45 AU is acceptable for $R_\mathrm{int}$, because of the very steep decrease of the surface density profile for this high negative value of $\gamma \simeq -2$. In essence, this means $\gamma$ is constrained by the apparent sharp decrease of the emission near 120 AU, and not by the central deficit.

\subsubsection{Sources with holes}
\label{sec:holes}

For \object{DM Tau}, modeling the near and mid-IR SED \citep{Calvet+etal_2005} indicates an inner hole of about 3 AU.
Although this small hole is below the detectability limit of our observations, we used it in our analysis.

A deficit of emission at the center of the disk of \object{LkCa 15} was discovered by \citet{Pietu+etal_2006}, who interpreted it as a 45 AU radius hole.  This central dip was also observed at lower resolution by \citet{Isella+etal_2009}, but they suggested that it could be due to a negative value of $\gamma$.
Our higher angular resolution data allow us to test which hypothesis best represents the observed brightness distribution. Results are reported in Table \ref{tab:lkca15}. The no-hole hypothesis is rejected
at the $3 \sigma$ level, and the best fit is obtained with an inner hole of $38 \pm 4$ AU.
The near-IR imaging of \citet{Thalmann+etal_2010} confirms the sharp nature of the rim of the inner
hole and indicates a radius of 46 AU.
The transition radius $R_t$ remains relatively unaffected by the presence or absence of a hole, but the value found for $\gamma$  strongly depends on the hole size: the best-fit solution is compatible with $\gamma = 0$.

For \object{GM Aur}, the lack of 10 $\mu$m emission suggested a central hole of $R_\mathrm{int} = 25$\,AU \citep{Calvet+etal_2005}.
The hole has also been detected in the gas traced by CO, through spectro-imaging of the J=2-1 transitions of
the $^{12}$CO, $^{13}$CO, and C$^{18}$O isotopologues indicating very low gas surface densities in these regions: \citet{Dutrey+etal_2008} indicate a size of $R_\mathrm{int} = 19 \pm 4$~AU.
This size has been confirmed by direct imaging of the dust emission
at mm wavelengths \citep{Hughes+etal_2009}. Like for DM\,Tau, we thus assumed $R_\mathrm{int} = 20$ AU.
The strong dependence of $\gamma$ upon the possible existence of a central hole also exists for GM\,Aur.
Indeed, assuming no hole, we recover a very similar solution to that found by \citet{Isella+etal_2009} (see Table \ref{tab:valisella}), although it is somewhat worse (near the $2 \sigma$ level) than our nominal solution obtained for $R_\mathrm{int} = 20$ AU.

\begin{table} 
 \caption{Effect of the central hole on the derived parameters for Lk\,Ca\,15}
 \label{tab:lkca15}
\begin{tabular}{lrrrr}
\hline \hline
$R_\mathrm{int}$ & \multicolumn{1}{c}{$\gamma$} & $R_c$ & $R_t$ & \multicolumn{1}{c}{$\chi^2$} \\
(AU)  &  & (AU) & (AU) & \\
\hline
 $[1]$ & $-1.24 \pm 0.11$ & $111 \pm 2$ & $62 \pm 2$ & 108674 \\
 $[46]$ &  $0.12 \pm 0.19$ & $102 \pm 3$ & $51 \pm 2$ & 108679 \\
 $38 \pm 4$ & $-0.35\pm 0.30$ & $110\pm 4$   & $57 \pm 2$ &  108664\\
 \hline
  \end{tabular}
 \end{table}
In conclusion, with the exception of HH\,30 which was discussed in Sec.\ref{sec:multiple}, allowing for central holes offer better solutions,
and brings the surface density exponent $\gamma$ back to ``standard'' values between 0 and 1.5.

\subsubsection{Young sources}
\label{sec:young}


\object{HL Tau} is a Class II object, for which the central star is not directly visible.  Our measured position angle is consistent with that of the jet and of previous high-resolution studies of the mm and centimeter emission from this region
\citep{Looney+etal_2000,Anglada+etal_2007,Carrasco-Gonzalez+etal_2009}. The inclination of the source is more debated:
early work from \citet{Cohen_1983} assumed a nearly edge-on disk, while $i = 56\pm10^\circ$ can be derived from the 7 mm deconvolved size from \citet{Wilner+etal_1996}. Our result better agrees with the submm data obtained by \citet{Lay+etal_1997}, $42 \pm 5^\circ$, and is also consistent with the obscuration of the redshifted jet \citep{Pyo+etal_2006}. At the observed scale, the envelope that dominates the submm flux is filtered out \citep{Looney+etal_2000}. Our major finding is the substantial opacity at mm wavelengths in the inner 40 AU, which allows us to constrain the temperature, but this significant opacity does not prevent structures from becoming visible at longer wavelengths, 1.3 cm or 7 mm. Our angular resolution is insufficient to separate the possible enhancement reported near 65 AU at 1.4\,mm by  \citet{Welch+etal_2004IAUS} and 1.3\,cm by \citet{Greaves+etal_2008}, but not confirmed at 7 mm by \citet{Carrasco-Gonzalez+etal_2009}.

\object{DG Tau} is a bright embedded star driving an optical jet at PA 226$^\circ$ \citep{Eisloffel+Mundt_1998}. It is surrounded by a large $^{13}$CO disk orthogonal to the jet \citep{Sargent+Beckwith_1989,Kitamura+etal_1996}, whose kinematics indicate a stellar mass around $0.7~\Msun$ \citep{Testi+etal_2002}. Inclinations of $45^\circ$ and $38^\circ$ are found by \citet{Pyo+etal_2003} and \citet{Eisloffel+Mundt_1998}, respectively. Our measured inclination of $32\pm 3^\circ$ is in favor of lower values. The results quoted in Table \ref{tab:main}  only slightly depend on the assumed orientation and inclination: $\gamma$ can be decreased by 0.1 and $R_t$ increased to 19 AU for the best fit orientation. The higher resolution data of \citet{Isella+etal_2010} also give
lower inclinations and small (22 AU) $R_t$, but with a very different value for $\gamma$ (see Sec.\ref{sec:sub:surface}).


\object{DG Tau-b} is a young, totally obscured, star at the apex of a wide angle cavity seen in scattered light \citep{Padgett+etal_1999}. It drives an optical jet and a molecular outflow \citep{Mitchell+etal_1997}. Although the position angles derived from the jet and disk agree, we find the disk inclination to be only $64 \pm 1^\circ$, while the jet inclination is estimated to be higher than $> 75^\circ$ from proper motion measurements \citep{Eisloffel+Mundt_1998}.
We also note that the DG\,Tau-b disk is best fitted with a higher flaring index $h$ than assumed for the other objects of our sample. We used $h = 1.35$, for which a scale height $H_{100} = 27 \pm 8$ is required to reproduce the observed continuum emission. The high flaring index is consistent with the fairly flat temperature distribution ($q = 0.3 \pm 0.1$) also found in this source.


\textbf{HERE HERE HERE}

\subsection{Radial dependency of the dust properties}
\label{sub:radial:dust}

Most previous studies assumed that the dust properties were uniform across the disk.  The dual-frequency resolved images allow us to test the validity of this hypothesis, and eventually constrain the variations of dust properties as a function of radius.

\subsubsection{Emissivity Index $\beta$}

In Table \ref{tab:comparewave} smaller transition radii $R_t$ are found from 2.7\,mm data than from 1.3\,mm data for four out of eight sources: \object{CI Tau}, \object{CY Tau}, \object{DL Tau} and \object{DM Tau}. For the other sources, the combination of sensitivity and resolution at 2.7\,mm data is insufficient to distinguish. Equivalently, in the truncated power law analysis (Table \ref{tab:comparewave}) the slope of the surface density $p$ is systematically steeper at 2.7 mm than at 1.3 mm. A similar result was recently obtained for CQ\,Tau by \citet{Banzatti+etal_2010}.

A possible cause for this effect is contamination by free-free emission, which adds a point-like source at lower frequencies. However, none of these sources have sufficient free-free emission to significantly contaminate the 2.7\,mm flux (see \cite{Rodmann+etal_2006} for the measurements). From \citet{Rodmann+etal_2006}, the contamination does not exceed $3 \%$  near 2.7\,mm. Removing a point source of this intensity from our 2.7\,mm data does not affect our results.

Thus, the different solutions found at the two wavelengths indicate a change of dust properties, at least in the spectral index of the emissivity $\beta$, with radius.
The larger $p$ and smaller $R_t$ at 2.7\,mm than at 1.3\,mm imply that the ratio of $T_b$(1.3mm)/$T_b$(2.7mm) increases with radius, hence $\beta$ increases with radius. The apparent $\beta(r)$ as a function of radius can be derived from
\begin{equation}
\label{eq:beta}
    \beta(r) = \beta_0 + \log{(\Sigma_a(r)/\Sigma_b(r))} / \log{(\nu_a/\nu_b)} ,
\end{equation}
where $\beta_0$ is the constant value used to derive the apparent surface densities $\Sigma_a(r)$ and $\Sigma_b(r)$ at both wavelengths, i.e. $\beta_0 = \beta_m$ (see also \citet{Isella+etal_2010}).

The increase of $\beta(r)$ with radius is most easily understood in the framework of the truncated power law analysis, because it simply turns into a logarithmic dependence of $\beta(r)$ as a function of radius
\begin{equation}\label{eq:betapower1}
    \beta(r) = \beta_0 + \log {((\Sigma_a (r/r_0)^{-p_a}) / (\Sigma_b (r/r_0)^{-p_b}))} / \log{(\nu_a/\nu_b)}
\end{equation}
\begin{equation}\label{eq:betapower2}
    \beta(r) = \beta_0 + \frac{ \log{(\Sigma_a/\Sigma_b)} }{ \log{(\nu_a/\nu_b)} } + (p_b-p_a) \frac{\log{(r/r_0)}}
    { \log{(\nu_a/\nu_b)}}
\end{equation}
$\Delta p = p(2.7\,\mathrm{mm})-p(1.3\,\mathrm{mm})$ is systematically positive in our sample (see Table \ref{tab:comparewave}).
However, the apparent significance level is low for each source, as $\Delta p$ apparently exceeds its $2 \sigma$ uncertainty in only two sources (CY\,Tau and DM\,Tau). Better constraints can be obtained by fitting the logarithmic dependence
of $\beta(r)$ directly to the data
\begin{equation}\label{eq:betapower3}
    \beta(r) = \beta_i + \beta_r \log{(r/r_0)} .
\end{equation}
The values of $\beta_r$ are reported in Column 2 of Table \ref{tab:betaradius} (for Model 1, but similar values are obtained for Model 2). It becomes now clear that the radial dependence is highly significant, because the weighted mean value is $\beta_r = 0.34 \pm 0.04$ (ignoring FT\,Tau, which has a negative $\beta$ everywhere). A Student's T-test applied to the distribution of values of $\beta_r$ reported in Table \ref{tab:betaradius} (including FT Tau) indicates less than 2 \% chances of being compatible with
$\beta_r = 0$.

For the softened-edge model, the $\beta(r)$ function implied by Eq.\ref{eq:beta} is more complex, and an illustration of the shape of this function is given in Figure \ref{fig:betaradius}, which displays  this  apparent $\beta(r)$ for two of the sources, CI\,Tau and DL\,Tau. The hatched areas indicate the approximate range of allowed values, obtained by adding and subtracting 1 $\sigma$ to each of the parameters defining the opacity function at the two wavelengths ($R_c$, $\gamma$ and $\Sigma_0$ from Eq.\ref{eq:edge}). Because some of these parameters are actually correlated, this is only an estimate of the error on the profile. The apparent index $\beta$ is large ($> 1.7$) in the outer disk parts ($r > 150 $--$250$ AU), while it is smaller than about 0.6 near 50 AU.

\begin{figure}
  \includegraphics[angle=270,width=8.0cm]{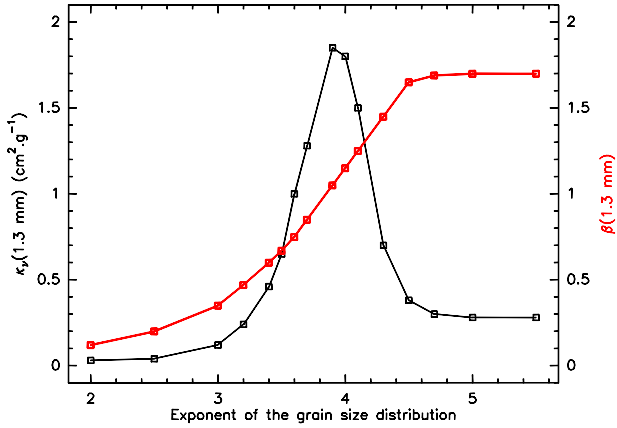}
  \caption{Dust emissivity  $\kappa$ and emissivity index $\beta$ at 1.3 mm from Isella et al 2009, as a function of the exponent of the size distribution $g$.} \label{fig:dust-isella}
\end{figure}

\begin{figure*}
  \includegraphics[angle=270,width=8.0cm]{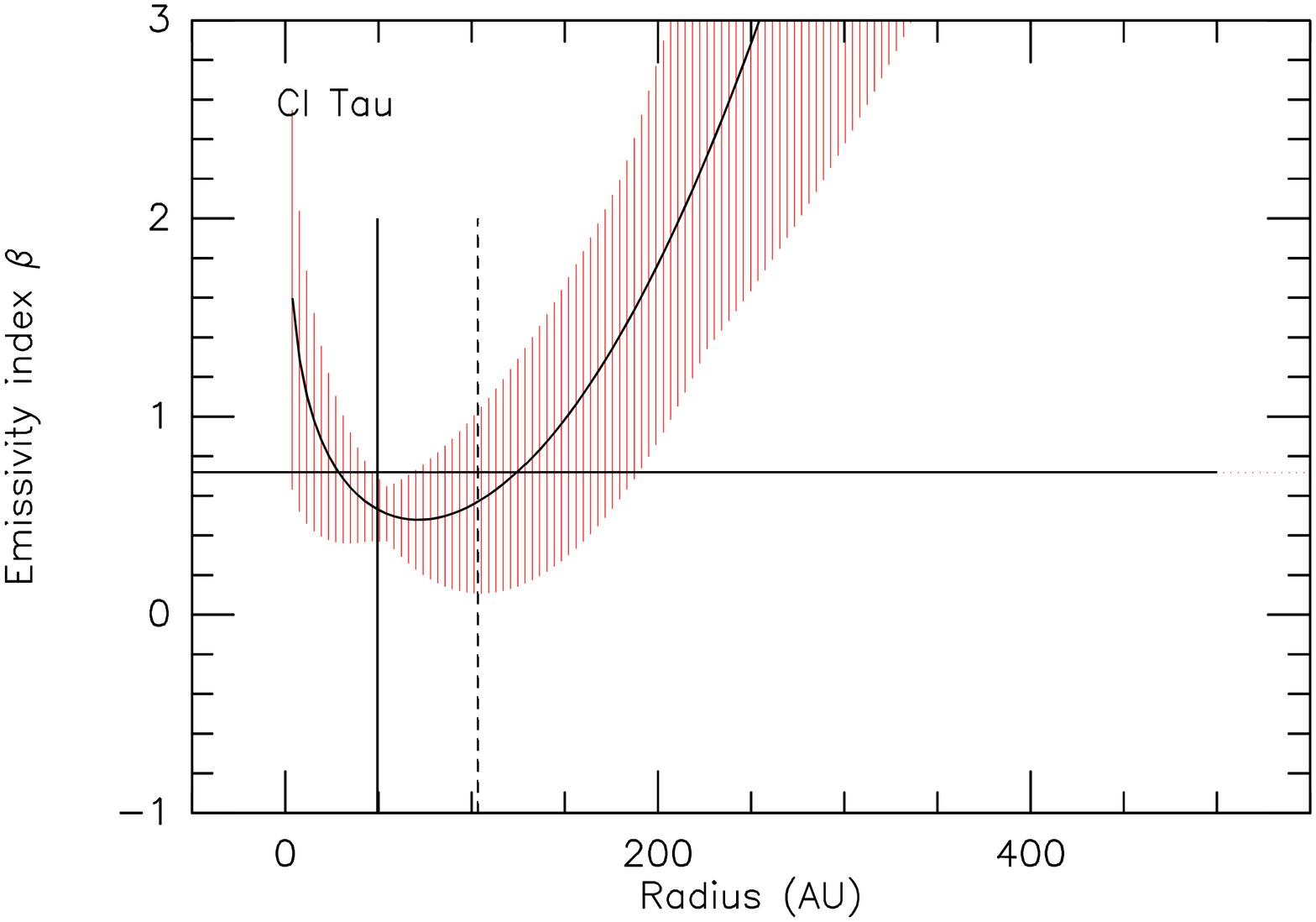}
  \includegraphics[angle=270,width=8.0cm]{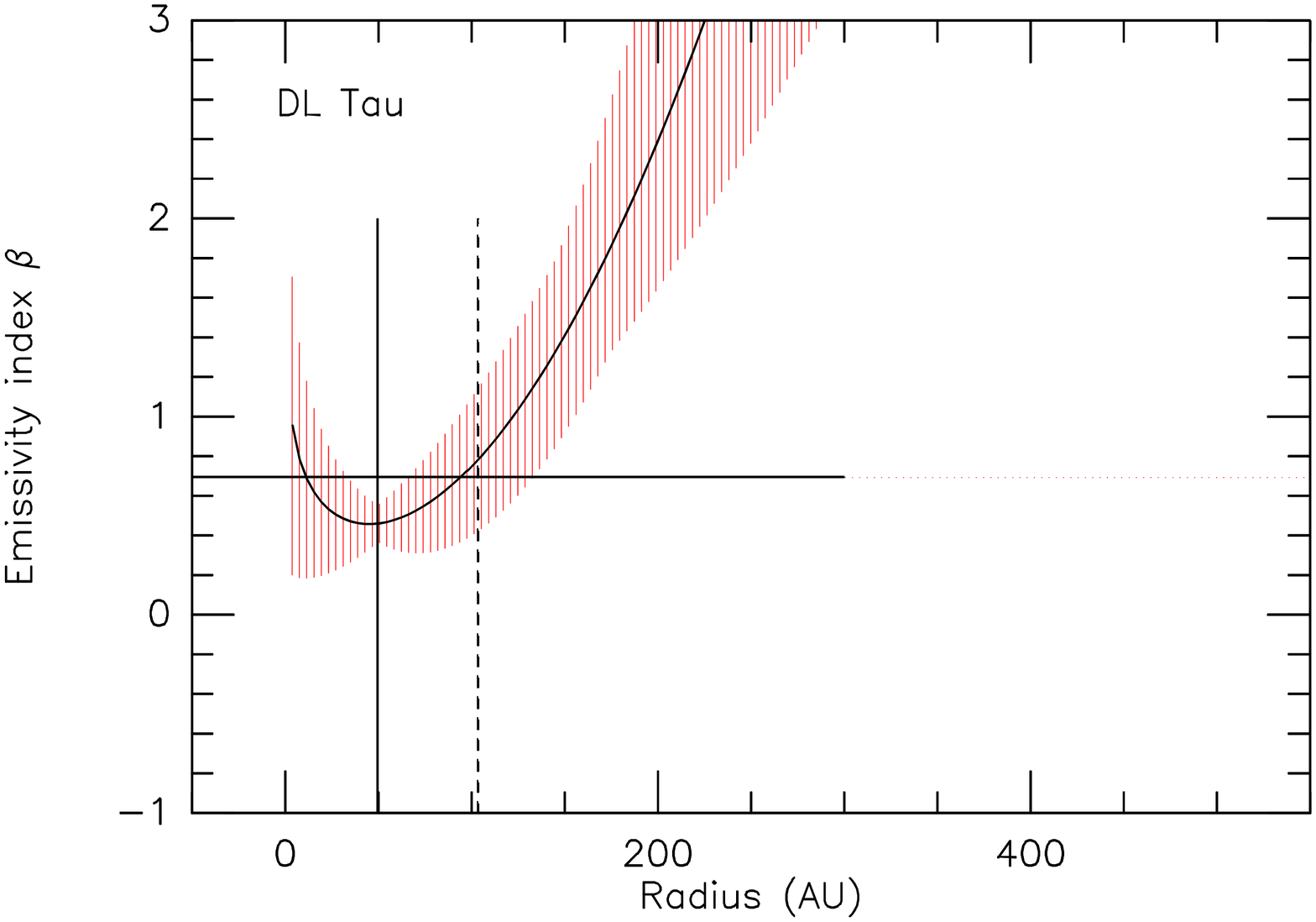}
  \caption{Apparent values of the emissivity index as a function of radius for CI\,Tau and DL\,Tau for Model 2}
  \label{fig:betaradius}
\end{figure*}

\begin{figure*}
  \includegraphics[angle=270,width=8.0cm]{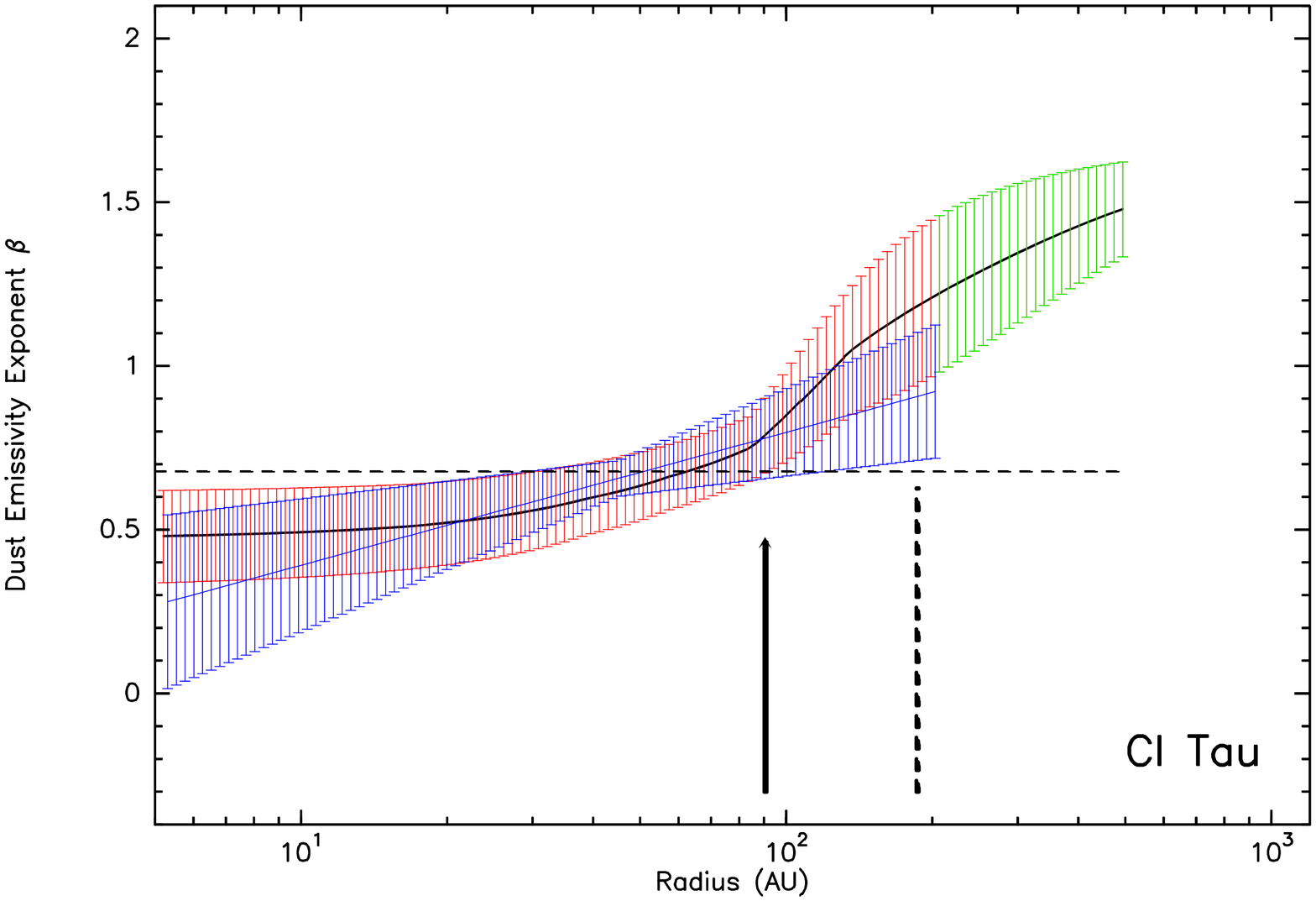}\hspace{1.0cm}
  \includegraphics[angle=270,width=8.0cm]{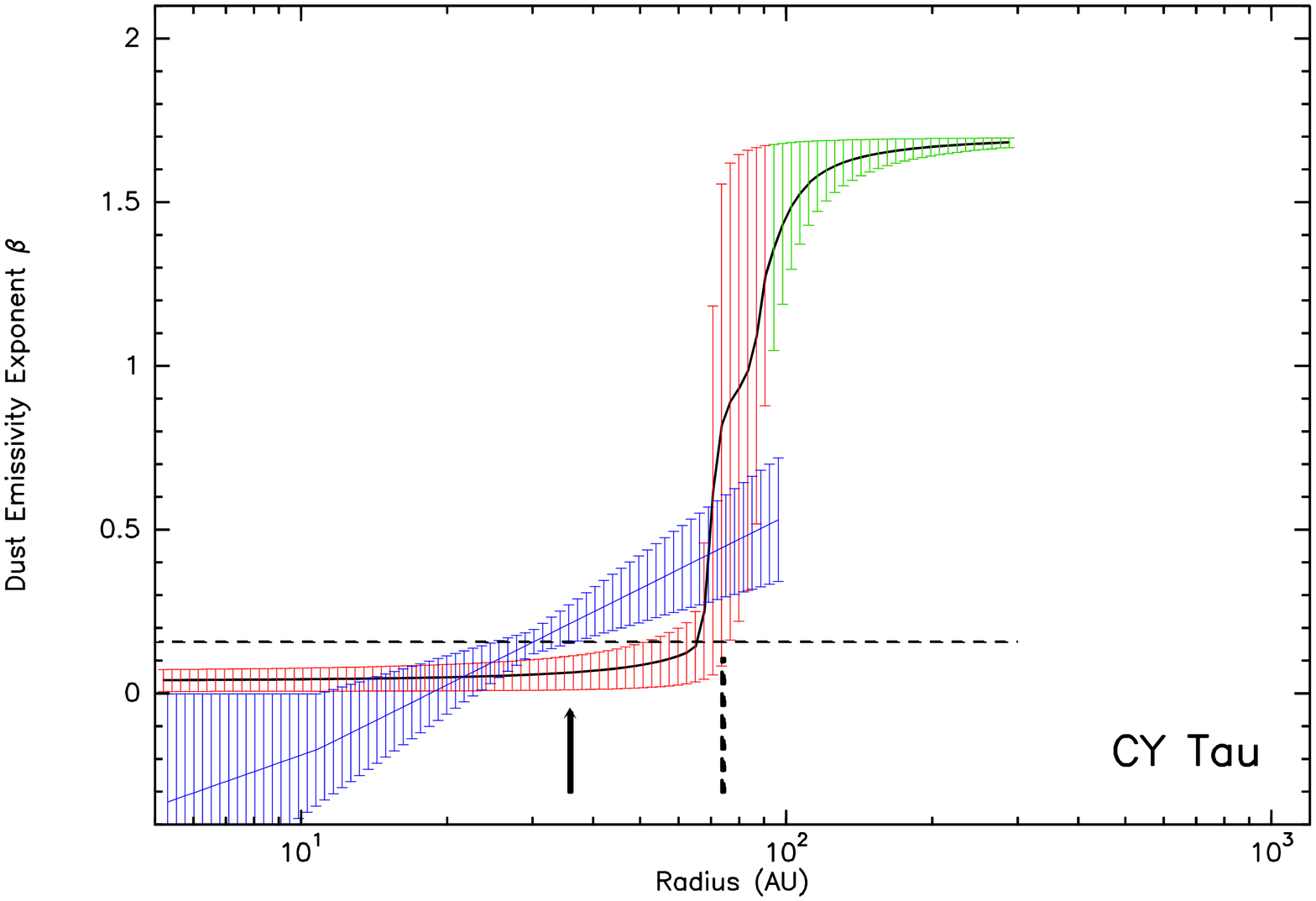}\vspace{0.3cm}\\
  \includegraphics[angle=270,width=8.0cm]{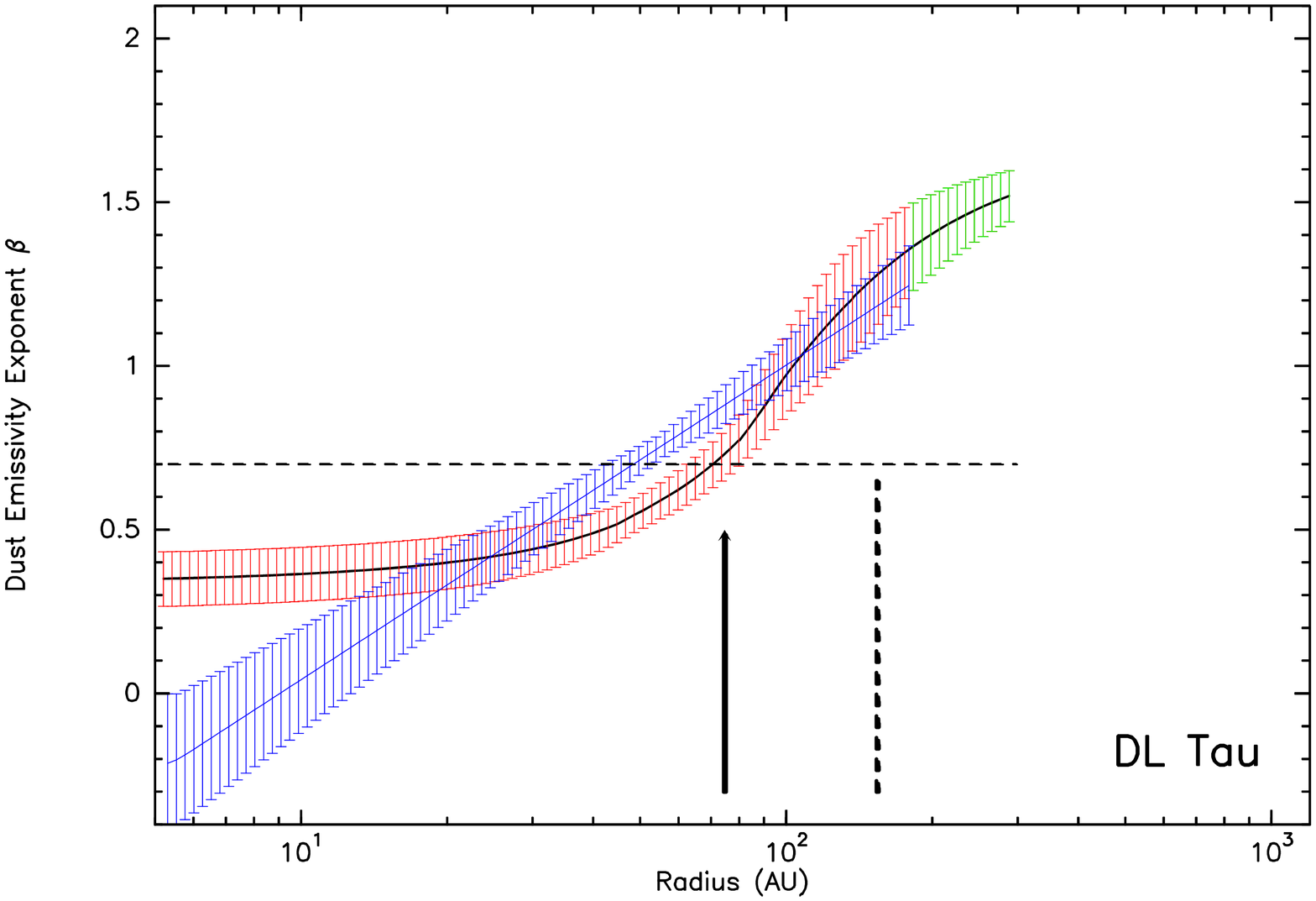}\hspace{1.0cm}
  \includegraphics[angle=270,width=8.0cm]{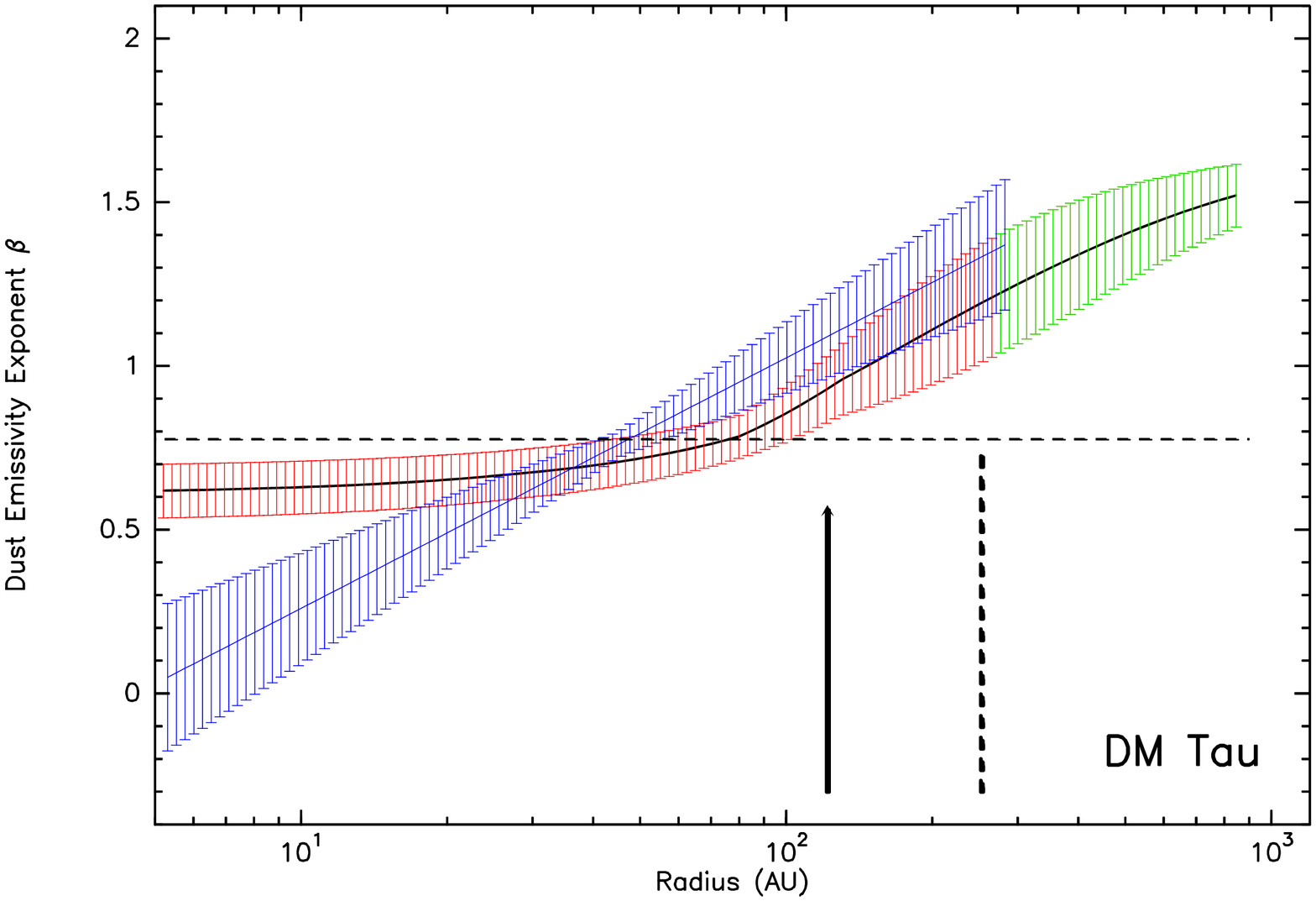}\vspace{0.3cm}\\
  \includegraphics[angle=270,width=8.0cm]{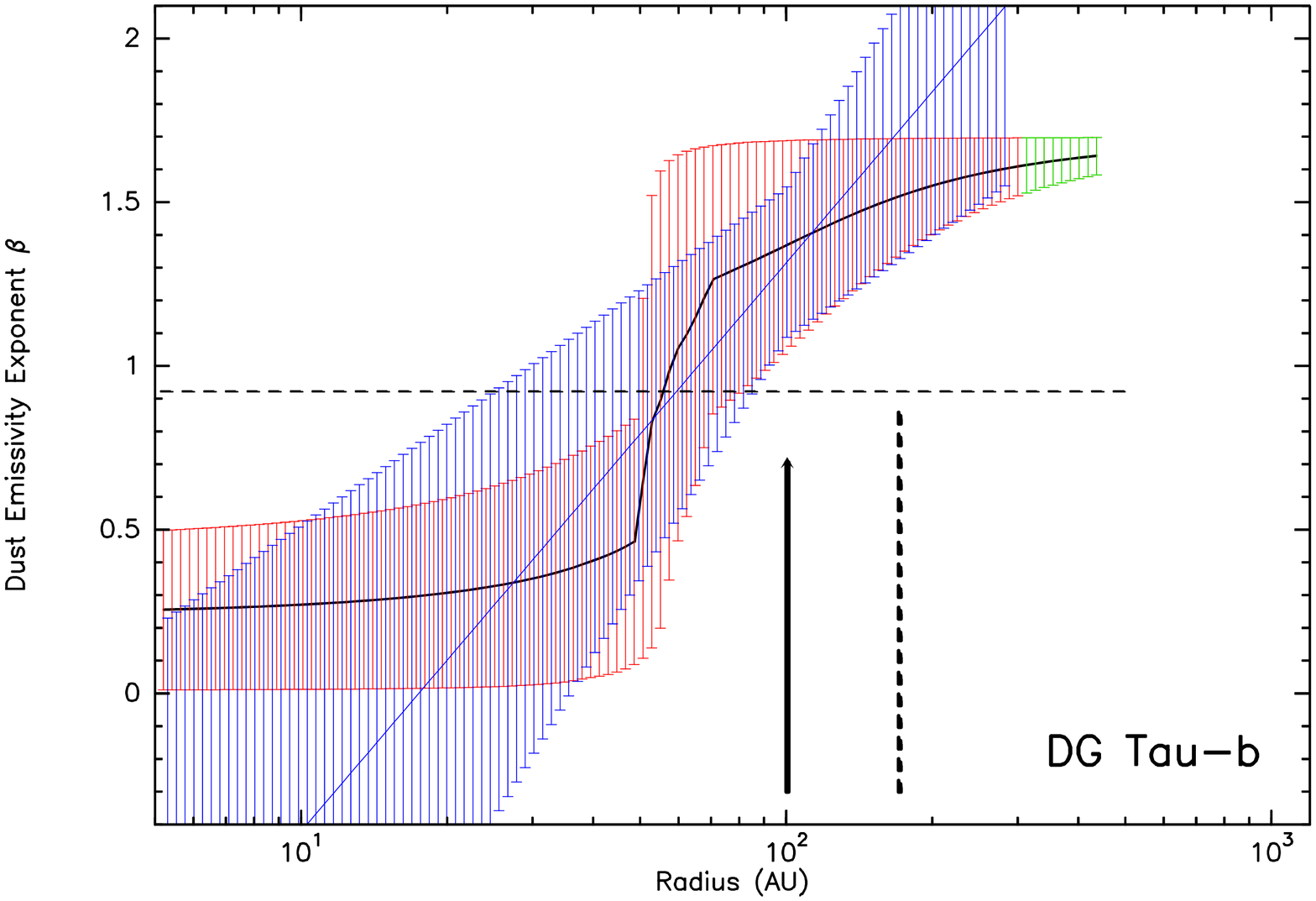} \hspace{1.0cm}
  \includegraphics[angle=270,width=8.0cm]{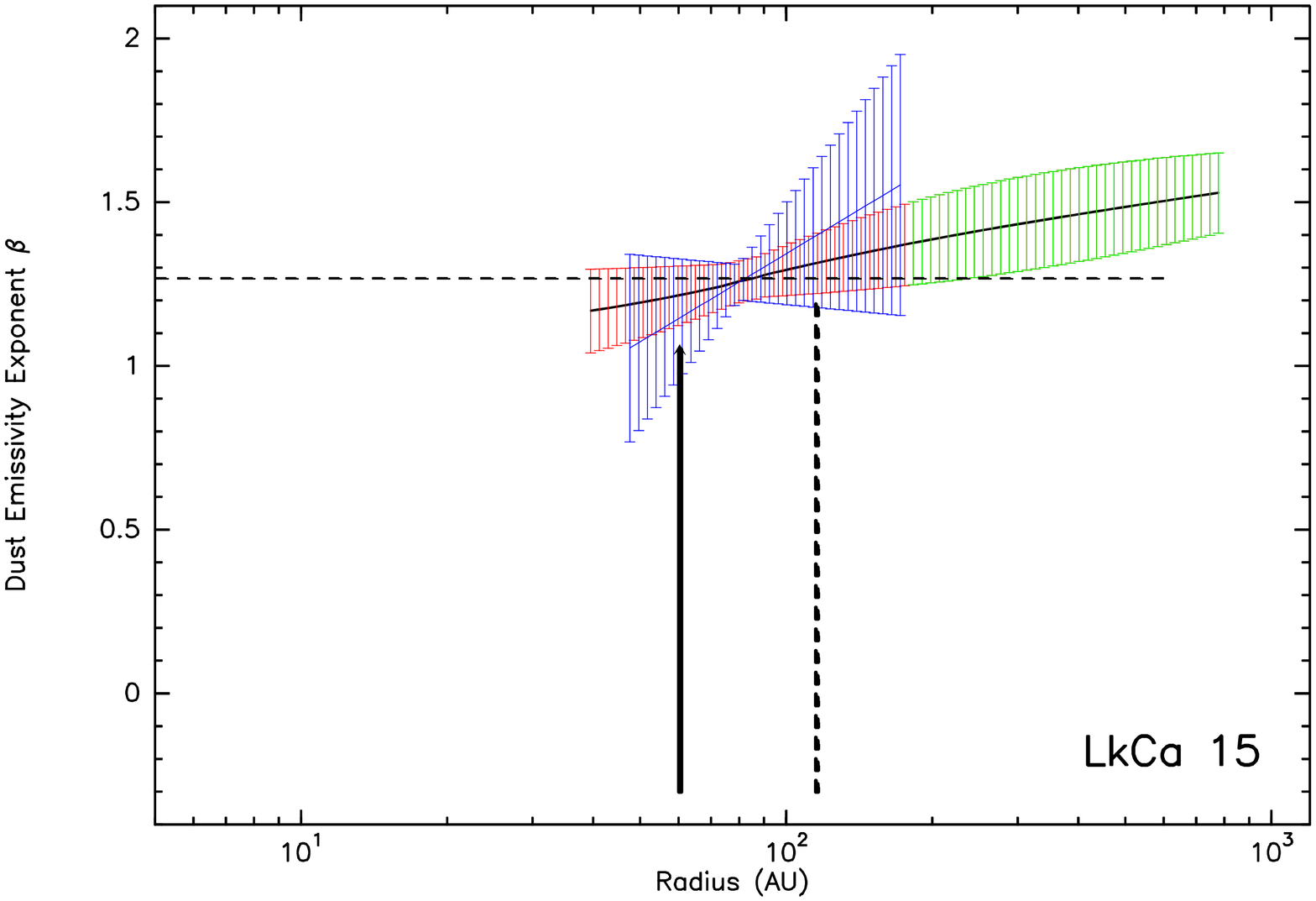} \vspace{0.3cm}\\
  \includegraphics[angle=270,width=8.0cm]{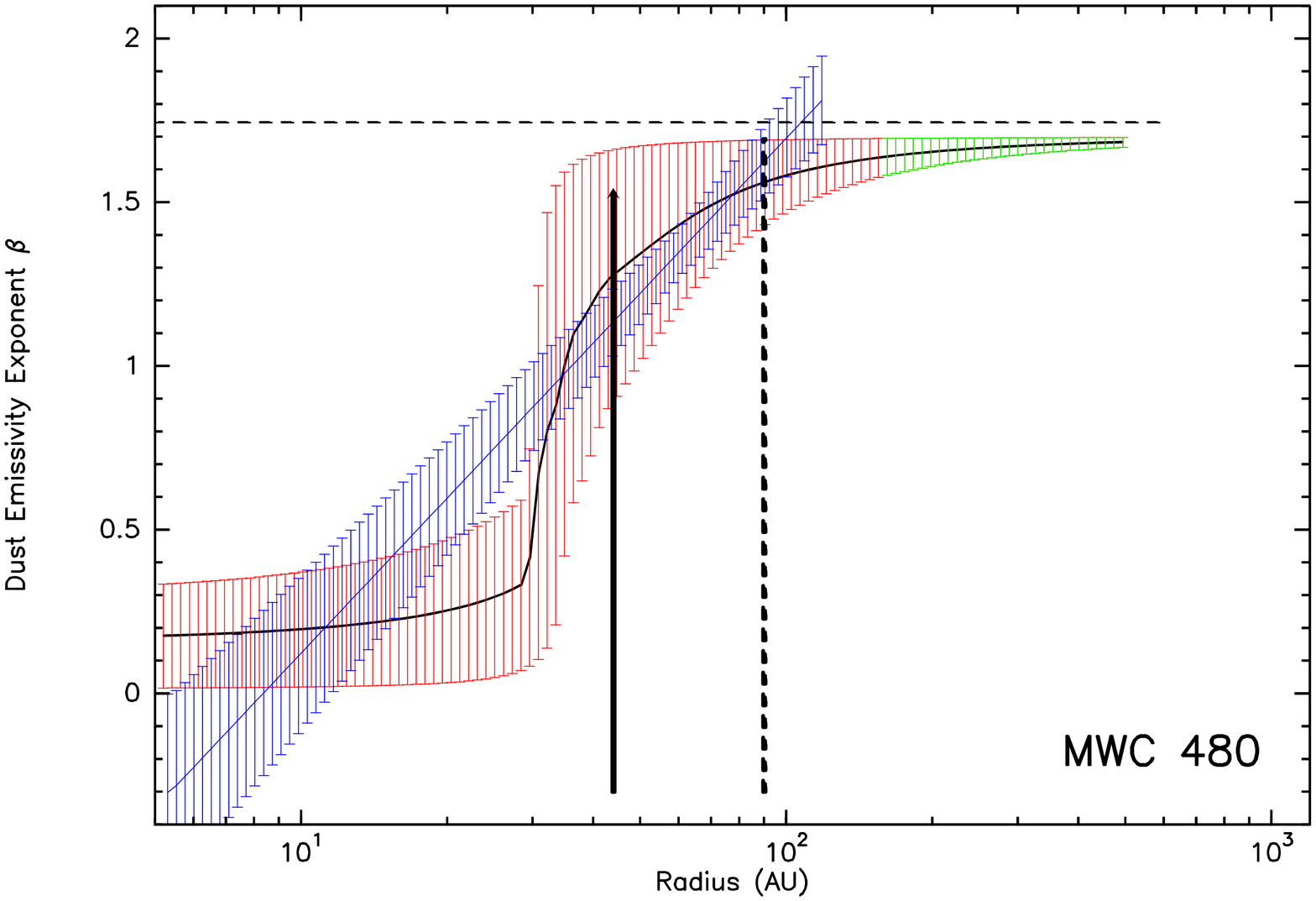} \vspace{0.3cm}\hspace{1.0cm}
  \includegraphics[angle=270,width=8.0cm]{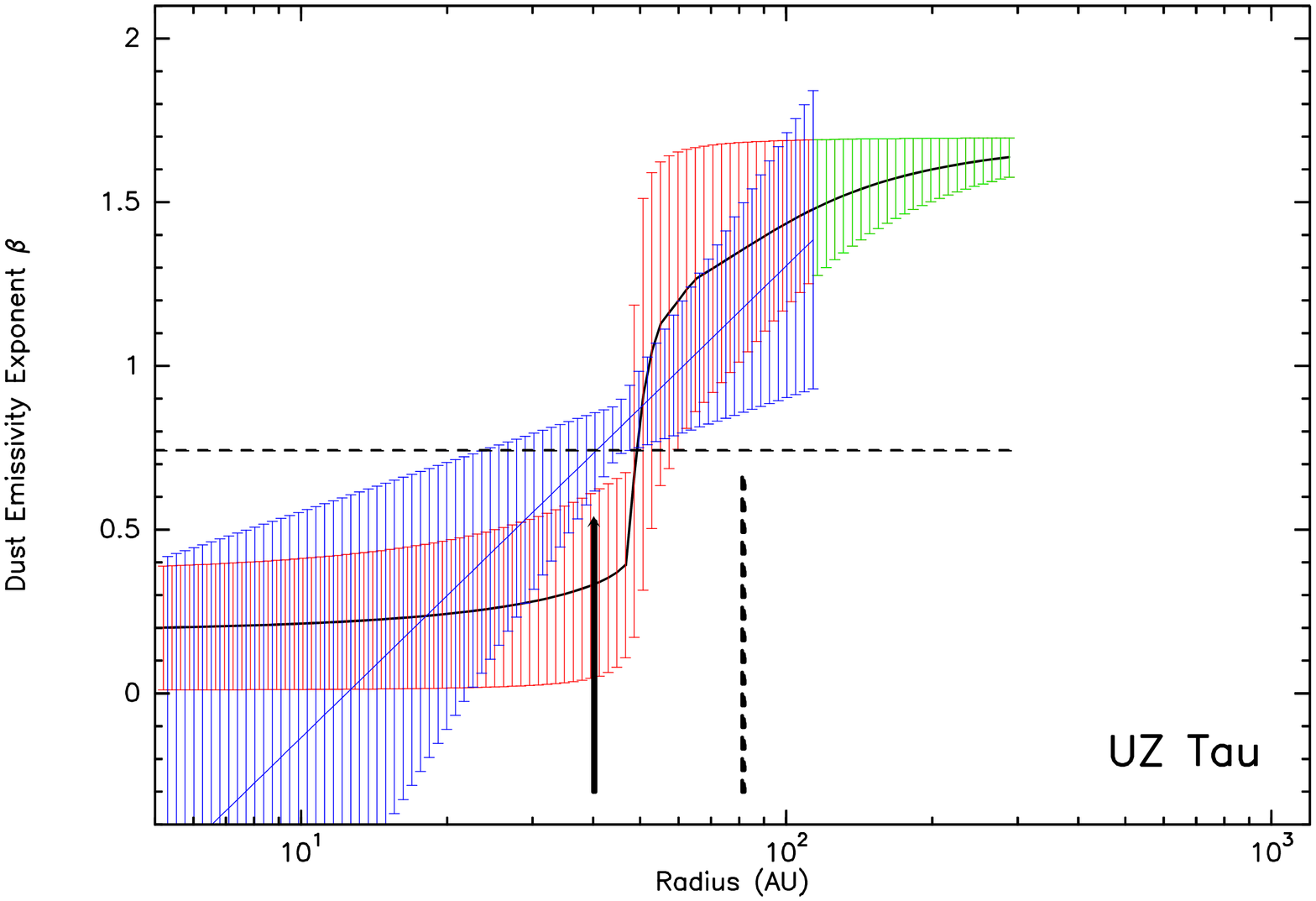} \vspace{0.3cm}\\
  \caption{Constraints on the variations of the dust emissivity index $\beta$ with of radius. The red hatched area indicate the allowed range of values using the prescription of Eq.\ref{eq:betaatan}. The blue hatched area uses the power law
  prescription of Eq.\ref{eq:betapower3} and is truncated at the outer radius found in the power law model. The thick vertical line
  indicates $R_c$, while the dashed line is $R_t$.}
 \label{fig:betaratan}
\end{figure*}

\begin{table*} 
\caption{Radial dependence of dust emissivity}
\label{tab:betaradius}
\begin{tabular}{l|c|ccccc}
\hline
\hline
(1) & (2) & (3) & (4) & (5) & (6) & (7) \\
Source & $\beta_r $ & Pivot $R_b$ (AU)  & Width $R_w$ (AU)  & $\Delta \chi^2$ &  $R_t$ & $\gamma$ \\
CI Tau & $0.18 \pm 0.10$ & $110\pm 25$ & $150 \pm 90$  & 13 & $90 \pm 5$ & $0.33\pm0.04$ \\ 
CY Tau & $0.32 \pm 0.13$ & $90 \pm 10 $ & $5 \pm 5$ &  17 &  $22 \pm 1$ & $0.17 \pm 0.07$ \\ 
DL Tau & $0.42 \pm 0.07$ & $90 \pm 9$ & $65 \pm 20$ & 63 &  $75 \pm 5 $ & $ 0.37 \pm 0.03$ \\ 
DM Tau & $0.33 \pm 0.09$ & $110 \pm 25$ & $245 \pm 140$ & 14 &  $125 \pm 11 $ & $0.48 \pm 0.05$ \\ 
DG Tau-b & $0.75 \pm 0.31$  & $60 \pm 11$   & $27 \pm 27$ &  35   &  $72 \pm 45$ &  $1.25 \pm 0.27$ \\
\hline
DG Tau & $0.27 \pm 0.22$ \\
FT Tau & $-0.38 \pm 0.30$ & $> 60$ &  & -4 & \\
LkCa 15 & $0.39 \pm 0.44$ & $-60 \pm 160$ & $160 \pm 360$ & 2 \\
MWC 480 & $0.30 \pm 0.19$ & $45 \pm 10$ & $130 \pm 16$ & -60 & $> 70$  & $1.50\pm0.10$ \\
\textit{MWC 480} & - & $36 \pm 3$ & $6 \pm 5$ & 0 & $41\pm1$ & $0.38\pm0.03$ \\
UZ Tau & $0.63 \pm 0.41$ & $55 \pm 8$ & $14 \pm 26$ & 10 & $ 25 \pm 7$  & $ 1.05 \pm 0.25 $ \\
\hline
\end{tabular}\vspace{0.1cm}\\
{$\beta_r$ as defined in Eq.\ref{eq:betapower3}. $R_b$ and $R_w$ as defined in Eq.\ref{eq:betaatan}. $\Delta \chi^2$ is the $\chi^2$ offset (positive means better fit) of the fit using Eq.\ref{eq:betaatan} compared to the constant $\beta(r)=\beta_m$ hypothesis. $R_t$ and $\gamma$ in Col.6-7
are the parameters of the softened edge surface density distribution derived assuming the dust properties from \citet{Isella+etal_2009}, see Fig.\ref{fig:dust-isella}.
For MWC\,480, the line in italics is under the assumption of $\kappa_\nu(1.3\,\mathrm{mm}) = 2$~cm$^2$g$^{-1}$.}
\end{table*}

The shape of the radial dependence of $\beta$ in Fig.\ref{fig:betaradius}, and the logarithmic dependence in Eq.\ref{eq:betapower3}, are simple results of the choice of shape of the surface emissivity distribution, and have no physical constraints attached. In particular, apparent values of $\beta$ below 0 or above $1.7$ can result from  such an analysis.

Because of the limits in angular resolution and sensitivity, some prescription of the evolution of the dust properties as a function of radius, assuming realistic conditions, must be specified to obtain better insights on the dust properties versus radius. A poor choice could make the radial dependence apparently non significant. To illustrate the problem, we used
\begin{equation}
\beta(r) = 0.85 + \frac{1.7}{\pi} \mathrm{atan}\left(\frac{r-R_b}{R_w}\right) ,
\label{eq:betaatan}
\end{equation}
which varies between 0 (large grains) and 1.7 (small ISM-like grains). With this functional, we obtain significantly better fits, at least by $3 \sigma$, but up to $8 \sigma$ in DL\,Tau (see Table \ref{tab:betaradius}).  Furthermore, the improvement does not depend on the assumed shape of the surface density: power laws or tapered edges yield identical results for the pivot $R_b$ and width $R_w$, although the errorbars on these parameters are typically 30\% lower in the power law hypothesis. Note that there is a fairly strong correlation between $R_b$ and $R_w$, and their errorbars are in general not symmetric. To better illustrate the variations of $\beta(r)$, the resulting range of allowed values for $\beta(r)$ for each source is given in Fig.\ref{fig:betaratan}. The logarithmic dependence found from Eq.\ref{eq:betapower3} is also indicated. Both functionals give approximately the same values in the regions where $\beta(r)$ is actually constrained, that is from 30 AU to the $R_\mathrm{out}$ of the power law. However, the log dependence fails to characterize the sharpness of the transition from low to high values of $\beta$.

Finally, although our analysis of $\beta$ excludes the flux calibration uncertainty, it is worth emphasizing that
this does not affect the radial variations of $\beta(r)$, but only the mean value $\beta_m$. It also does not affect
the relative differences in $\beta_m$ between sources, because all observations were made in an homogeneous way, with
all spectral index measurements based on an assumed index of 0.6 for MWC 349.

\subsubsection{Absorption coefficient $\kappa_\nu{1.3\,\mathrm{mm}}$}

If grains vary in size with radius, the absorption coefficient $\kappa(\nu,r)$ will also vary. The surface density laws that were derived so far were derived assuming $\kappa(\nu,r) = \kappa(\nu_0) (\nu/\nu_0)^{\beta(r)}$ with $\nu_0= 230$~GHz.
In practice, there is no physical justification for any value for $\nu_0$, because for essentially all models of grain growth the absorption coefficient and the apparent emissivity index vary simultaneously in a more complex way. One can attempt to use a more physical approach to the grain properties, using dust absorption coefficients derived from a physical model \citep[e.g.][and references therein]{Draine_2006}. For example,
\citet{Isella+etal_2009} \citep[see also][]{Natta+etal_2004}
derived  the absorption coefficient from a fixed grain composition, with a size distribution controlled by a single variable parameter. The size distribution is a power law with a fixed minimum and maximum radius and an exponent $g$. The absorption coefficient $\kappa$ and apparent emissivity index $\beta$ at 1.3\,mm are plotted as a function of $g$ in Figure \ref{fig:dust-isella}.  From this dust model, we can derive a function $\kappa(\beta)$, which can be used in our model with the same assumptions about the radial dependency of $\beta(r)$ as previously done.

With the prescription of the opacity law described by Fig.\ref{fig:dust-isella}, and $\beta(r)$ as in Eq.\ref{eq:betaatan}, the pivot radii $R_t~/~R_c$ are not changed very significantly. The largest changes are for CY\,Tau and UZ\,Tau, where $R_c$ decreases by 50\%, DM\,Tau, where it increases by 50\%, and MWC\,480. Effects on $\gamma$ are negligible except for MWC\,480 and UZ\,Tau (see Table \ref{tab:betaradius}). The relatively small effect on $R_c,$ and $\gamma$  is explained  because  $\beta(R_c)$ is close to 1 in most of the sources studied, and for this value $\kappa_\nu(230\,\mathrm{GHz})$ has an extremum. Thus the variations of $\kappa_\nu(r)$ around $R_c$ are relatively moderate, and accordingly the shape of the derived surface density is mildly affected by the radial variations of $\kappa_\nu(r)$.

However, these small apparent changes may be misleading, because they implicitly depend on the assumed shape of the surface density law. As $\beta(r)$ is getting close to 0 in the disk center, and thus the absorption coefficient $\kappa(1.3\,\mathrm{mm})$ could be much smaller at small radii, it is also completely possible to have a much steeper surface density gradient in the inner 40 AU. This remains hidden from our study because of the angular resolution, but also because the inner 20 -- 40 AU become optically thick in
some sources.  If  steep gradients like this exist, longer wavelength images should be able to reveal them. The strong changes observed in $R_c,\gamma$ for MWC\,480 and UZ\,Tau are also manifestations of this effect, although at larger scales.

In our sample, only HL\,Tau was studied with sufficiently high angular resolution at 7\,mm and 1.3\,cm to confront images with the above prediction. Although surrounded by a diffuse halo, the 7\,mm image of \citet{Carrasco-Gonzalez+etal_2009} is indeed very centrally peaked, but a quantitative comparison with our results is not directly possible because it displays complex structures.

\subsection{MWC 480 revisited}
\label{sec:sub:mwc480}

In the simple $\beta(r)=\beta_m$ hypothesis, the disk of \object{MWC 480} appears sufficiently optically thick
at 230 GHz to allow the derivation of the dust temperature \citep{Pietu+etal_2006}. The optically thick region is even
large enough to constrain the exponent $q$ to some extent.
Leaving both $T_0$ and $q$ as free parameters, Model 1 and Model 2 give different best fits
for the temperature, because in the best fit for Model 2, the radius at $\tau(1.3\,\mathrm{mm})=1$ is much larger (80 AU) than for Model 1 (35 AU, see Fig.\ref{fig:allmwc480}).  In addition, $\beta_m$ is larger by about 0.3 in Model 2 than in Model 1, because the optically thick core is much larger. Furthermore, since the extrapolated temperatures in the best Model 2 are very low (7 K at 100 AU, and 2.7 K at 400 AU), the emission is no longer in the Rayleigh-Jeans domain, and $\beta_m$ increases because the corrections are larger at 1.3\,mm than at 2.7\,mm. 
In practice, Model 2 finds a  low temperature with a steep exponent ($\simeq 0.6$) because of two effects: i) the brightness
is identical at 2.7\,mm and 1.3\,mm in the inner 40 AU, and ii) the imposed shape of the surface density is too flat in the inner 80 -- 100 AU (in order to provide sufficient opacities beyond 100 AU). To account for these two constraints, an optically thick core of 80 AU is fitted, with a  steeply decreasing temperature. High temperatures can only be found by allowing the surface density to fall faster than Model 2 allows between 40 and 80 AU.

Clearly, in this case, although low temperatures are needed in the inner regions, extrapolating the same  power law introduces non-physical biases on the disk mass and on $\beta_m$. Leveling the temperature to a minimum value of 12~K beyond 45 AU provides a better fit to the observations, and allows us to bring back $\beta$ below 2. This may be an indication of the temperature rise with radius that is expected to begin when the opacity for re-emission drops below 1. However, the very low apparent temperatures in this object are surprising because of the  luminous central star. This may be linked to the geometry of that source. From its IR SED,  MWC\,480 is a Group II Ae disk, which is interpreted as a self-shadowed disk with small flaring \citep{Meeus+etal_2001}. Indeed, it has never been detected in scattered light, despite a fairly favorable inclination.  Yet, the temperature derived from $^{13}$CO is $\simeq 23$ K at 100 AU, with an exponent $q=0.4 \pm 0.1$ \citep{Pietu+etal_2007}, and if the disk remains optically thick even
at 3\,mm, we would expect dust and gas to be thermalized at the same temperature.

Allowing $\beta(r)$ to change with radius also offers a much more attractive solution to the continuum emission of MWC\,480.
The flattening of the emission in the inner 50-80 AU is no longer ascribed to an optically thick core at low temperatures, but to a flattening of the surface density distribution, while the ratio of 2.7 to 1.3 mm emission is matched by allowing $\beta(r)$ to become small in the inner 30 AU. Although it is equivalent in $\chi^2$ to the constant $\beta(r) = \beta_m$ solution, this new model agrees with less extreme dust temperatures. In fact, the dust emission is largely optically thin in this case, and there is a substantial degeneracy between the dust temperature and the derived disk mass / surface density. A lower limit to the dust temperature at 100 AU is 23 K (assuming $q=0.4$), which is consistent with the temperature derived from $^{13}$CO line emission by \citet{Pietu+etal_2007}. This lower limit was used to derive the surface density.

If we use $\kappa(\beta)$ as implied in Fig.\ref{fig:dust-isella}, the fit quality is slightly degraded, but most importantly, the derived shape for the surface density and the temperature profile are significantly affected (see Table \ref{tab:betaradius}). We find $\gamma \simeq 1.5$, and a large transition radius $R_t > 70$ AU, much like for GM\,Aur. The best-fit temperature profile is flat, $q = 0.0 \pm 0.1$, with $T > 25$ K.  The higher $\chi^2$ value derived under these assumptions may be related to an oversimplified temperature profile, as in the simpler analysis $q\simeq 0.5$ was found in the inner regions.

\section{Discussion}
\label{sec:discussion}

\subsection{Dust properties}
\label{sec:discussion:dust}

From the above results, we find large grains ($\beta < 0.5$) in the inner 60 to 100 AU, and small grains beyond for seven sources
(CI Tau, CY Tau, DL Tau, DM Tau, DG Tau-b, MWC,480 and UZ\,Tau-E).  Two other sources in our sample have very low $\beta$:  the FT Tau disk is truncated at 60 AU, while DQ Tau has not been observed with sufficient resolution at 2.7 mm, so that $\beta(r)$ is not constrained in the outer regions. A third source may have low $\beta$ up to 60 AU: T\,Tau N, although
we interpreted it as being optically thick.
On the other hand, $\beta$ is not constrained in the inner region for the two other sources observed with sufficiently high resolution in our sample, because in DG\,Tau, the inner 50 AU may be optically thick, while for LkCa 15, the inner 50 AU are (largely) devoid of dust. For HH\,30 we find a low $\beta$ below 120 AU, while it is known from the scattered light images that small grains exist at least up to 250 AU or even 430 AU \citep{Burrows+etal_1996}, the outer radius of the gas distribution \citep{Pety+etal_2006}. Finally, in HL\,Tau, large grains exist in the inner 20 AU, as shown by the 1.3\,cm and 7\,mm images \citep{Carrasco-Gonzalez+etal_2009}.
Thus, in essence, \textit{all} sources in our high-resolution sample show large grains (low $\beta$) below 100 AU and small grains beyond,
although the detailed shape of the radial dependence cannot be characterized by our data.

The apparent variations of $p$ with wavelength observed by \citep{Banzatti+etal_2010} for CQ\,Tau also points out towards an increase of $\beta(r)$ with radius in that source. Moreover, although they considered it to be insignificant, the same trend is found in
RY Tau and DG\,Tau by \citet{Isella+etal_2010}. Thus, the radial dependency of $\beta(r)$ appears to be a general property of disks.
%
%

Our findings that $\beta$ is low in the inner 60 to 100 AU of all disks in which we can constrain the radial dependency
also sheds new light on the results quoted by \cite{Ricci+etal_2010}. \cite{Ricci+etal_2010} found a lower average value for the
spectral index $\alpha$ for disks with low 1.3\,mm flux than in disks that show strong emission. A possible interpretation is that these weaker disks are optically thick and very small, like those surrounding the binary Haro 6-10. These weak disks may just miss the extended, low brightness parts with high values of $\beta$ that we found in bright sources. In our sample, a clear example for this behaviour is FT Tau. Given our measured $R_b$, a testable prediction is that these faint disks should be smaller than about 100 AU in radius. Note that this does not address the origin of the small size for these disks, although tidal truncation is an obvious candidate. On the other hand,  in AB Aur, which has an inner hole around 100 AU \citep{Pietu+etal_2005}, the mm emission is coming from the small grain regions, which
results in a mean $\beta_m=1.4 \pm 0.2$, which is different from all other sources. Such a high $\beta_m$ is not
an indication of different grain growth in this source, but just a side effect of the radial dust distribution.
We further stress that the $\beta_m$ values derived in all previous analyses represent an ill-defined average over the disk structure.

\subsection{The shape of the surface density distribution}
\label{sec:discussion:density}
Given the high resolution and sensitivity, can we decide which model  fits the data better? The lowest $\chi^2$ is the usual indicator, but care must be taken that the $\chi^2$ is not affected by different biases between the two models owing to numerical effects in the model computation.
The precision required for this is always higher than the precision required to obtained converged parameters and errors within a given model, because the discretization effects impact models differently (see Appendix \ref{app:sampling}, available online only). For the models considered, the problem is somewhat relaxed because they both derive from a generic one (see Eq.\ref{eq:edge}).
We nevertheless checked by using oversampled grids that the $\chi^2$ results were converged. 

From Table \ref{tab:main}, the softened-edge model does not appear superior to the power law
model to represent the observations. In this process, the compact optically thick sources should be ignored. For these sources, the data are insensitive to the true shape, but can be significantly affected by small instrumental effects. For example, the seeing that results in flux spreading because of atmospheric phase variations tends to produce a small halo around the compact core. In our sample of 23 individual objects, this may affect five sources. Of the remaining objects, four sources are best represented by a power law: DG\,Tau,  DQ\,Tau, HL\,Tau and (marginally) DM\,Tau.  On the other hand, six sources are better fitted by the exponential-edge model: CI\,Tau, CY\,Tau, DL\,Tau, UZ\,Tau E, and marginally LkCa\,15. Both models fit equally well the last seven sources, which were observed with lower angular resolutions except for DG\,Tau b.

Despite the high resolution (projected baselines above $500 k\lambda$) and sensitivity, the shape of the surface density remains difficult to constrain.
The truncated power law was initially used because it provides the simplest parametric model. It is furthermore not linked to any specific physical disk model, a property which can be seen either as an advantage (by providing no specific bias) or handicap (as having no physical ground). Its principal failure was its inability to represent continuum \textit{and} spectral line emission with the same outer radii \citep{Pietu+etal_2007}. The softened-edge model has recently gained favor because, as suggested by \citet{Hughes+etal_2008}, it may provide a framework that can explain both the continuum and optically thick CO emission. The exponential taper is often referred to as having a physical background, because viscosity is expected to spread out initially small disks. However, the exponential taper is only a specific solution of self-similar evolution of a viscous disk with a power law distribution of the viscosity (with a constant in time exponent). In practice, self-similarity and time independence are unlikely to strictly apply to real disks, so the resulting specific shape is also an approximation. Any core+halo structure would essentially yield the same result, provided the halo is just dense enough to explain the molecular emission, but tenuous enough to have little continuum emission from dust. This core+halo structure was invoked by \citet{Dutrey+etal_1994} and \citet{Guilloteau+Dutrey+Simon_1999} to interpret the circumbinary environment of GG\,Tau.

In the strict framework of a viscous disk model, we find values of $\gamma$ that are somewhat larger (and with higher dispersion) than those derived in the previous studies by \citet{Isella+etal_2009} and \citet{Andrews+etal_2009}.
The discovery of a radial dependence of the dust properties brings additional complexity to the problem.
Clearly, the surface density of the gas is not well traced by the continuum emission at a specific frequency in this picture. It all depends on how the dust emissivity $\kappa(\nu,r)$ changes as a function of radius, so that the derived $\gamma$ is expected to  depend
on the assumed dust properties.

It is also important to realize that the derived dust masses of the disk may be significantly affected by the variations of the dust properties with radius. Table \ref{tab:masses} indicates the disk masses obtained for Model 2 using Eq.\ref{eq:betaatan} with
the two different hypotheses on the dust absorption coefficient ($\kappa_\nu$(1.3\,mm) constant or tied to $\beta(r)$ as from Fig.\ref{fig:dust-isella}). In our sample, although the effect is small for the other sources (about 20 \%), the masses of the CY Tau and DM Tau disks are strongly modified when using the dust properties from \citet{Isella+etal_2009}. In particular, the mass of the DM\,Tau disk becomes quite significant (0.2 \Msun) compared to the stellar mass  \citep[0.5 \Msun][]{Dartois+etal_2003}. Such a large mass would have significant effect on the rotation curve of the gas, while it is known to be Keplerian with high accuracy \citep[velocity exponent $0.50 \pm 0.01$,][]{Pietu+etal_2007}.

\begin{table}
\caption{Disk masses with variable dust properties}
\label{tab:masses}
\begin{tabular}{c|cc|c}
\hline
\hline
Source &  \multicolumn{1}{c}{$M_c$ ($10^{-3}$ \Msun)} & \multicolumn{1}{c}{$M_v$ ($10^{-3}$ \Msun)} & \multicolumn{1}{c}{Ratio} \\
\hline
CI Tau&       43 $\pm$        4 &       51 $\pm$        5 &      1.18$\pm$      0.18\\
CY Tau&       18 $\pm$        1 &       46 $\pm$        2 &      2.52$\pm$      0.09\\
DG Tau&       36 $\pm$        5 &       42 $\pm$        6 &      1.18$\pm$      0.27\\
DG Tau-b&    151 $\pm$       59 &      179 $\pm$       60 &      1.19$\pm$      0.73\\
DL Tau&       51 $\pm$        1 &       60 $\pm$        1 &      1.18$\pm$      0.04\\
DM Tau&       32 $\pm$        8 &      192 $\pm$       49 &      6.07$\pm$      0.52\\
UZ Tau&       24 $\pm$        1 &       32 $\pm$        2 &      1.33$\pm$      0.14\\
\hline
\end{tabular}\\
{$M_c$ is the disk mass for Model 2 (tapered edge) for $\kappa(1.3 \mathrm{mm}) = 2$~cm$^2$.g$^{-1}$, while
$M_v$ is for $\kappa(1.3 \mathrm{mm})$ as in Fig.\ref{fig:dust-isella}. A gas-to-dust ratio of 100
was assumed. Ratio = $M_c/M_v$.}
\end{table}

This result is under the assumption of a ``normal'' gas-to-dust ratio of 100. However, the gas-to-dust ratio itself is expected to change as a function of time and radius in the disk. For DM\,Tau, the potentially large dust mass suggests that the gas-to-dust ratio must be decreased. Molecular tracers may help to constrain  the gas surface density more directly, but then a good understanding of the chemistry is required to recover the hydrogen content from the (very few) trace molecules that display strong enough lines to be observable: CO and its isotopologue $^{13}$CO,  HCO$^+$, CN, HCN, CS and H$_2$CO \citep{Dutrey+etal_1997}.

A simpler alternative is that the adopted dust properties are inappropriate. For example, with similar grain size distributions, but using a different dust composition (in particular porous grains), \cite{Ricci+etal_2010} derive dust opacities on the order of $3 - 20$~cm$^2$.g$^{-1}$ instead of $0.4 - 2$~cm$^2$.g$^{-1}$ from \citet{Isella+etal_2009}. The overall dependencies of $\kappa$ and $\beta$ upon the grain size distribution display the same characteristic behavior. Note, however, that it is possible to obtain $\beta$ values above 2, provided the grain size distribution as a relatively steep cutoff near $a_+ = \lambda/2\pi$, i.e. about 0.5 mm for $\lambda = 3$~mm, because the emissivity of a grain size $a$ has a pronounced maximum
at wavelengths $\sim 2\pi a$ \citep[see e.g.][]{Natta+etal_2004}, before dropping as $1/a$ at longer wavelengths. In the following, we scale down the surface densities of sources analyzed with radial dust opacity gradients by a factor 3, to avoid a different bias in the comparison with sources for which this analysis was not possible.

\begin{figure}
   \includegraphics[angle=270,width=\columnwidth]{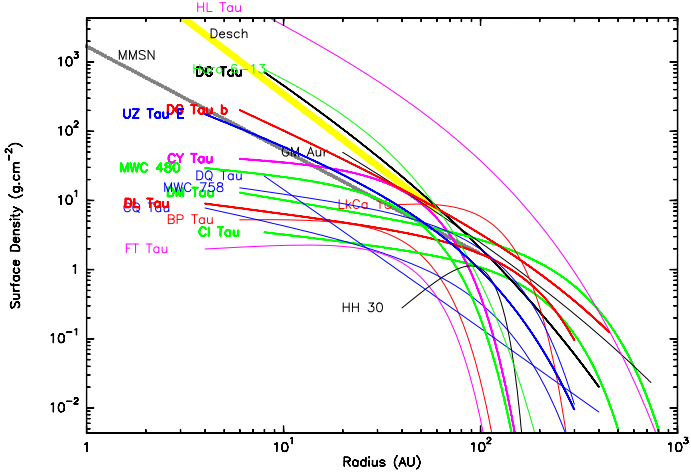}
\caption{Surface densities of observed sources. Thick lines are for sources in which a variation of $\beta$ and thus $\kappa$
with radius was derived. Thin lines are for sources for which we assumed $\kappa(1.3\,\mathrm{mm}=2$ cm$^2$.g$^{-1}$. The gray line is the MMSN, while
the yellow area indicates the Solar Nebula from Desch (2007).}
\label{fig:surface}
\end{figure}

Figure \ref{fig:surface} displays the resulting surface densities (of gas+dust) for the sources in the sample.  Uncertainties were omitted for clarity in this figure, but can be recovered from the Figs.\ref{fig:allbptau}-\ref{fig:alluztauw}.  We note that the younger sources have higher surface densities in the inner 50 AU than other objects. They are also more centrally peaked, on average. This picture is qualitatively similar to the predictions from viscous disk evolution.
Figure \ref{fig:surface} also displays the profiles derived for the Solar Nebula, the MMSN \citep[gray line]{Hayashi_1981} and the solution proposed by \citet{Desch2007}, which accounts for the early planet migration as proposed by the Nice model \citep{Tsiganis+etal_2005,gomes+etal_2005} (yellow range). The solution proposed by \citet{Desch2007} (see Appendix \ref{app:others}, available online only) for the Solar Nebula is a steady-state solution, which allows for sufficient time (few Myr) for the giant planets to reach isolation mass. In comparing with our results, it is important to realize that our observations constrain essentially the slope and surface density between 50 to 150 AU, while the other regions are obtained by extrapolation of the analytically prescribed shape. In our sample, only the youngest objects have sufficiently high surface densities to be compatible with the MMSN.

\subsection{Towards an evolutionary model ?}

\subsubsection{Viscous evolution of the gas disk}
\label{sec:visc:gas}

\begin{figure}
   \includegraphics[angle=270,width=\columnwidth]{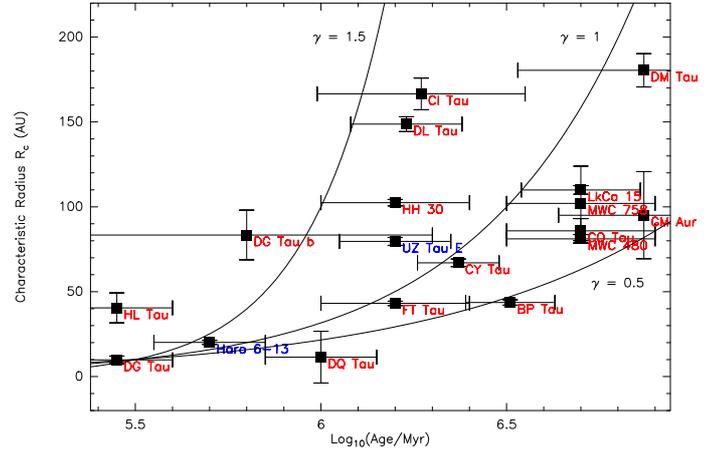} 
\caption{Characteristic radius $R_c$ (in AU) as a function of estimated stellar ages (in Log$_{10}$ of $10^6$ years).}
\label{fig:age-pivot}
\end{figure}

Figure \ref{fig:age-pivot} displays the characteristic radius $R_c$ as a function of estimated stellar ages. The figure apparently suggests an increase of $R_c$ with age. Performing a Spearman rank-order correlation test indicates a correlation coefficient of 0.60, and a small probability of random distribution (0.7 \% only). This correlation study does not include the error bars on age and $R_c$, however. Furthermore, the correlation coefficient is heavily influenced by the two youngest objects, DG\,Tau and HL\,Tau and the two oldest ones, GM\,Aur and DM\,Tau, all sources for which the power law model gives a better fit than the softened-edge model.
Nevertheless, taken at face value, our data seem to confirm the trend suggested by \cite{Isella+etal_2009}, which they have interpreted as evidence for the viscous evolution of disks.

\begin{figure}
   \includegraphics[angle=270,width=\columnwidth]{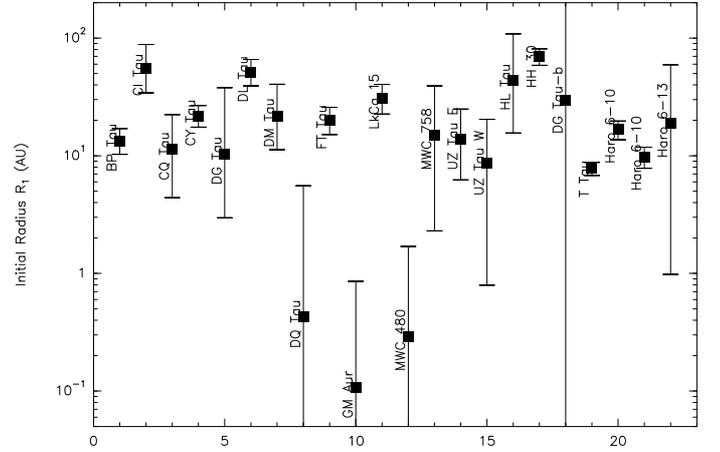}
\caption{Initial disk radii (AU).}
\label{fig:r1}
\end{figure}

\begin{figure}
   \includegraphics[angle=270,width=\columnwidth]{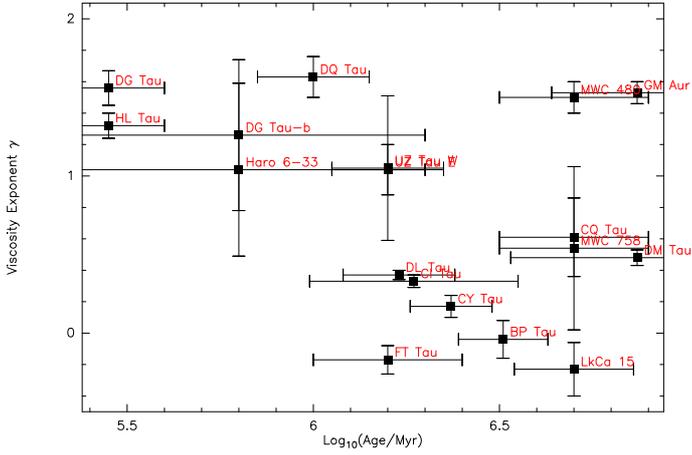}
\caption{Surface density exponent $\gamma$ as a function of estimated stellar ages (in Log$_{10}$ of $10^6$ years).}
\label{fig:age-gamma}
\end{figure}

\begin{figure}
   \includegraphics[angle=270,width=\columnwidth]{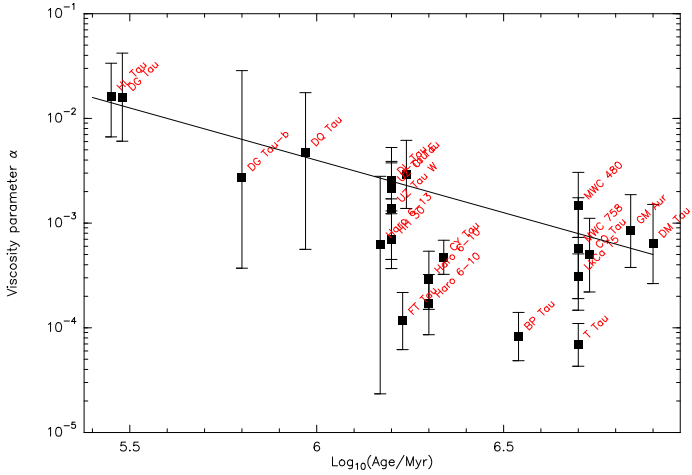}
\caption{$\alpha$ parameter as a function of estimated stellar ages.}
\label{fig:age-alpha}
\end{figure}

In the framework of self-similar viscous evolution, the surface density depends on five intrinsic parameters:
a normalization constant $C$, the initial disk radius $R_1$, a normalized age $T$, the viscosity $\nu_1$ at radius $R_1$ and its radial exponent $\gamma$.
We have three measurements from our study ($M_d~\mathrm{or}~\Sigma_0, R_c, \gamma$), the stellar age $t_*$ from evolutionary tracks as quoted in Table 1,
and the mass accretion $\dot{M}$, usually derived from the accretion luminosity \citep{Gullbring+etal_1998}. Appendix \ref{app:viscous} (available online only) details the relationship between these observable quantities and the primary parameters of the radial surface density evolution.

A perfect correlation between disk sizes and age is not expected. The initial characteristic sizes of disks will add significant scatter. In this respect, the most significant fact is perhaps the envelope of allowed $R_c$ versus ages, which places an upper limit on these initial sizes. This limit is related to the initial specific angular momentum. Larger disks would fragment and lead to binary and/or multiple systems. In this respect, it may be relevant that the young object with the largest $R_c$ is UZ\,Tau E, a spectroscopic binary member of a hierarchical quadruple system.  Another source of scatter resides in the exponent of the viscosity $\gamma$. Fig.\ref{fig:age-pivot} displays evolutionary curves of $R_c = R_1 T^{1/(2-\gamma})$ for three values of $\gamma$, starting with a common initial radius $R_1 = 10$ AU (see Appendix Eq.\ref{app:4} for a derivation of the evolution of $R_c$ versus time). Although the extreme values of
$\gamma = 0.5$ and $1.5$ appear, at first glance, to provide a good fit to the envelope of the distribution of $R_c$ versus ages, the actual picture is more complex. In particular, a number of stars close to the $\gamma =0.5$ curve have in fact $\gamma = 1.5$ from our data set, while the reverse is also true.

The viscous timescale is given by (see Eq.\ref{app:10})
$$  t_* + t_s = \frac{M_d}{ 2 (2-\gamma) \dot{M}} .$$
Unfortunately, because the stellar ages $t_*$ are very uncertain and we expect in general $t_s \ll t_*$, $t_s$ remains largely unconstrained by the observations. Rather, Eq.\ref{app:10} provides a loose constraint on the allowed range of disk masses and ages. An alternate way
to constrain the viscous timescale is to look at the younger objects, for which the viscous evolution may not have had time to erase the initial conditions. In our sample, younger objects are better represented by power laws. This suggests that young disks are still influenced by the history of infall from the original proto-stellar cloud, and that the viscous timescale is on the order of a few $10^5$ years, the age of these youngest objects.

With this rough estimate for $t_s$, we can in principle derive $T$ and recover the distribution of $R_1$ in our sample from Eq.\ref{app:4}, but the propagation of errors leads to large uncertainties (see Fig.\ref{fig:r1}). This is to be expected, because the viscous evolution has largely erased the memory of the initial conditions.

Self-similarity would also imply that the exponent $\gamma$ remains constant over age. The distribution of $\gamma$ vs age is given in Fig.\ref{fig:age-gamma}. Our distributions of $\gamma$ are somewhat different from those derived by \citet{Andrews+etal_2009} and \citet{Isella+etal_2009}. The former is centrally peaked around 0.9. The latter exhibits values lower than 0.8; however, we have argued in Sec.\ref{sec:holes} that some of the derived values are affected by the interpretation of the central deficit of emission.
In our sample, although there is no obvious correlation, stars of ages 1-3 Myr have on average lower $\gamma$ ($\simeq 0.3$) than either younger or older objects. Note that from Eq.\ref{app:7}, for $\gamma = 1.5$, we expect $\dot{M}(t) \propto t^{-2}$ in good overall agreement with the empirical relation found by \citet{Hartmann+etal_1998}. On the other hand,
$\gamma = 0.5$, which corresponds to the so-called $\beta$ prescription of the turbulence (see Appendix \ref{app:others} for details), yields $\dot{M}(t) \propto t^{-4/3}$ only, somewhat too small to explain \citet{Hartmann+etal_1998} correlation. Note however that large values of $\gamma$ are unlikely to apply to the whole lifetime of the disks: if we assume $\gamma$ has been constant with time, the two old disks with large $\gamma$ (GM\,Aur and MWC\,480) would have started with exceptionally small radii ($< 2$~AU, see Fig.\ref{fig:r1}). Thus, invoking some evolution of the viscosity exponent with age seems required.

An alternate vision on the viscosity is to look at its value at some arbitrary fixed radius, for example at $R_{100} = 100$ AU. Using the $\alpha$ prescription of
the viscosity, the $\alpha$ parameter at 100 AU is given by (Eq.\ref{app:alpha6})
\begin{equation}\label{disc:alpha6}
    \alpha(R_{100}) = \frac{R_c^{(2-\gamma)} R_{100}^\gamma}{3(2-\gamma)^2 c_s(R_{100}) H(R_{100}) t_*} ,
\end{equation}
where $c_s H$ scales as $(L_*/M_*^2)^{1/4}$ to first order (see Appendix \ref{app:viscous} for the derivation).
Using $L_*/M_*^2$ from Table 1, and our adopted values of $T_g = 15$~K and $H = 16$ AU at 100 AU for the median value
of $L/M^2 = 3 \Lsun/\Msun^2$, the resulting $\alpha$ are displayed in Fig.\ref{fig:age-alpha}. There is substantial scatter, but the measurements suggest an overall decrease of $\alpha$ versus time, roughly as $1/t_*$.

\subsubsection{Evolution of the dust}
\label{sec:visc:dust}

The radial dependence of $\beta(r)$ and the behavior of $R_c$ as a function age may be understood in a more complex scheme where viscous spreading plays a significant role. Indeed, only small dust grains are efficiently coupled to the gas, while the larger ones should drift quickly inward \citep[e.g.][]{Weidenschilling_1977}. Hence, one naturally expects that large grains will remain confined to the inner regions, which leads to an apparent increase of $\beta(r)$ with radius. Simulations of the grain-growth, dust-gas
coupling and fragmentation processes have been performed by \citet{Brauer+etal_2008}, and further expanded by \citet{Birnstiel+etal_2010} to include the disk accretion phase and viscous evolution. There is no specific prediction for the evolution of the shape of the grain size distribution with radius which could be compared to our data.  However, from Fig.10 of \citet{Birnstiel+etal_2010}, the smaller grains have outward net velocities beyond about 80 AU. A similar result was found by \citet{Garaud_2007}, although her approach neglects the fragmentation processes. This radius is similar to the transition radius between low and high values of the emissivity exponent $\beta$ found in our study.
\begin{figure}
   \includegraphics[angle=270,width=\columnwidth]{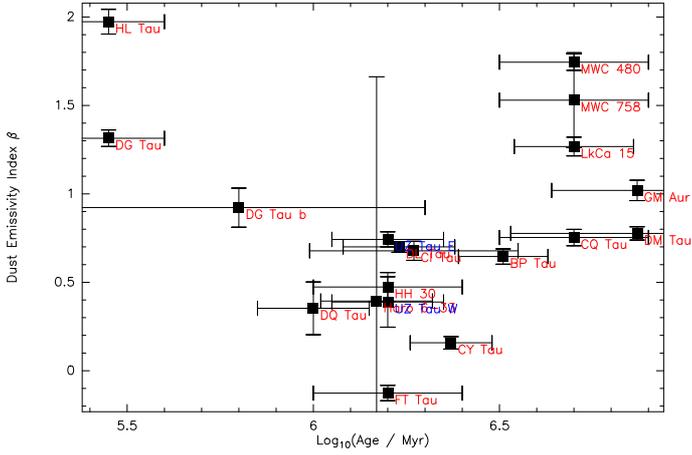}
\caption{Dust opacity index $\beta_m$ as a function of estimated stellar ages.}
\label{fig:age-beta}
\end{figure}

Despite being a rather ill-defined quantity,  the average $\beta_m$ has been used to characterize disks in most previous studies. Figure \ref{fig:age-beta} displays $\beta_m$ as a function of estimated stellar ages. Very young sources have high values of $\beta_m$, comparable to the value found for ISM grains, which could indicate that the dust grains have not yet significantly evolved in these objects, at least at the characteristic distances that we sample in these sources (100 -- 300 AU). Typical T Tauri disks have $\beta_m < 0.7$, which indicates significant grain growth. However, we also find that the older disks display high values of $\beta_m$ too, well above the characteristic value for the ``middle-aged'' T Tauri stars.
The radial dependency of $\beta$ provides an explanation for this distribution of average $\beta_m$ with stellar ages. As disks get larger with time, the apparent average $\beta_m$ increases, which leads to the secondary increase of  $\beta_m$ for older objects as shown in Fig.\ref{fig:age-beta}.

\citet{Birnstiel+etal_2010b} evaluate the effect of the grain growth and fragmentation on the apparent spectral index $\alpha_{1-3\mathrm{mm}}$ for disk masses ranging from 0.005 to 0.1 \Msun~ and compared them to the observed distribution obtained by \citet{Ricci+etal_2010}. They use a fixed disk model with $R_c = 60$ AU and $\gamma =1 $.
The growth and fragmentation model predicts an increase of $\beta(r)$ at radii ranging from 40 to 100 AU (their Fig.3),
which broadly agrees with our finding.  However, in their analysis the distribution of average $\alpha_{1-3\mathrm{mm}}$ vs observed 1.3\,mm flux density only roughly matches the strongest sources. This effect is related to the dependence of the ``fragmentation barrier'' grain size on the surface density: less massive disks are expected to have smaller grains, and thus larger
$\alpha_{1-3\mathrm{mm}}$.  As mentioned in Sec.\ref{sec:discussion:dust}, a simple way to provide a better agreement with the data is to assume that the faint disks are actually not less dense than the strong ones, but truncated to smaller sizes. In this case, $\alpha_{1-3\mathrm{mm}}$ is expected to decrease with source flux because of the radial dependence of $\beta(r)$.

Finally, the discussion of the viscous properties of the disk presented in Section \ref{sec:visc:gas} used $\gamma$ and $R_c$ derived from the dust content, i.e. they implicitly assume a constant dust-to-gas ratio. In reality, the dust-to-gas ratio is expected to change with radius because of the concurrent effects of accretion, viscous spreading, grain drift, growth and fragmentation. In general, it is expected to decrease with radius, because the coupling between dust and gas increases with density, see for example Fig.6 of \citet{Birnstiel+etal_2010}. The average dust-to-gas ratio is also expected to decrease with time, as the largest particles are drifting inward, being eventually advected onto larger bodies, either embryos or the central star. Accordingly, the discussion on possible changes of viscosity ($\alpha$ parameter and/or $\gamma$) presented in Section \ref{sec:visc:gas} should be taken with some additional care.

\section{Conclusions}
\label{sec:conc}

We report here the results of the first dual-frequency and high-resolution study of dust disks in the mm domain where the dust is mostly optically thin.
\begin{itemize}
\item Independent data sets allowed us to verify the robustness of the derived parameters and of their error bars.
The geometric parameters (inclination and orientation)  agree well with determinations from other constraints, such as scattered light images, optical jets, and the Keplerian rotation of the disks.

\item We derived proper motions for 10 sources in our sample.

\item Tidal truncation is found to affect the disk sizes in binary systems.

\item Despite the combination of high angular resolution and sensitivity, we found that the viscous disk model does not generally provide a significantly better fit of the continuum data only than the simple truncated power law description. Baselines well above 300 k$\lambda$ are required to distinguish between these two descriptions.  In very young sources, the simple power law model appears to work somewhat better, while the exponential edge is marginally better for evolved objects.

\item Inner holes also appear to provide a better explanation than negative values of $\gamma$ for sources
     showing a deficit of emission at the center like GM Aur and LkCa 15.

\item We have strong evidence for radial dependence of the dust emissivity exponent $\beta$ with radius. In all  cases, $\beta$ is found to increase with radius, i.e. we find grain size which decreases with distance from the star. High $\beta$ values (1.7 --2, typical for ISM grains, or even possibly higher) are found beyond 100 AU, while the inner regions may display values down to nearly 0. This result is obtained whatever disk model has been adopted (surface density shape and temperature profile).

\item We have possible evidence for optically thick cores in a few sources, which provide a direct estimate of the temperature
  of large grains. However, in some cases, inner regions with  $\beta = 0$ may be misinterpreted as thick cores at low temperatures.
\item Despite the ambiguities introduced by the variable dust properties, the characteristic size of the disk appears
to increase with stellar ages, which broadly agrees with the viscous evolution. A more detailed comparison with the models suggests
a decrease of the $\alpha$ viscosity parameter with time, as well as changes in the exponent of the viscosity.

\end{itemize}
These observations provide the first evidence for the expected effect of the dust grain evolution in circumstellar disks resulting from grain growth, fragmentation and, viscous transport.  The comparison with model predictions is limited by the angular resolution obtained at the longest wavelengths, about 100 AU, which requires some parametric approach to constrain the radial dependence of $\beta(r)$ (and by inference, $\kappa(r)$ using some specific dust model).
With the advent of ALMA and e-VLA, a direct inversion of the $\beta(r)$ profile at linear resolutions of order 10-20 AU will become possible, enabling us to derive much more accurate constraints on the dust properties as a function of radius. This will be possible not only using two wavelengths as here, but over more than a decade in frequency.

\listofobjects

\begin{acknowledgements}
  We acknowledge the IRAM staff at Plateau de Bure and Grenoble for
  carrying out the observations. We also thank Rowan Smith who
  made a part of the numerical tests during a training period at LAB in
  July 2005. We thank the national programmes PCMI "Physico-Chimie du Milieu Interstellaire"
  and PNPS "Physique Stellaire" from INSU / CNRS  for providing funding.
  This research has made use of the SIMBAD database, operated at CDS, Strasbourg, France, and
  of the NASA ADS Abstract Services.
\end{acknowledgements}

\bibliographystyle{aa} %
\bibliography{all}

\begin{appendix}

\section{Optical depth vs variable $\beta$}
\label{app:fitting}

\begin{figure}
\includegraphics[angle=270,width=\columnwidth]{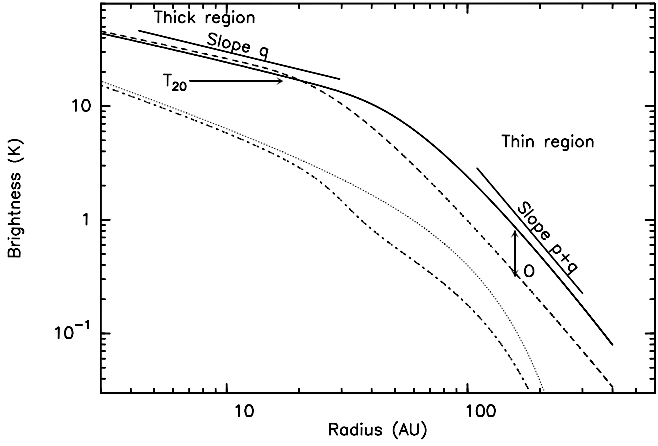}
\label{fig:app:brightness}
\caption{Sample result illustrating the shape of the brightness
distribution for our disk models. Thick line: constant $\beta$
power law model at 1.3\,mm, dashed line: same model at 2.7\,mm. Dotted
line: constant $\kappa$(1.3\,mm) but variable $\beta(r)$, tapered-edge model at 1.3\,mm;
dash-dotted line: same model at 2.7\,mm.}
\end{figure}

Because a direct inversion of the brightness temperature profile is impossible,
the determination of the parameters is fully implicit.
Figure \ref{fig:app:brightness}
illustrates two possible types of brightness temperature profiles that can occur in our analysis.
The continuous and dashed lines represent brightness at 1.3\,mm and 2.7\,mm for
a typical power law distribution, with constant dust properties $\beta(r) = \beta_m$.
The outer region is optically thin, and the slope constrains $p+q$. $\beta_m$ is derived
from the brightness ratio $O$. The inner region is optically thick, and constrains
the exponent $q$ as well as the temperature at 20 AU, $T_{20 }$. The small brightness
difference between the two frequencies is caused by the Rayleigh-Jeans correction.
The dotted and dot-dashed lines represent an optically thinner disk at 1.3\, and 2.7\,mm respectively, with
a viscous type profile with $R_c=150$ AU. In addition, $\beta(r)$ is assumed to vary with radius
following Eq.\ref{eq:betaatan} with $R_b = 60$ AU and $R_w=20$ AU, $\kappa$(1.3\,mm) being
constant.  Here, the inner region is optically thin, and its slope is $q+\gamma$. Note
that if the temperature of that disk would be 4 times higher, it would mimic reasonably
well the previous power law, optically thick case, provided $\gamma$ is not too large.
Accordingly, sources displaying a wavelength-independent flattened (apparent exponent $\simeq 0.4 -- 0.7$)
inner brightness distribution can be interpreted either as optically thick sources,
or as variable $\beta(r)$ with $\beta(r) \simeq 0$ in the inner region.  Steeper apparent
exponents are not realistic for the temperature dependence.
Note that the typical noise level is around 0.05 -- 0.1 K in our observations at both wavelengths.

\section{Sampling effects and best model}
\label{app:sampling}

Because of the fully implicit derivation of the model parameters, an objective determination of the ``best'' model is difficult.
The same source may be (nearly) equally well represented by either Model 1 or Model 2. We use a $\chi^2$ criterium to determine
the best matching model. However, it is important to realize that our data consists in a large (several $10^4$) number of statistically
independent visibilities, each with very little (essentially zero) signal-to-noise. The $\chi^2$ is given by
\begin{equation}
\chi^2 = \Sigma (O_i-M_i)^2  * W_i ,
\end{equation}
where $O_i$ are the (complex) observed visibilities ($O^2$ actually being used to note $O \times O^*$, $O^*$ being
the complex conjugate of $O$), $M_i$ the modeled visibilities.
The weights $W_i = 1/\sigma_i^2$ are derived from the theoretical noise using the system temperature, antenna gain,
observing bandwidth and integration time. In general, $\sigma_i >> M^b_i$, where $M^b$ is the best-fit model,
so even the null model $M_i = 0$ yields a $\chi^2$ on the order of $N$, the number of visibilities, as $W_i$ is
the inverse of the variance of $O_i-M^b_i$. Thus, the reduced $\chi^2$, $\chi^2_r = \chi^2/N$ is a poor evaluation of
the fit quality, which is close to 1 even for a very poor (null) model. Only the relative differences $\Delta \chi^2$ between
models of equivalent number of parameters can reveal whether one is better than the other.

Another subtle effect in comparing absolute values of $\chi^2$ is the impact of discretization. A numerical model $M$ is
an approximation of the theoretical model $T$, $M = T+E$, where $E$ is a numerical error term. So
\begin{eqnarray}
\chi^2 & = & \Sigma (O_i-M_i)^2 W_i \\
  &  = & \Sigma (O_i-T_i)^2 W_i + \Sigma E_i^2 W_i - 2 \Sigma E_i (O_i-T_i) W_i .
\end{eqnarray}
Because the model fit the observations and the numerical errors are not correlated with the
observations, the last term is negligible, consequently the final $\chi^2$ is a sum of the true (no numerical
errors) term plus an offset cause by numerical effects. To make numerical errors negligible requires $\Sigma_i E_i^2 W_i$ to
be much less than 1. This is especially important when comparing different theoretical models. However, within
a given model, the best-fit parameters may be determined with sufficient precision even if the numerical error
term is not small.

\section{Impact of the assumed temperature law}
\label{app:temperature}

\begin{figure*}
\includegraphics[angle=270,width=14.0cm]{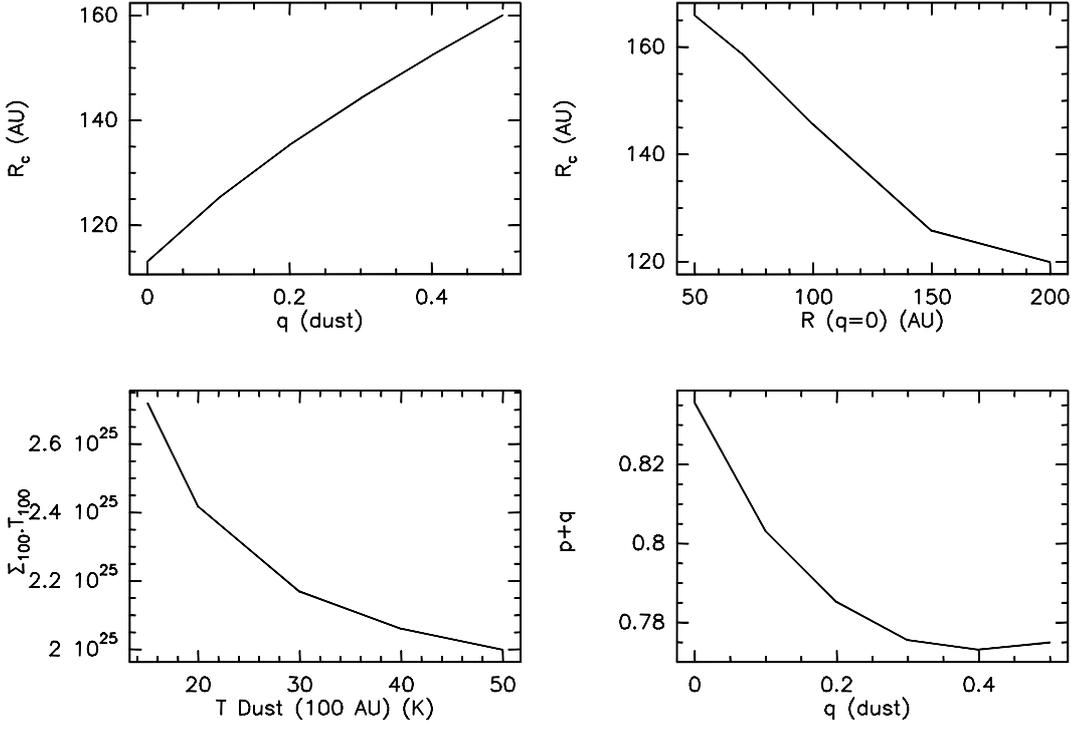}
\caption{Sample results illustrating the main
dependency of the surface density profile on
the temperature law. Top left: $R_c$ versus $q$. Top right:
$R_c$ versus $R_f$; bottom left: $\Sigma.T$ verus $T$;
bottom right: $p+q$ versus $q$. The observed source is DL\,Tau. }
\label{fig:templawone}
\end{figure*}

In this appendix, we investigate the impact of the dust temperature
profile on the derived disk parameters. We consider two different
profiles. Profile (i) is a power law $T(r) = T_{100} (r/100 \mathrm{AU})^{-q}$. Profile
(ii) is a broken power law: it has  a constant temperature between $R_i = 40$ AU
and $R_f$, $R_f$ being a variable parameter, while for $r<R_i$ or $r>R_f$, the temperature
is a power law with exponent q=0.5, with to $T(r) = T_1$  at $r=1$ AU. The temperature law
is continuous as a function of $r$, and we used $T_1= 200$ K by default.
We analyzed the observations of a few sources (DL\,Tau, DM\,Tau and MWC\,480)
to explore the dependency of the derived surface density parameters on
$T_0, q$ and $R_f$. Figure \ref{fig:templawone} illustrates the main impact of
the temperature law on the surface density parameters, which is applicable to
all optically thin sources. Figure \ref{fig:templawone} is for Model 2
(so $p$ is to be interpreted as $\gamma$), but similar results are
obtained for Model 1.

For Profile (i) :\\
- $\Sigma_0$ is nearly proportional to 1/T, with small corrections at low T
owing to deviations from the Rayleigh-Jeans behavior.\\
- Similarly, $p+q$ is nearly constant. This is equally valid for Model 1
(power law) and Model 2 (tapered edge).\\
- In Model 1, $R_\mathrm{out}$ is only weakly affected by the changes in $p$\\
- In Model 2, $R_c$ increases by 20 to 30 \% when $q$ increases from 0 to 0.5.

For Profile (ii) :\\
- In Model 1, $R_\mathrm{out}$ slightly decreases with $R_t$ (by about 10 \%), and $p$
changes by about 0.1.  Variations are not fully monotonic, however.\\
- In Model 2, $R_c$ decreases by about 20 to 30 \%, when $R_t$ goes from 50 to 200 AU.
This is similar to the effect of $q$ in Profile (i), as increasing $R_t$ flattens
the temperature distribution.

For more optically thick sources, like MWC\,480, the effect on $p$ is larger, because
of the opacity corrections. However, in this case, $q$ can be determined from
the observations, because the $\chi^2$  significantly depends on its value. Restricting
the range of $q$ to within its typical uncertainty limits the impact on $p$ to about 0.2.

Except for the absolute scaling of the density as $1/T_{100}$ (or $1/T_1$ in Profile (ii)),
the derived density distribution are thus not significantly affected by the assumed temperature law.

\begin{figure*}
\includegraphics[angle=270,width=14.0cm]{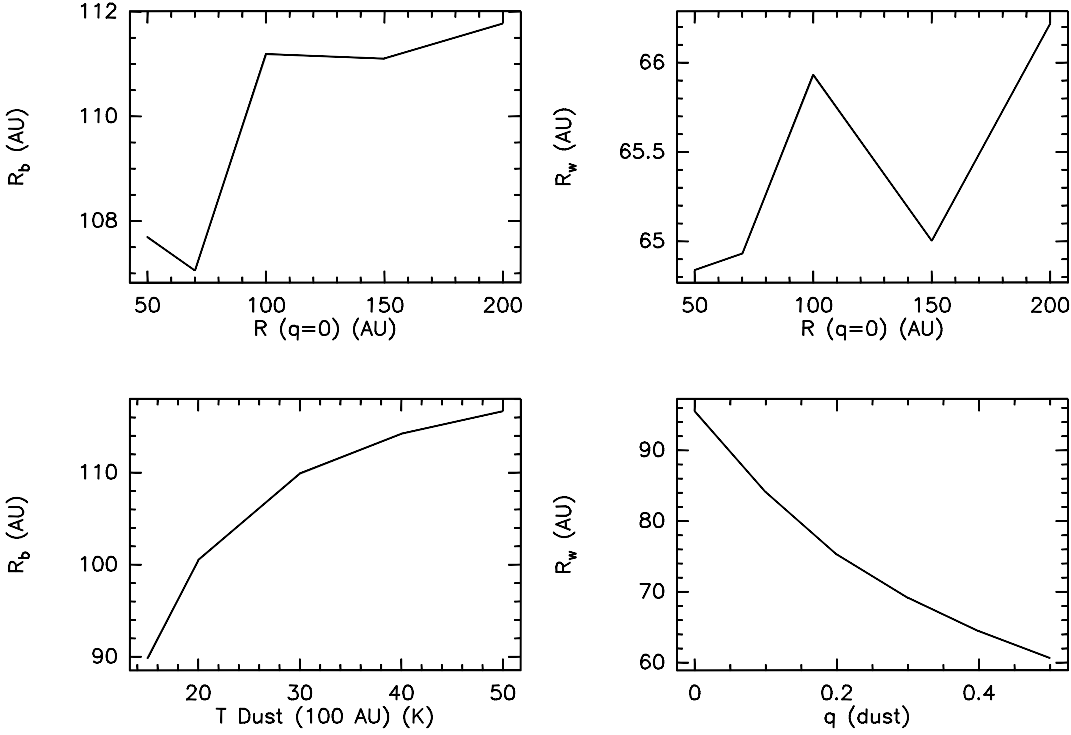}
\caption{Sample results illustrating the weak dependency
of the dust parameters $R_b$ and $R_w$ on the assumed
temperature law.}
\label{fig:templawtwo}
\end{figure*}

More importantly, $R_c$ and $p$ are affected in the same proportions at both wavelengths. Thus, the
uncertainties on the temperature law have no significant effect on the derivation of the
radial dependence of $\beta(r)$ (see Figure \ref{fig:templawtwo}). Incidentally, we
note that in DL\,Tau, a better fit to the observations is obtained using Profile (ii) with
$R_f = 100$ AU.

\section{Disk parameters from observable quantities in the viscous model}
\label{app:viscous}

The shape of the surface density profile used in Model 2 corresponds to the self-similar solution of the viscous evolution of a disk under the assumption that the viscosity is constant in time and a power law of radius \citep[see][]{LyndenBell+Pringle_1974,Pringle_1981}. Under these assumptions, the surface density as a function of time and radius is given by (Eq.17 of \citet{Hartmann+etal_1998})
\begin{equation}\label{app:1}
    \Sigma(R,t) = \frac{C}{3 \pi \nu_1 r^\gamma} T^{-(5/2-\gamma)/(2-\gamma)} \exp\left(-\frac{r^{(2-\gamma)}}{T}\right) ,
\end{equation}
where $r = R/R_1$, $T = (1+t_*/t_s)$ is a dimensionless time, $t_*$ the disk/star age and $t_s$ is the viscous timescale at $R_1$, defined by
\begin{equation}\label{app:2}
   t_s = \frac{R_1^2}{3(2-\gamma)^2 \nu_1} .
\end{equation}
Our observations (at unknown time $T$) are characterized by the surface density law described by our Eq.\ref{eq:edge}
\begin{equation}\label{app:3}
\Sigma(r) = \Sigma_0 \left(\frac{r}{R_0}\right)^{-\gamma}  \exp\left(-(r/R_c)^{2-\gamma}\right) .
\end{equation}
So by identification, we obtain
\begin{equation}\label{app:4}
    R_c = R_1 T^{1/(2-\gamma)}
\end{equation}
and
\begin{equation}\label{app:5}
    \Sigma_0 = \frac{C T^{-(5/2-\gamma)/(2-\gamma)}}{3 \pi \nu_1} \left( \frac{R_1}{R_0} \right)^\gamma ,
\end{equation}
which, eliminating $R_1$ usinq Eq.\ref{app:4}
\begin{equation}\label{app:6}
    \Sigma_0 = \frac{C T^{-(5/(2(2-\gamma))}}{3 \pi \nu_1} \left( \frac{R_c}{R_0} \right)^\gamma .
\end{equation}
A time derivative of Eq.\ref{app:1}  (taken for r=0) further indicates that the mass accretion rate is
\begin{equation}\label{app:7}
    \dot{M} = C  T^{-(5/2-\gamma)/(2-\gamma)} .
\end{equation}
We have in principle five unknowns ($C, R_1, T, \nu_1, \gamma)$, and five measurements: three from our study ($M_d~\mathrm{or}~\Sigma_0, R_c, \gamma$), the stellar age $t_*$ from evolutionary tracks and the mass accretion $\dot{M}$, usually derived from the accretion luminosity
\citep[see][]{Gullbring+etal_1998}.
Although this formally yields a solution, it is nearly degenerate when one considers the uncertainties on the measured quantities.
This can be realized by noting that the mass accretion rate can be rewritten as \citep[Eq.14 from][]{Isella+etal_2009}
\begin{equation}\label{app:8}
 \dot{M} = \frac{M_d(t=0)}{2(2-\gamma)t_s} T^{-(5/2-\gamma)/(2-\gamma)} ,
\end{equation}
while from Eqs. \ref{eq:mass} and \ref{app:5}, the time dependency of the disk mass is simply \citep[Eq.A7 from][]{Andrews+etal_2009}
\begin{equation}\label{app:9a}
 M_d(t) = M_d(t=0) T^{-1/2(2-\gamma)} ,
\end{equation}
so, by simple elimination
\begin{equation}\label{app:9}
   \dot{M} = \frac{M_d}{2 (2-\gamma) T t_s} ,
\end{equation}
which simply gives
\begin{equation}\label{app:10}
    t_* + t_s = \frac{ M_d}{ 2 (2-\gamma) \dot{M}} .
\end{equation}
This is the only equation involving $t_s$. $R_c$, and thus $R_1$ does not appear in this expression because
$R_1$ only reflects the initial condition of disk size, not its future evolution.

The (time independent) viscosity at any arbitrary radius is given by
\begin{equation}\label{app:alpha0}
\nu(r) = \nu_1 (r/R_1)^\gamma ,
\end{equation}
which, using the expression of $R_1$ in Eq.\ref{app:4}, can be expressed in terms of the observable
quantities as
\begin{equation}\label{app:alpha1}
\nu(r) = \frac{R_c^{(2-\gamma)} r^\gamma}{3(2-\gamma)^2 (t_*+t_s)} .
\end{equation}
It is customary to express it in terms of the $\alpha$ parameter, $\nu(r) = \alpha(r) c_s(r) H(r)$,
where $c_s$ is the sound speed, and $H(r)$ the scale height
\begin{equation}\label{app:alpha2}
    \alpha(r) = \frac{R_c^{(2-\gamma)} r^\gamma}{3(2-\gamma)^2 c_s(r) H(r) t_*} .
\end{equation}
In hydrostatic equilibrium,
\begin{equation}\label{app:alpha4}
     c_s(r) H(r) =  \frac{k T_g(r)}{ \mu m_h \sqrt{G M_*}} r^{3/2} ,
\end{equation}
$T_g$ being the gas temperature in the disk mid-plane. Approximating $T_g(r)$ by a power law of exponent $-q$ ($q\simeq 0-0.6$), we derive
\begin{equation}\label{app:alpha5}
    \alpha(r) = \alpha(R_0) (r/R_0)^{\gamma-3/2-q/2} %
\end{equation}
\begin{equation}\label{app:alpha6}
    \alpha(R_0) = \frac{R_c^{(2-\gamma)} R_0^\gamma}{3(2-\gamma)^2 c_s(R_0) H(R_0) t_*} .
\end{equation}
$R_c$ and $\gamma$ are directly constrained by our observations, while $t_*$ is derived from evolutionary tracks.
The last term $c_s(R_0) H(R_0)$ depends on the stellar properties. We note from Eq.\ref{app:alpha4} that $(c_s H)^2 \propto T_g^2 / M_*$, and
thus scales to first order as $(L_* /M_*^2)^{1/4}$.

\section{Alternate disk models}
\label{app:others}

With the alpha prescription of the viscosity (radially uniform and constant in time $\alpha$) and a (time independent) power
law temperature $T_k=T_0 (r/r_0)^{-q}$, $\nu(r) = \alpha c_s^2 / \Omega$, so $\gamma = 3/2-q$,
the equation \ref{app:1} can also be written as
\begin{equation}\label{eq:pringle}
\Sigma(r,t) = S \left ( \frac{r}{r_0}\right )^{q-3/2}  T^{-(q+1)/(q+1/2)} \\
   \exp{\left ( \frac{-(r/r_0)^{(q+1/2)}}{T} \right )} .
\end{equation}
At long times, $T >> 1$, the density profile evolves as $p=3/2-q$, or $p+q=3/2$.

A similar formula can also be recovered for the $\beta$ prescription of the viscosity,
$\nu(r)= \beta' r^2\Omega = \beta' \sqrt{GM_*} r^{1/2}$ \citep{Richard_Zahn1999}.
It is equivalent to setting $q=1$ in equation \ref{eq:pringle}, and thus
corresponds to $\gamma=0.5$ .

The self-similar solutions of the evolution equation for the disk surface density were obtained
under several simplifying assumptions. \citet{Desch2007} pointed out that accounting for the early planet migration as predicted by the Nice model
\citep{Tsiganis+etal_2005,gomes+etal_2005}, the initial exponent of the surface density for the Solar Nebula would be very close to $p=2.2$.
To explain this slope, \citet{Desch2007} recovered a different shape for the surface density in {\em steady state} configuration.
The general form of the surface density in the \citet{Desch2007} solution is
\begin{equation}\label{eq:desch}
    \Sigma(r) = \frac{\Sigma_u}{1+x_u} \left(\frac{r}{r_u}\right)^{-(2-q)}\left[1+x_u \left(\frac{r}{r_u}\right)^{1/2}\right] ,
\end{equation}
where  $r_u$ is the radius at which the disk has an apparent slope $p$ and $x_u=(2-p-q)/(p+q-3/2)$.
For $p+q > 2$, $x_u <0$ and the surface density vanishes at radius $r_d = r_u/ x_u^2$.
Note that the classical steady-state result $ \Sigma(r) \propto r^{-(3/2-q)}$ corresponds to
the asymptotic limit $x_u \rightarrow \infty$, and is obtained by
imposing different boundary conditions on the evolution equation of angular momentum.

\section{Unresolved, possibly thick, sources}
\label{app:compact}

For unresolved sources, the outer radius can only indirectly be constrained from the observed flux.
Assuming uniform opacity $\tau$, and a standard power law for the temperature $T(r) = T_0 (r/R_0)^{-q}$,
the outer radius is given by ($i$ being the inclination)
\begin{equation}\label{eq:radius}
    R_\mathrm{out}(\tau) =  R_0 \left( \frac{(2-q) S_\nu D^2 \lambda^2}{4 \pi k_b R_0^2 T_0 \cos{i} (1-\exp(-\tau/\cos{i}))} \right)^\frac{1}{2-q} .
\end{equation}
A lower limit is recovered or $i = 0$ and $\tau \rightarrow \infty$
\begin{equation}\label{eq:minradius}
    R_\mathrm{min} > R_0 \left( \frac{(2-q) S_\nu D^2 \lambda^2}{4 \pi k_b R_0^2 T_0} \right)^\frac{1}{2-q} .
\end{equation}
The disk mass is given by
\begin{equation}\label{eq:minmass}
    M_d =  \pi R_\mathrm{out}(\tau)^2 \frac{\tau }{\kappa(\nu)} .
\end{equation}
With $q\sim 0 - 0.5$, a lower limit on $M_d$ is obtained for $\tau \simeq 0.5$. Solutions with density/opacity decreasing with radius will lead to higher masses.

\section{Figures for individual sources}

We display here the figures for individual sources.

\label{app:sources}
\clearpage
\begin{figure*}[h]
   \includegraphics[angle=270,width=18.0cm]{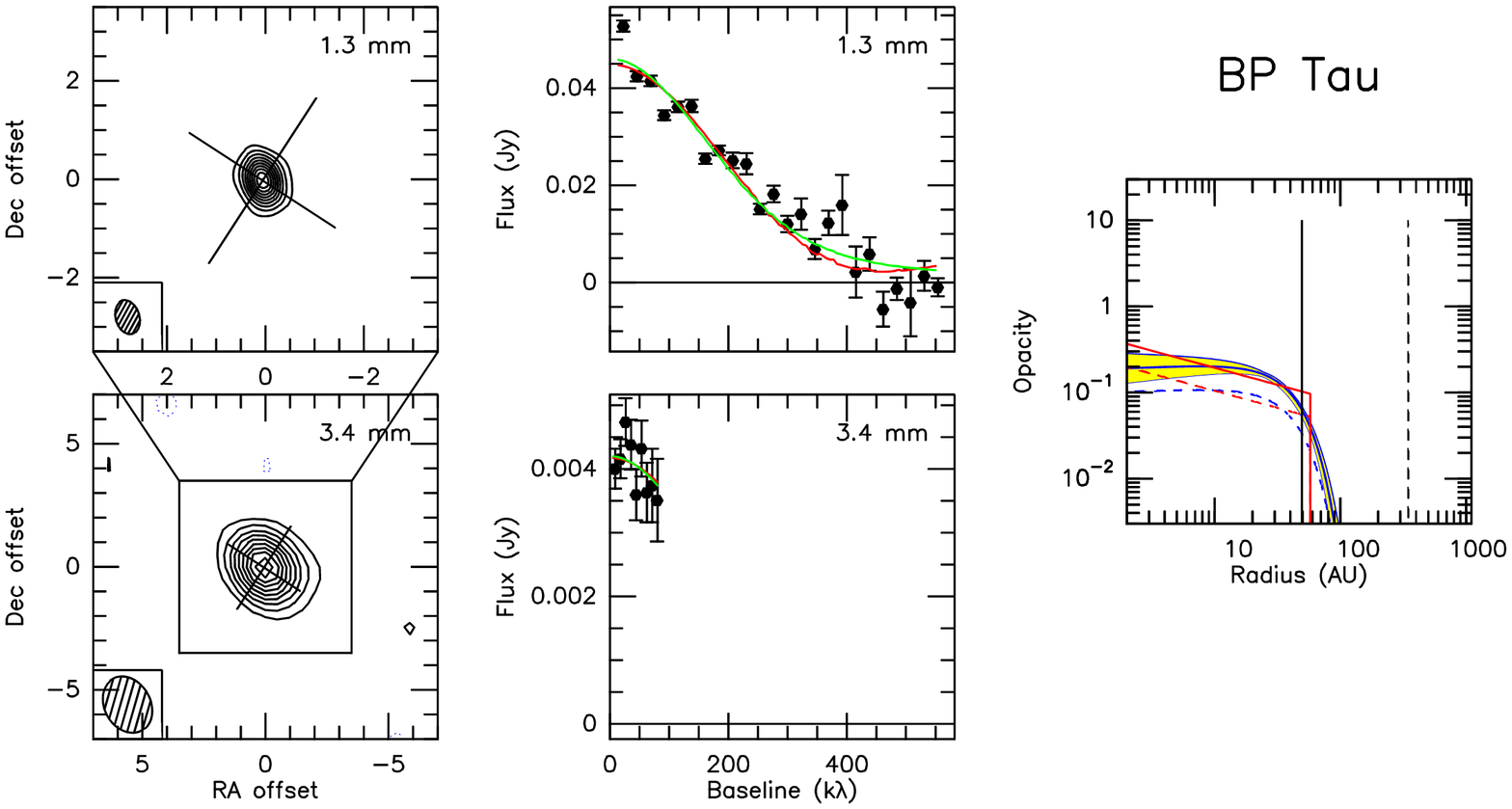}
\caption{As Fig.\ref{fig:alldmtau} but for \object{BP Tau}. Contour level is 3 mJy/beam ($6 \sigma$)
at 1.3 mm, and 0.4 mJy/beam ($3 \sigma$) at 3.4\,mm.}
   \label{fig:allbptau}
\end{figure*}

\begin{figure*}[h]
   \includegraphics[angle=270,width=18.0cm]{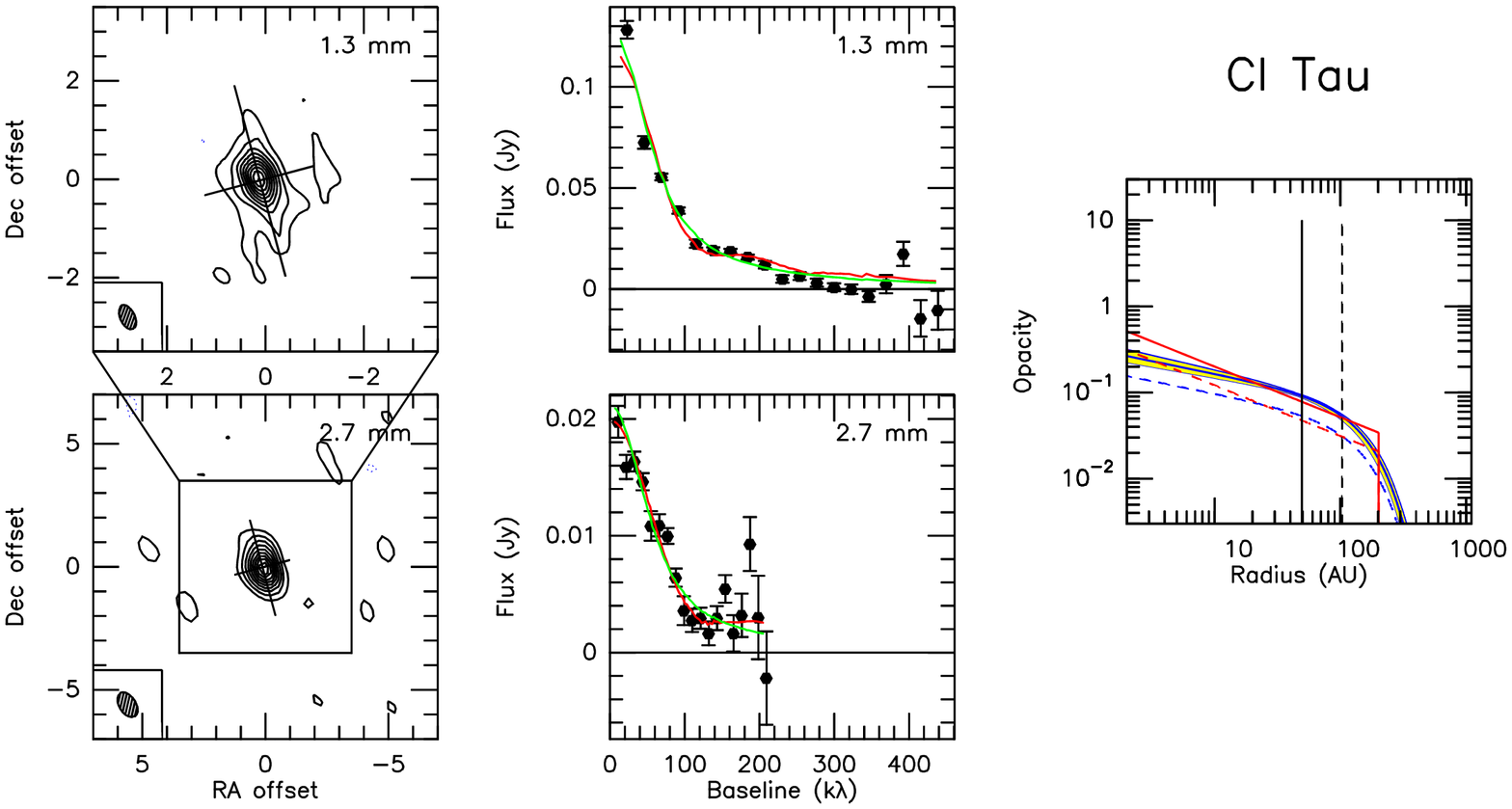}
\caption{As Fig.\ref{fig:alldmtau} but for CI Tau. Contour level is 2.2 mJy/beam ($3.5 \sigma$)
at 1.3 mm, and 0.86 mJy/beam ($2 \sigma$) at 2.7\,mm.}
   \label{fig:allcitau}
\end{figure*}

\clearpage

\begin{figure*}[h]
   \includegraphics[angle=270,width=18.0cm]{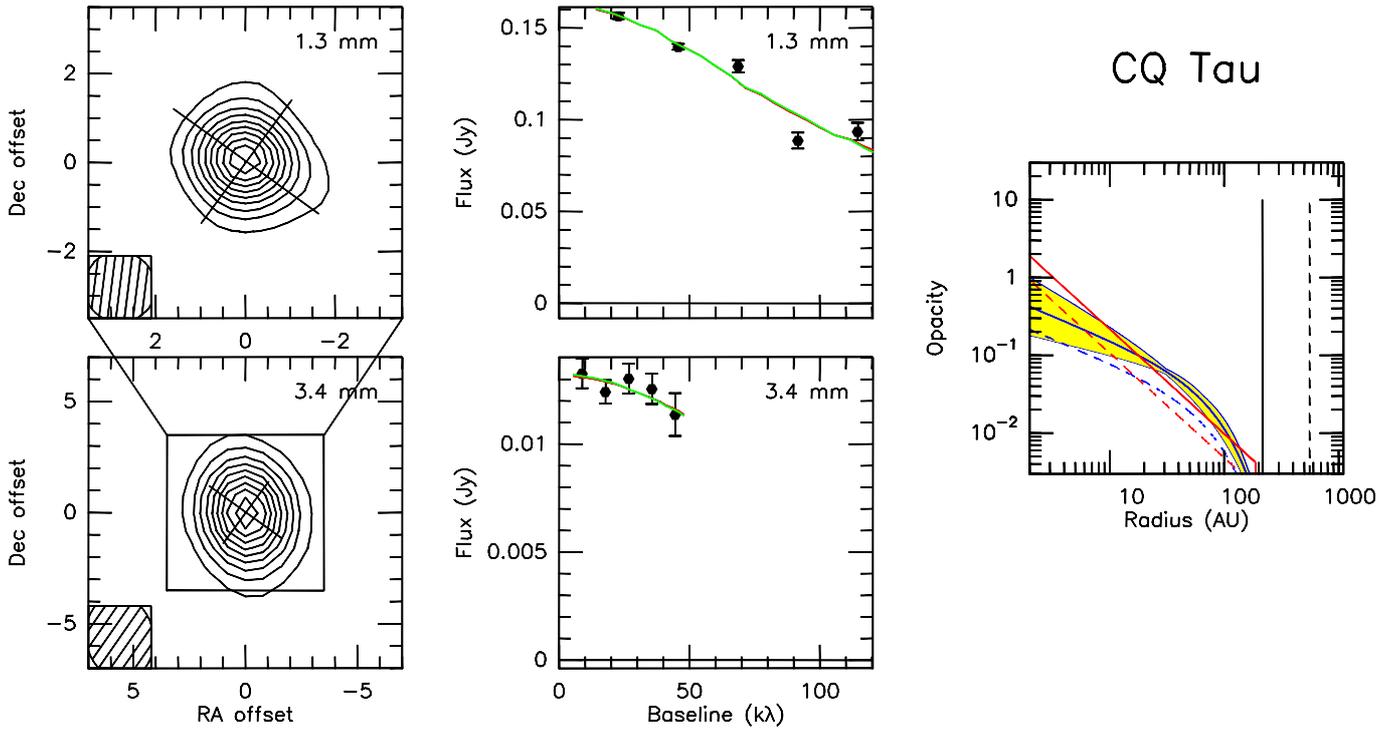}
\caption{As Fig.\ref{fig:alldmtau} but for CQ Tau.}
   \label{fig:allcqtau}
\end{figure*}

\begin{figure*}[h]
   \includegraphics[angle=270,width=18.0cm]{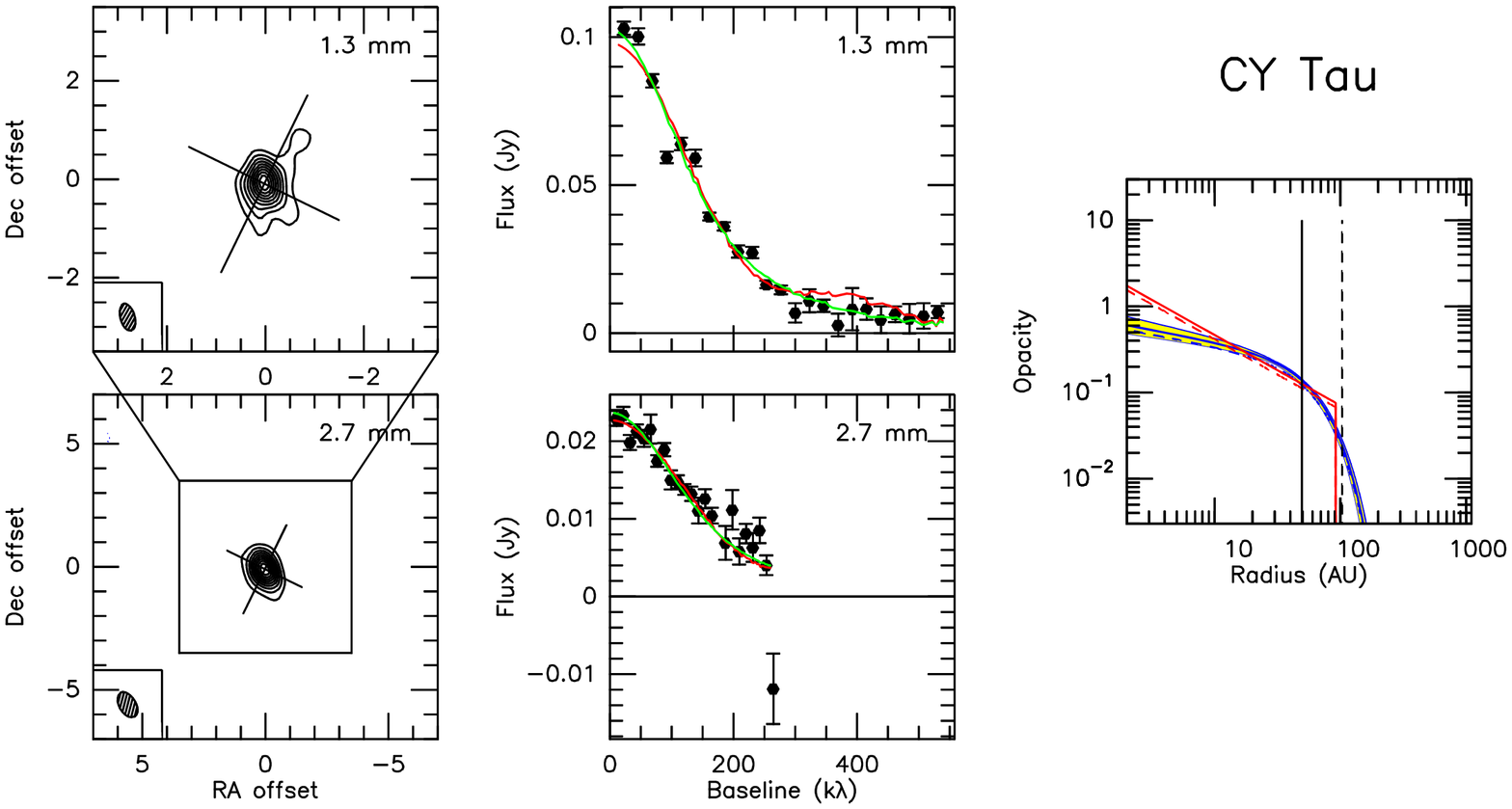}
\caption{As Fig.\ref{fig:alldmtau} but for CY Tau. Contour level is 3.3 mJy/beam ($4 \sigma$)
at 1.3 mm, and 1.6 mJy/beam ($4 \sigma$) at 2.7\,mm. }
   \label{fig:allcytau}
\end{figure*}

\clearpage
\begin{figure*}[h]
   \includegraphics[angle=270,width=18.0cm]{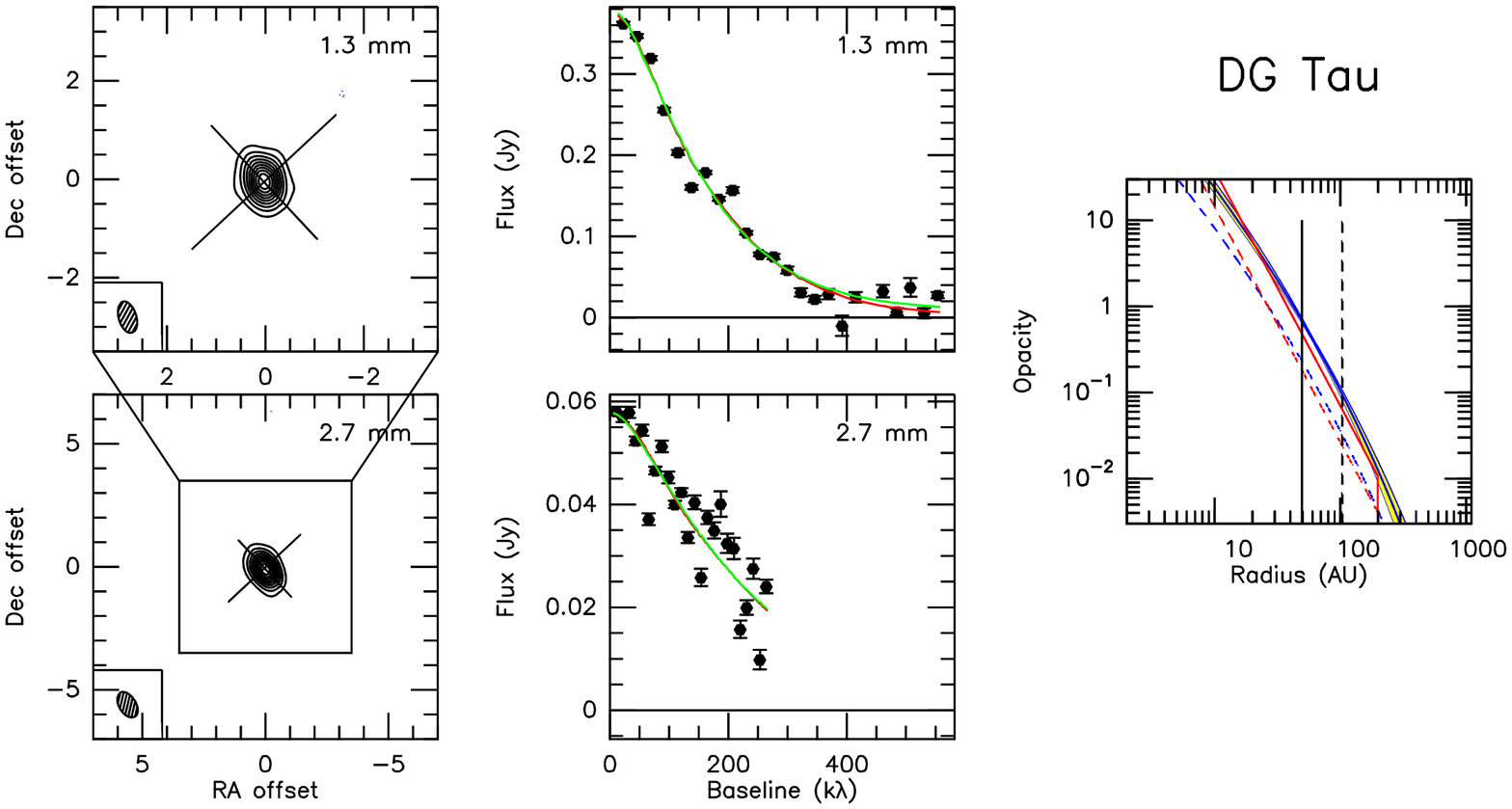}
\caption{As Fig.\ref{fig:alldmtau} but for DG Tau. Contour level is 16 mJy/beam ($5 \sigma$)
at 1.3 mm, and 4.3 mJy/beam ($4.3 \sigma$) at 2.7\,mm. }
   \label{fig:alldgtau}
\end{figure*}

\begin{figure*}[h]
   \includegraphics[angle=270,width=18.0cm]{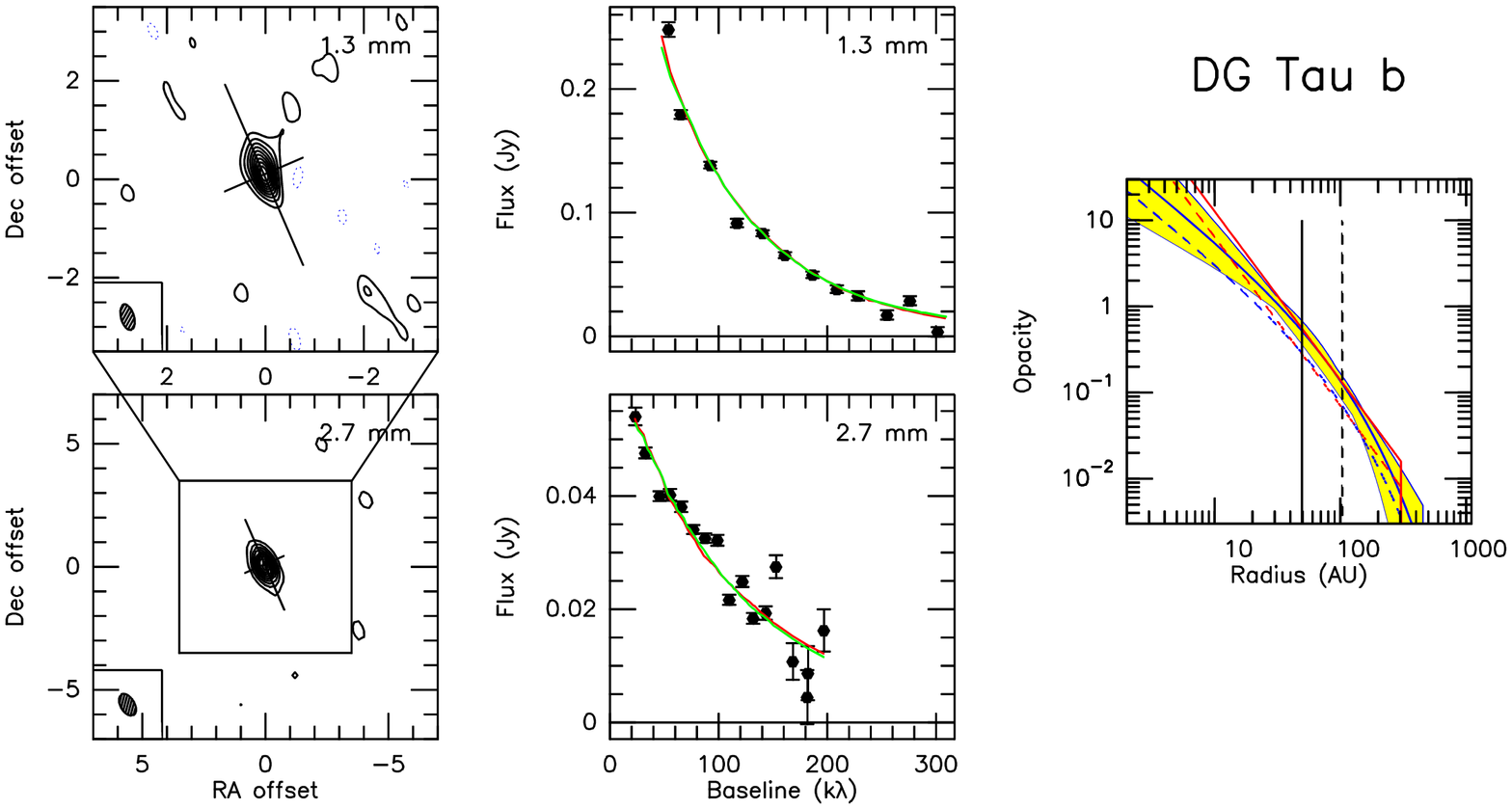}
\caption{As Fig.\ref{fig:alldmtau} but for DG\,Tau b. Contour level is 7.4 mJy/beam ($3.7 \sigma$)
at 1.3 mm, and 3.2 mJy/beam ($3.2 \sigma$) at 2.7\,mm. }
   \label{fig:alldgtaub}
\end{figure*}

\clearpage
\begin{figure*}[h]
   \includegraphics[angle=270,width=18.0cm]{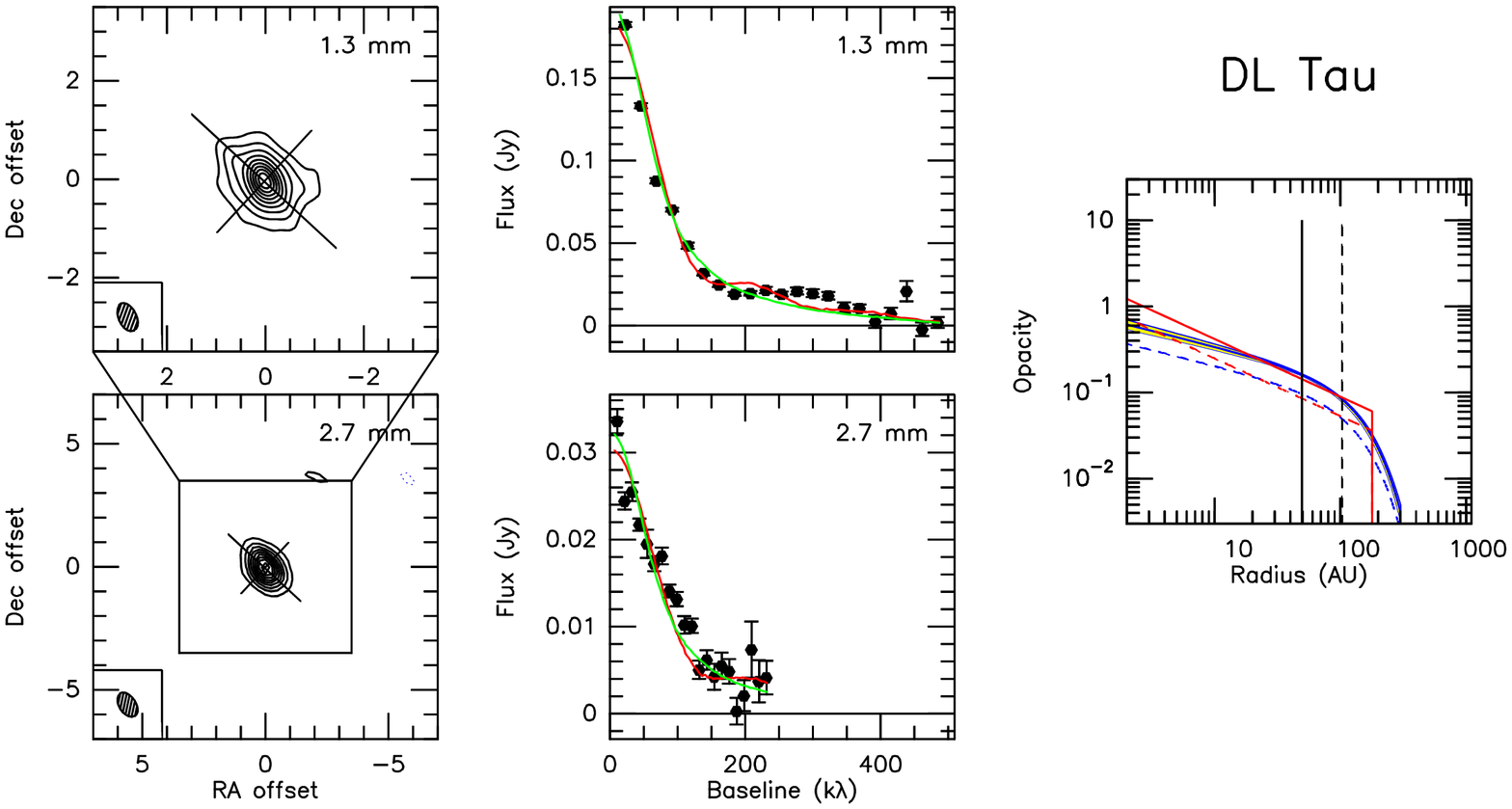}
\caption{As Fig.\ref{fig:alldmtau} but for DL Tau. Contour level is 4.3 mJy/beam ($5.5 \sigma$)
at 1.3 mm, and 1.4 mJy/beam ($3.5 \sigma$) at 2.7\,mm. }
   \label{fig:alldltau}
\end{figure*}

\begin{figure*}[h]
   \includegraphics[angle=270,width=18.0cm]{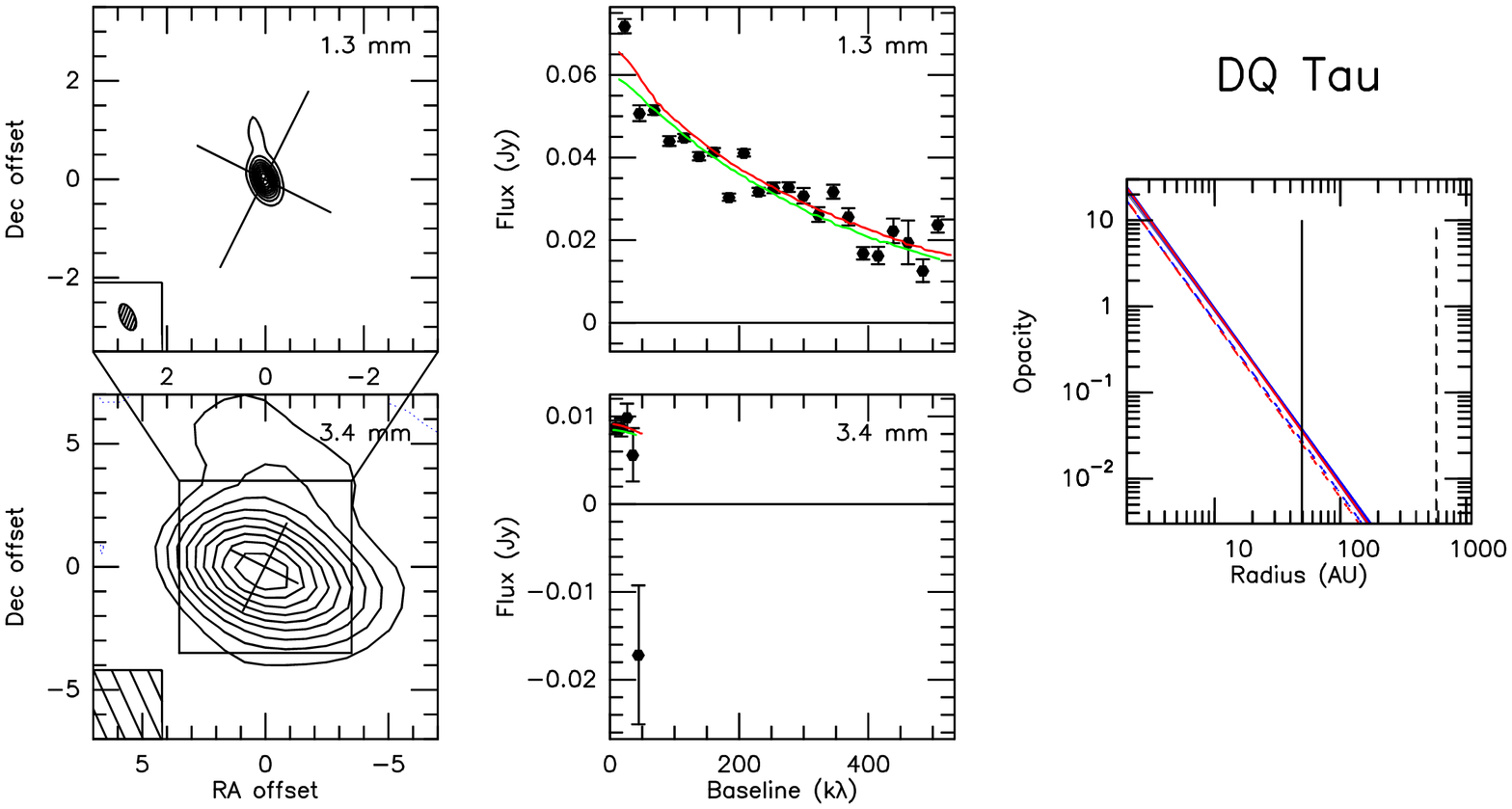}
\caption{As Fig.\ref{fig:alldmtau} but for DQ Tau. Contour level is 3.6 mJy/beam ($4.5 \sigma$)
at 1.3 mm, and 0.8 mJy/beam ($1.6 \sigma$) at 3.4\,mm. }
   \label{fig:alldqtau}
\end{figure*}

\clearpage
\begin{figure*}[h]
   \includegraphics[angle=270,width=18.0cm]{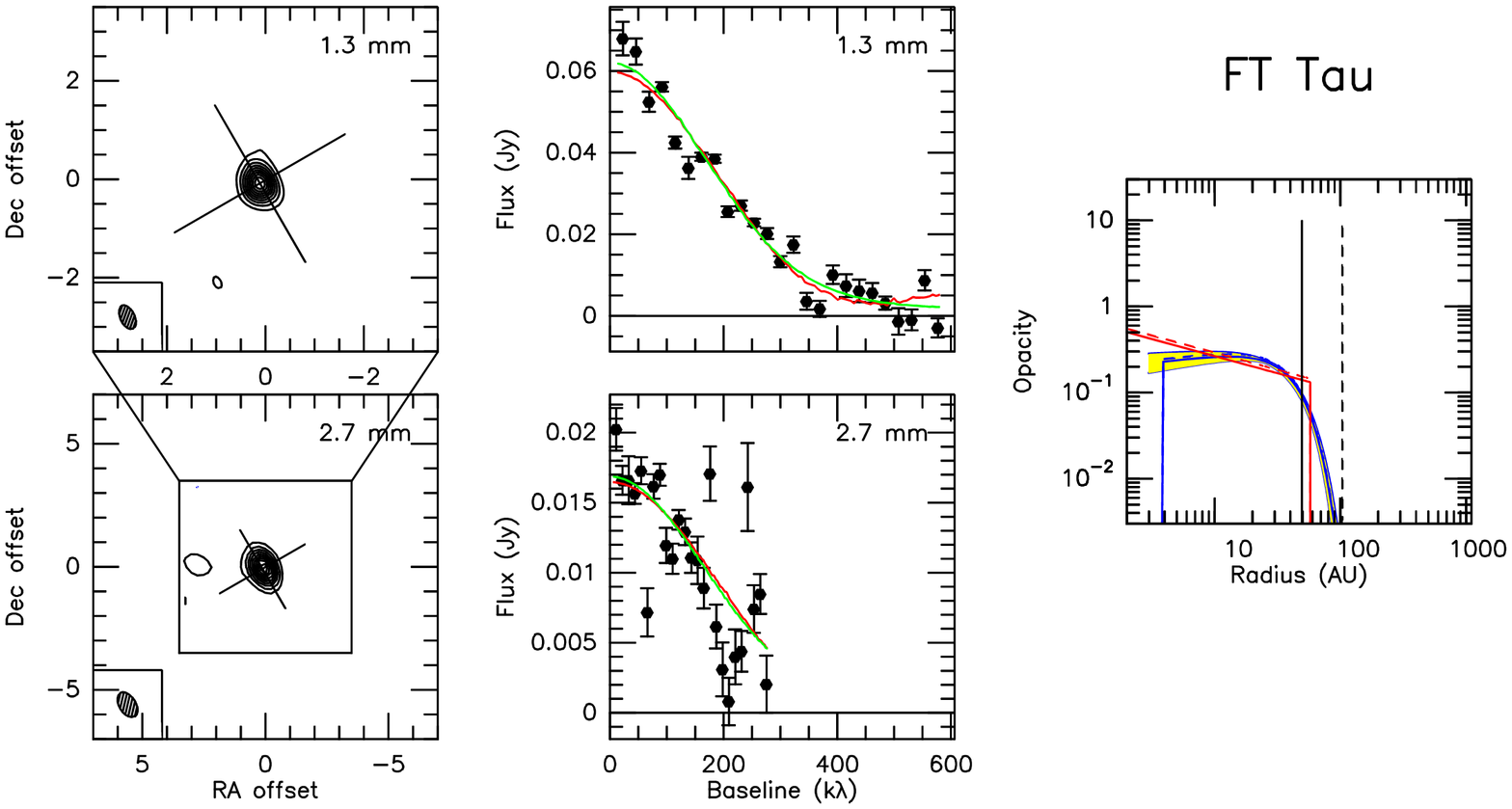}
\caption{As Fig.\ref{fig:alldmtau} but for FT Tau. Contour level is 2.6 mJy/beam ($8 \sigma$)
at 1.3 mm, and 1.3 mJy/beam ($6 \sigma$) at 2.7\,mm.}
   \label{fig:allfttau}
\end{figure*}

\begin{figure*}[h]
   \includegraphics[angle=270,width=18.0cm]{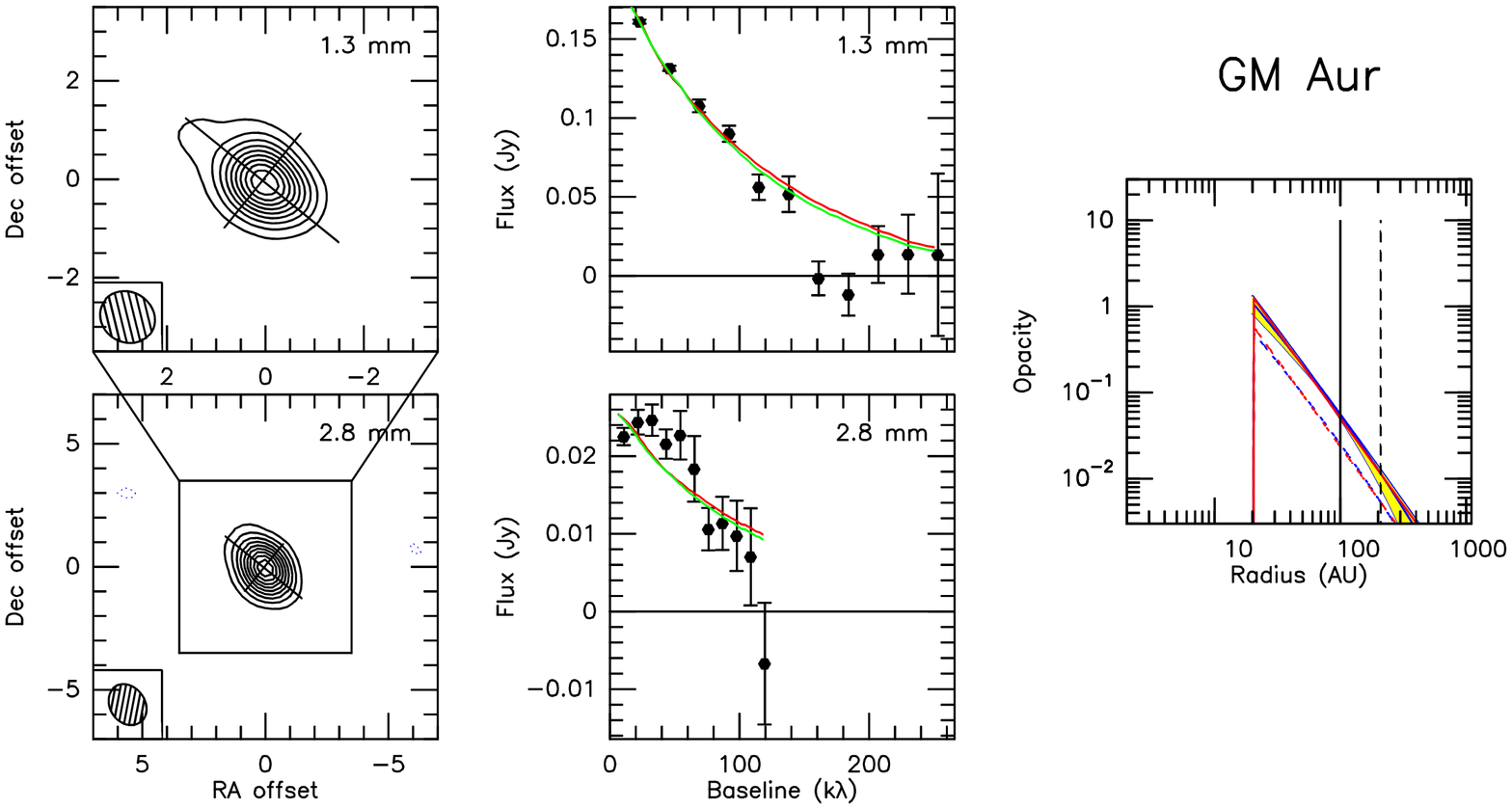}
\caption{As Fig.\ref{fig:alldmtau} but for GM Aur. Contour level is 10 mJy/beam ($65 \sigma$)
at 1.3 mm, and 1.9 mJy/beam ($3.2 \sigma$) at 2.8\,mm.}
   \label{fig:allgmaur}
\end{figure*}

\clearpage
\begin{figure*}[h]
   \includegraphics[angle=270,width=18.0cm]{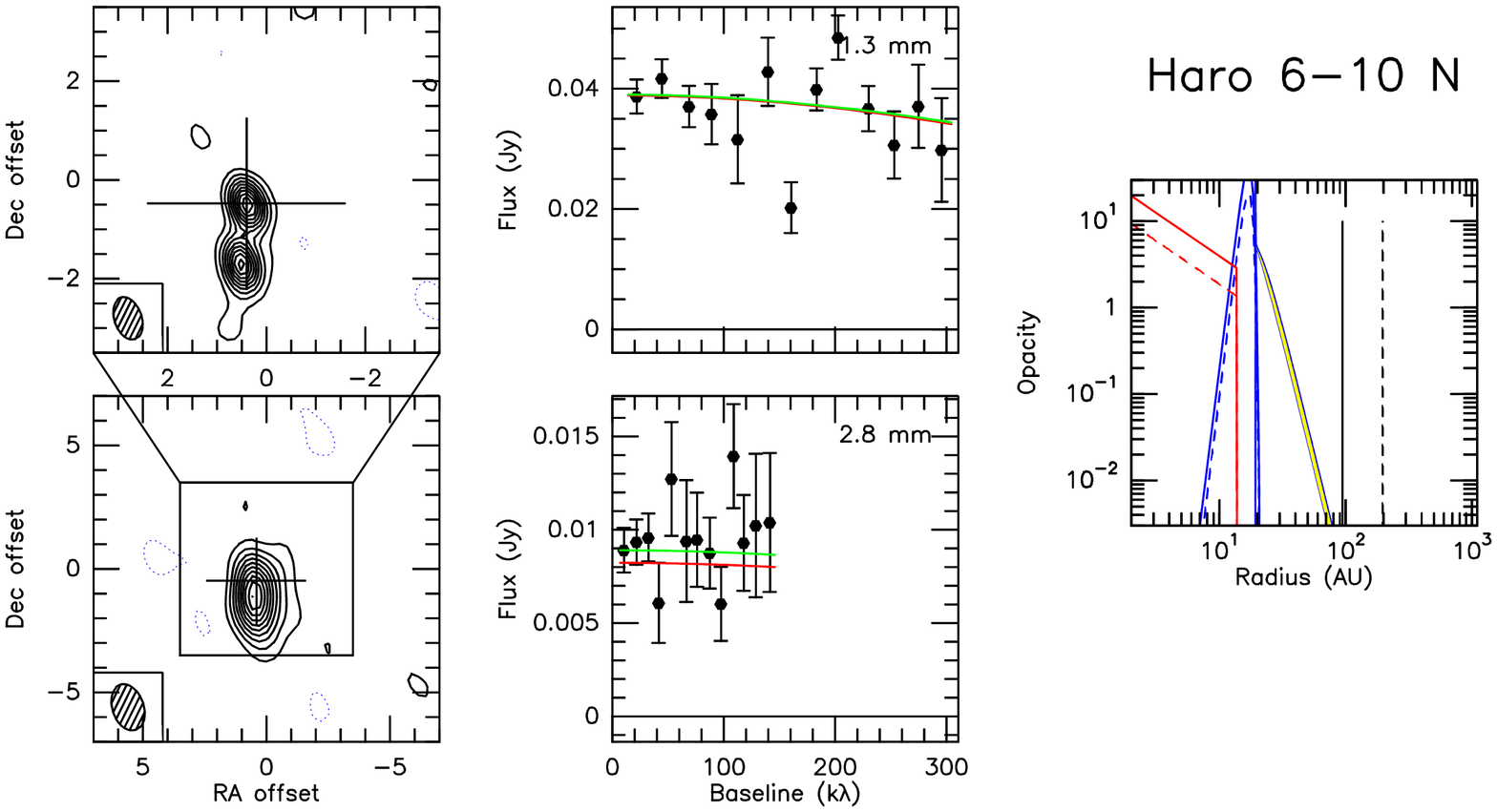}
\caption{As Fig.\ref{fig:alldmtau} but for Haro 6-10 N. Contour level is 3.5 mJy/beam ($4 \sigma$)
at 1.3 mm, and 1.2 mJy/beam ($3 \sigma$) at 2.8\,mm.}
   \label{fig:allharo610n}
\end{figure*}

\begin{figure*}[ht]
   \includegraphics[angle=270,width=18.0cm]{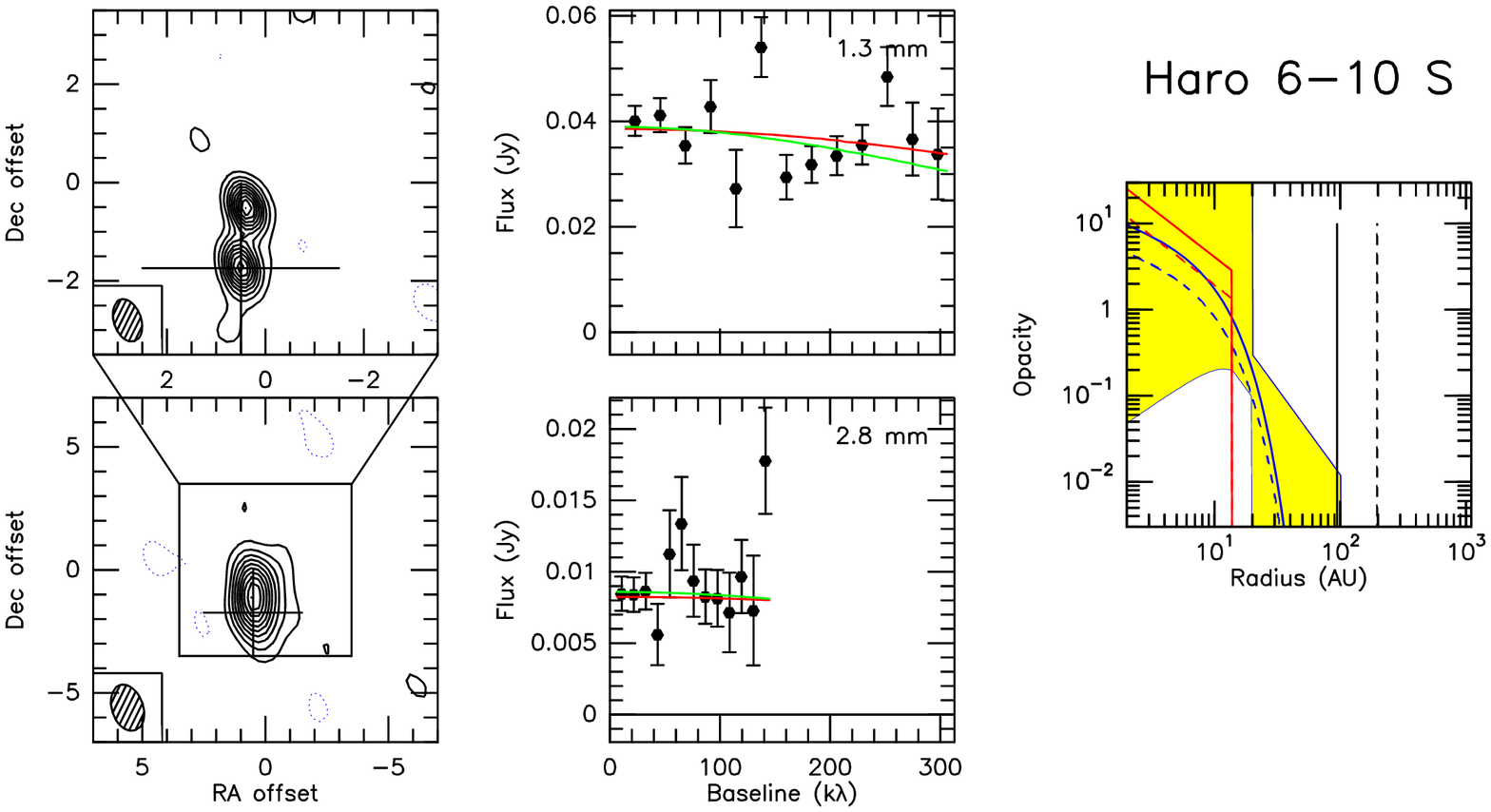}
\caption{As Fig.\ref{fig:alldmtau} but for Haro 6-10 S. Contour level is 3.5 mJy/beam ($4 \sigma$)
at 1.3 mm, and 1.2 mJy/beam ($3 \sigma$) at 2.8\,mm.}
   \label{fig:allharo610s}
\end{figure*}

\clearpage
\begin{figure*}[ht]
   \includegraphics[angle=270,width=18.0cm]{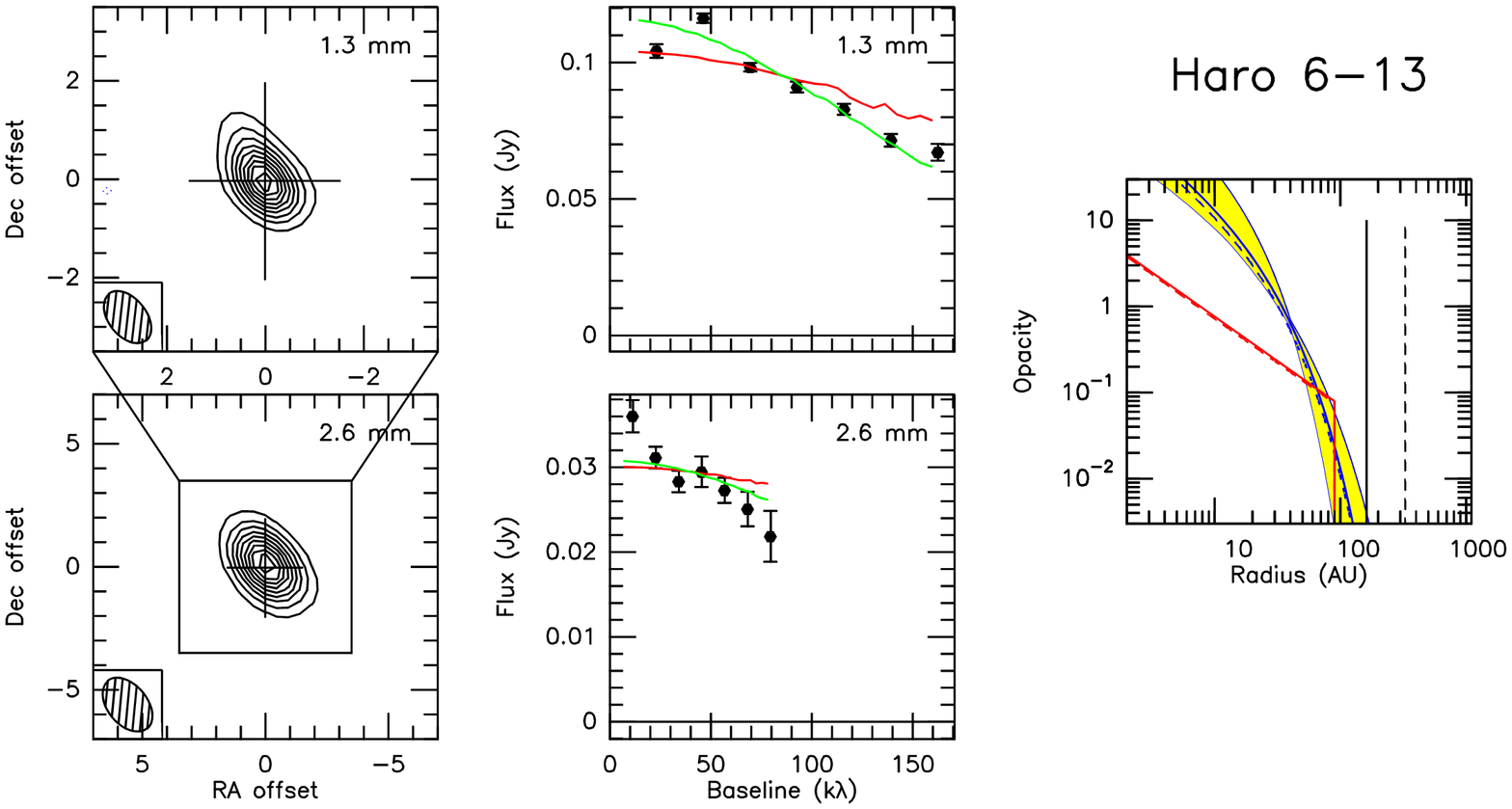}
\caption{As Fig.\ref{fig:alldmtau} but for Haro 6-13. Contour level is 8.3 mJy/beam ($5 \sigma$)
at 1.3 mm, and 2.8 mJy/beam ($5.6 \sigma$) at 2.6\,mm.}
   \label{fig:allharo613}
\end{figure*}

\begin{figure*}[h]
   \includegraphics[angle=270,width=18.0cm]{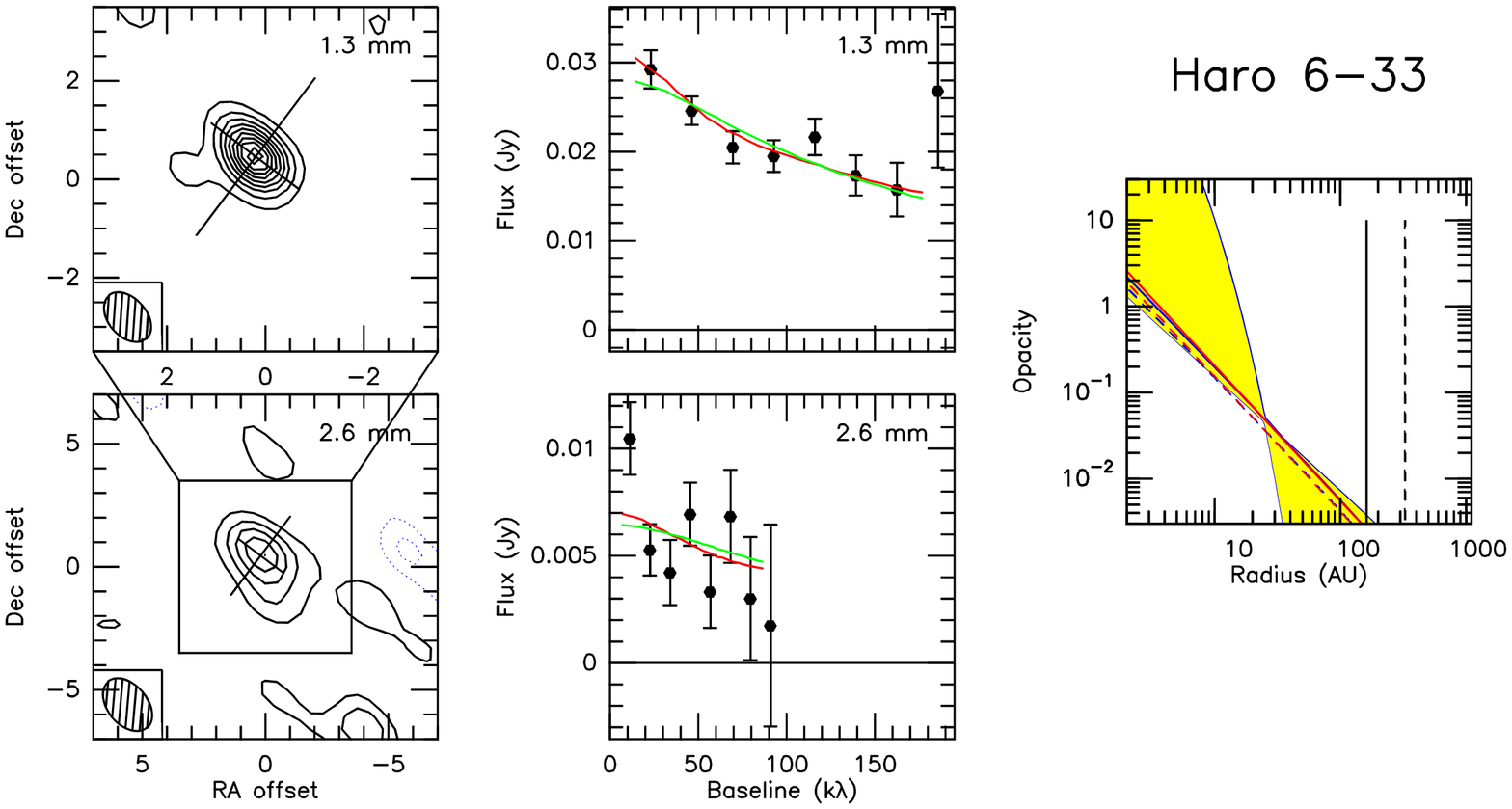}
\caption{As Fig.\ref{fig:alldmtau} but for Haro 6-33. Contour level is 2 mJy/beam ($3.3 \sigma$)
at 1.3 mm, and 1.5 mJy/beam ($1.7 \sigma$) at 2.6\,mm.}
   \label{fig:allharo633}
\end{figure*}

\clearpage
\begin{figure*}[h]
   \includegraphics[angle=270,width=18.0cm]{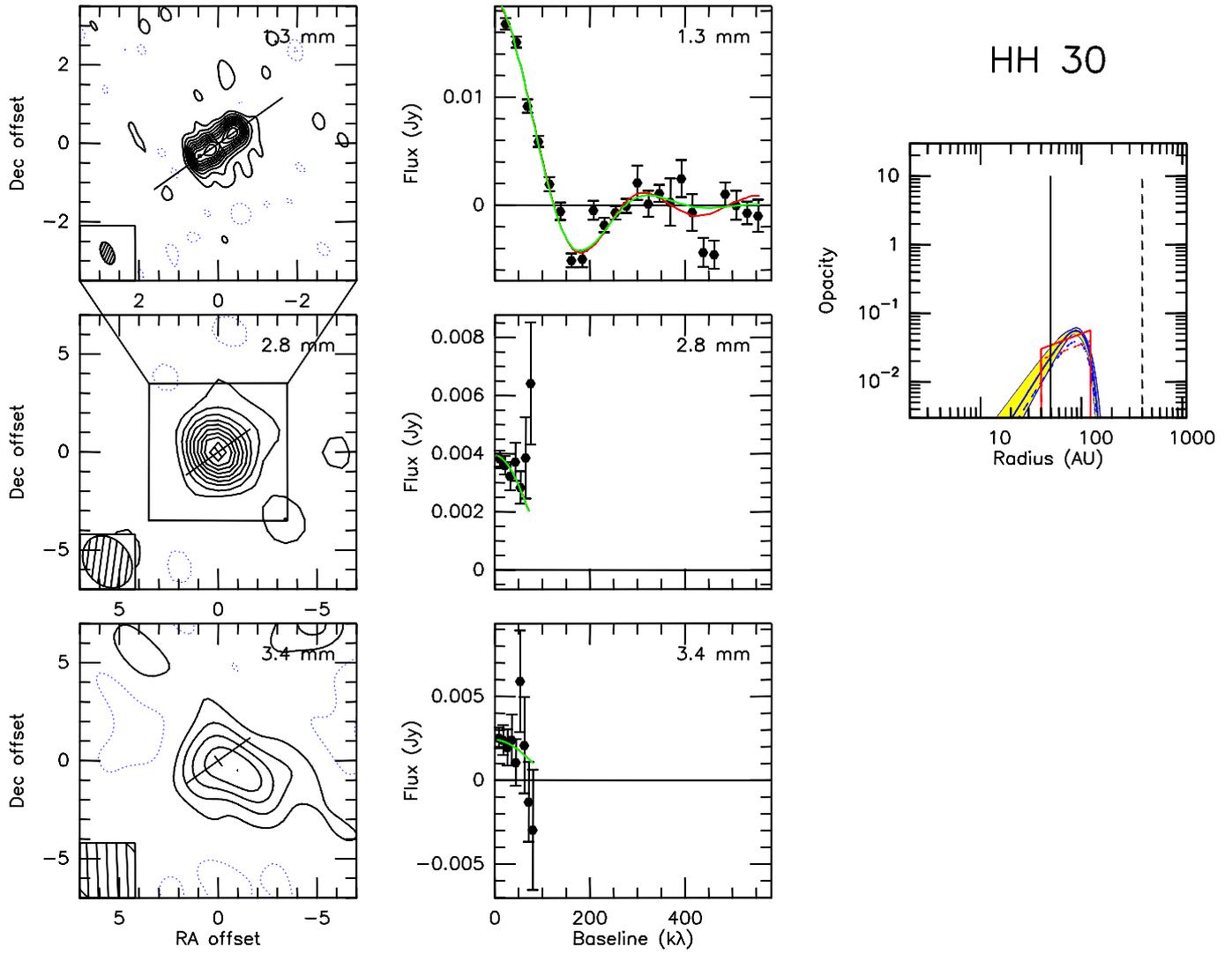}
\caption{As Fig.\ref{fig:alldmtau} but for HH\,30. Contour level is 0.4 mJy/beam ($2 \sigma$)
at 1.3 mm, 0.5 mJy/beam ($1.2 \sigma$) at 3.4\,mm, and 0.36 mJy/beam ($2.2 \sigma$) at 2.8\,mm.}
   \label{fig:allhh30}
\end{figure*}

\clearpage
\begin{figure*}[h]
   \includegraphics[angle=270,width=18.0cm]{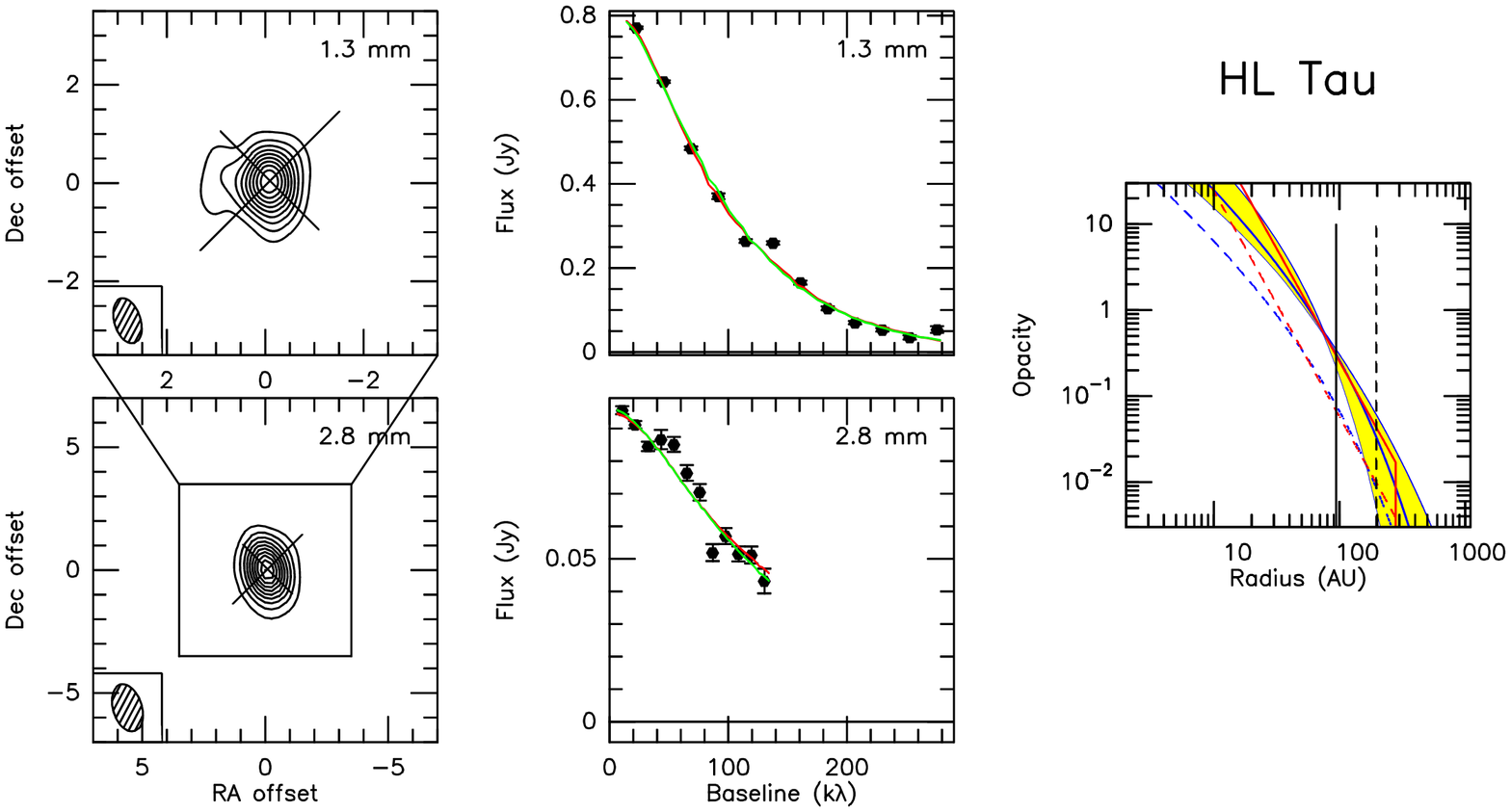}
\caption{As Fig.\ref{fig:alldmtau} but for HL\,Tau. Contour level is 32 mJy/beam ($4.5 \sigma$)
at 1.3 mm, and 7.3 mJy/beam ($9 \sigma$) at 2.8\,mm. }
   \label{fig:allhltau}
\end{figure*}

\begin{figure*}[h]
   \includegraphics[angle=270,width=18.0cm]{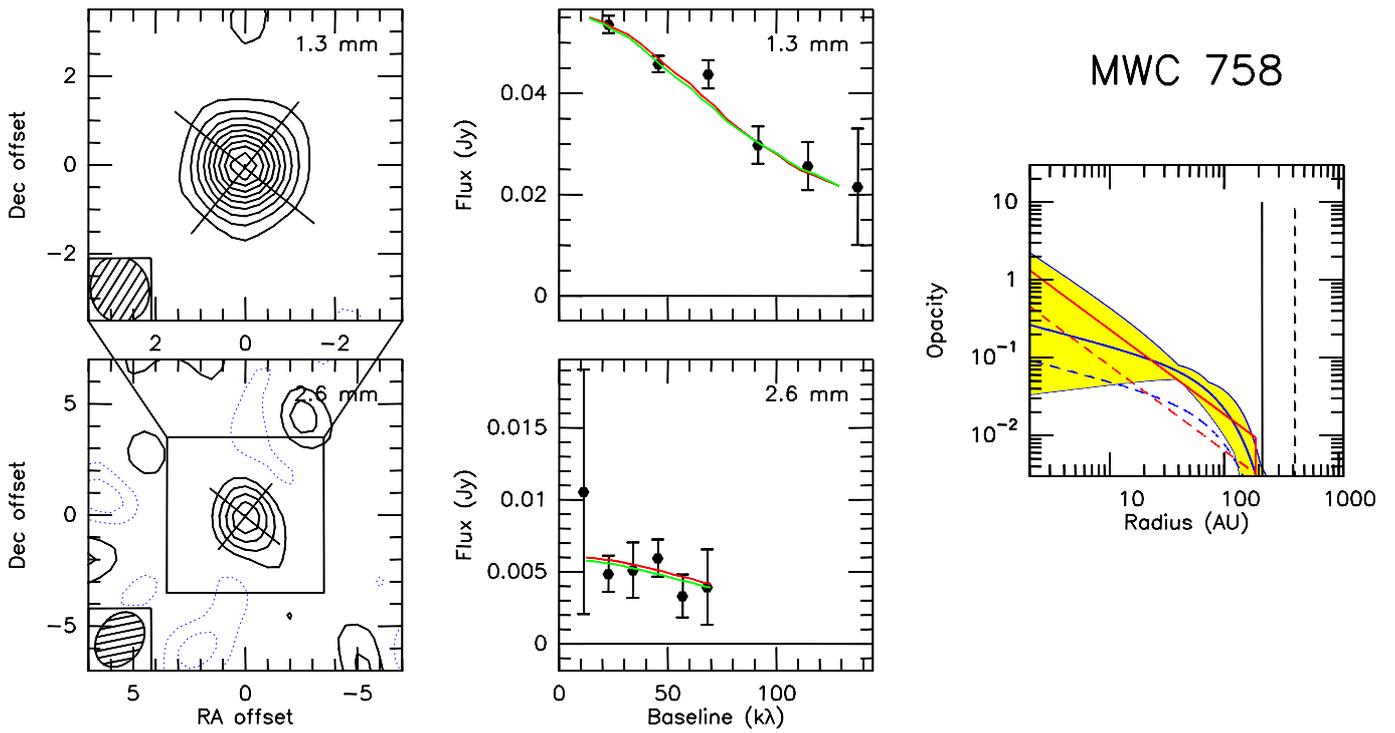}
\caption{As Fig.\ref{fig:alldmtau} but for MWC 758. Contour level is 4 mJy/beam ($2.7 \sigma$)
at 1.3 mm, and 0.8 mJy/beam ($1.3 \sigma$) at 2.6\,mm.}
   \label{fig:allmwc758}
\end{figure*}

\clearpage
\begin{figure*}[h]
   \includegraphics[angle=270,width=18.0cm]{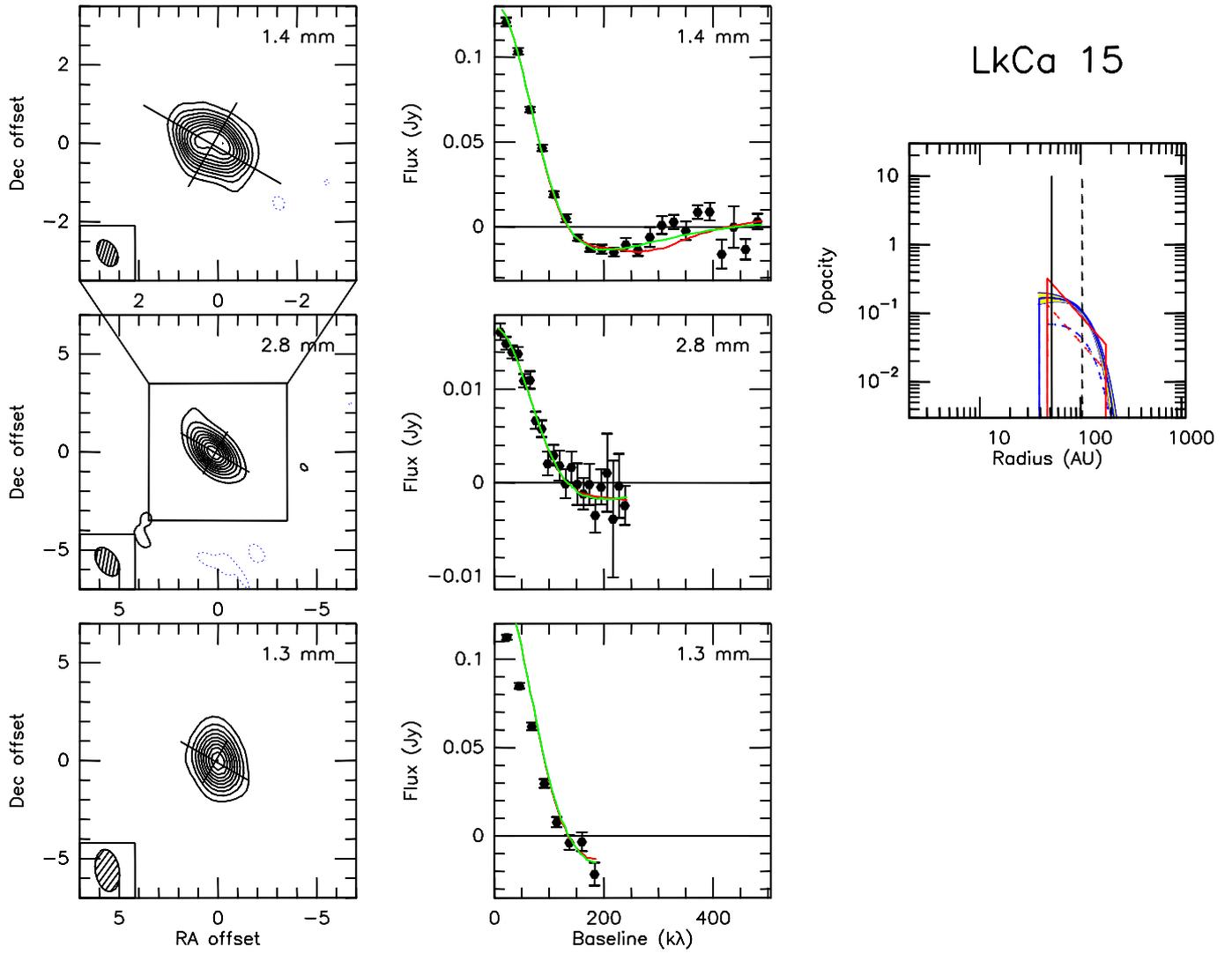}
\caption{As Fig.\ref{fig:alldmtau} but for Lk\,Ca\,15. Contour level is 2.6 mJy/beam ($4 \sigma$)
at 1.4 mm, 1.0 mJy/beam ($3 \sigma$) at 2.8\,mm, and 7.9 mJy/beam ($5 \sigma$) at 1.3\,mm. }
   \label{fig:alllkca15}
\end{figure*}

\clearpage
\begin{figure*}[h]
   \includegraphics[angle=270,width=18.0cm]{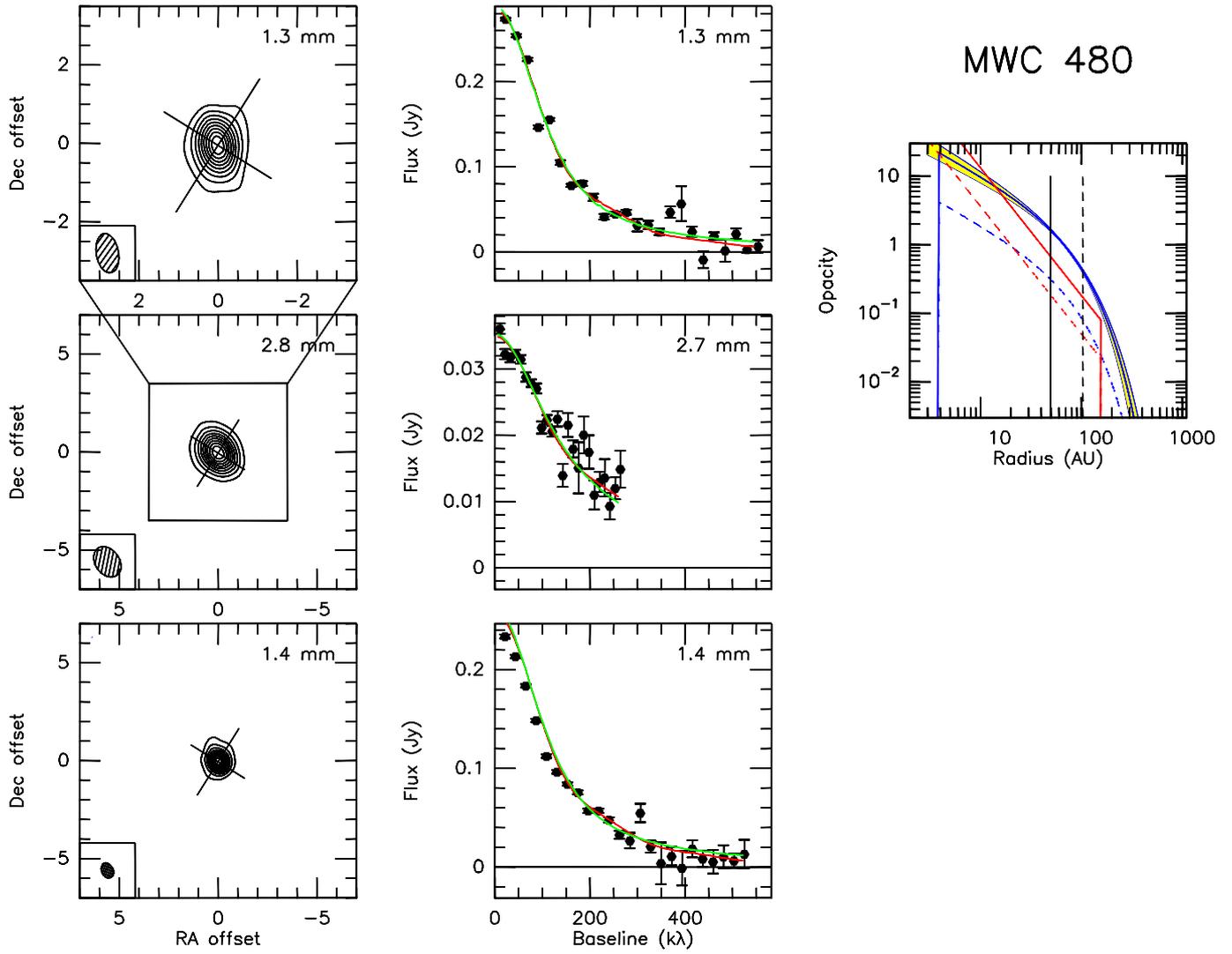}
\caption{As Fig.\ref{fig:alldmtau} but for MWC 480. Contour level is 15 mJy/beam ($5.8 \sigma$)
at 1.3 mm, 2.9 mJy/beam ($7 \sigma$) at 2.8\,mm, and 12 mJy/beam ($5.5 \sigma$) at 1.4\,mm. }
   \label{fig:allmwc480}
\end{figure*}

\clearpage
\begin{figure*}[h]
   \includegraphics[angle=270,width=18.0cm]{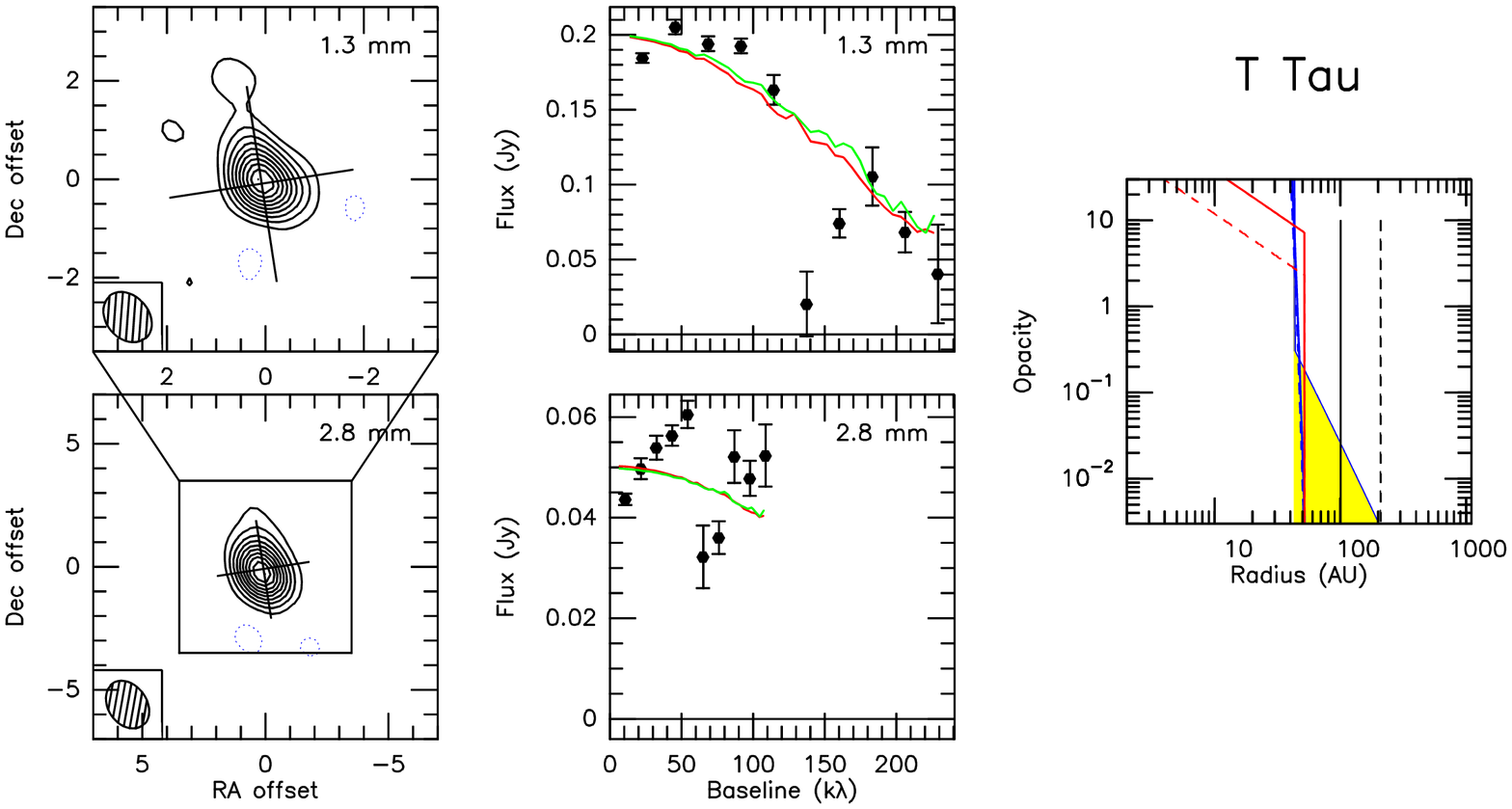}
\caption{As Fig.\ref{fig:alldmtau} but for T\,Tau. Contour level is 16 mJy/beam ($3 \sigma$)
at 1.4 mm, and 4.8 mJy/beam ($4.8 \sigma$) at 2.8\,mm.}
   \label{fig:allttau}
\end{figure*}

\clearpage
\begin{figure*}[ht]
   \includegraphics[angle=270,width=18.0cm]{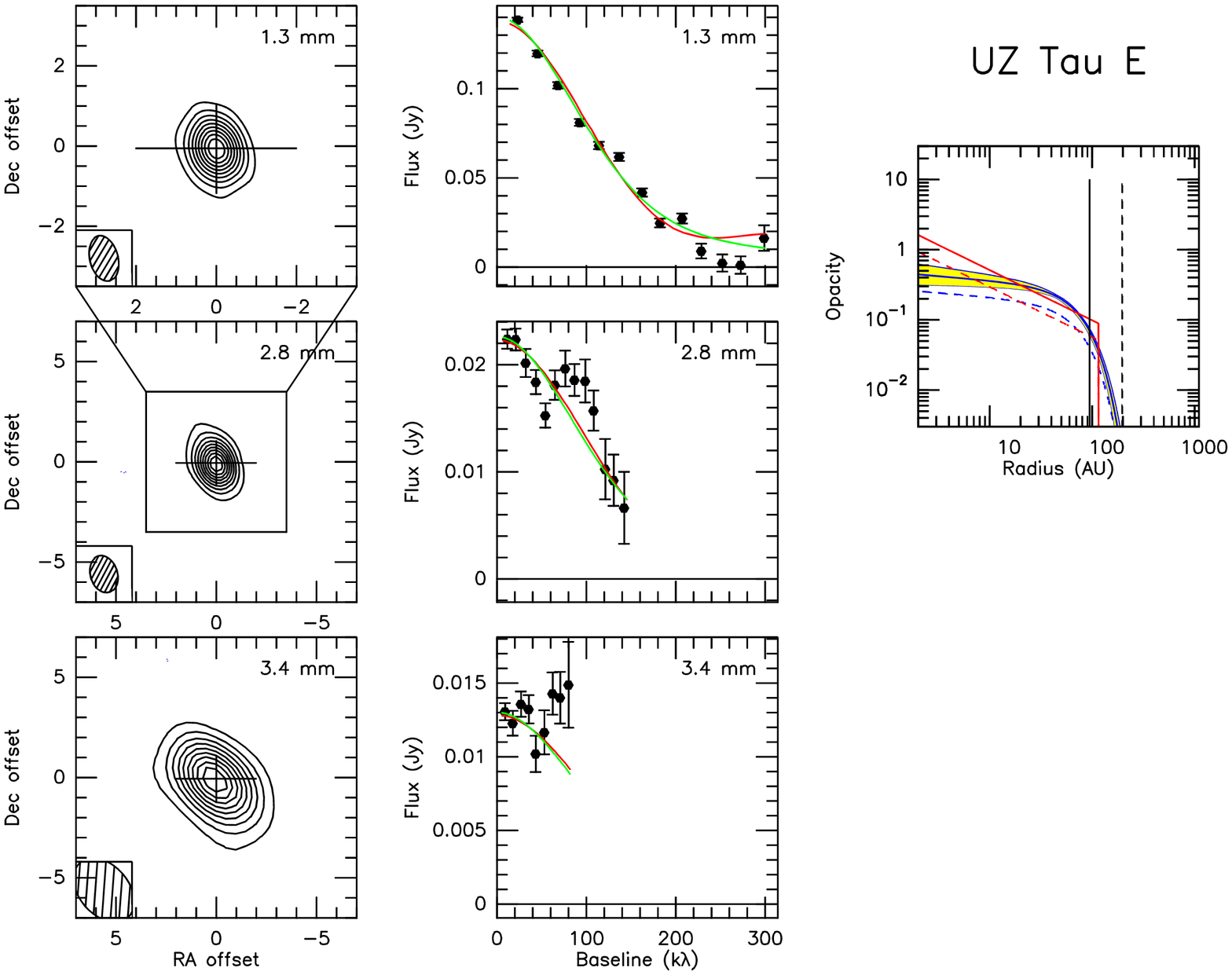}
\caption{As Fig.\ref{fig:alldmtau} but for UZ Tau E. Contour level is 8.6 mJy/beam ($6 \sigma$)
at 1.3 mm, 1.9 mJy/beam ($4.7 \sigma$) at 2.8\,mm, and 1.3 mJy/beam ($4 \sigma$) at 3.4\,mm. }
   \label{fig:alluztaue}
\end{figure*}

\begin{figure*}[ht]
   \includegraphics[angle=270,width=18.0cm]{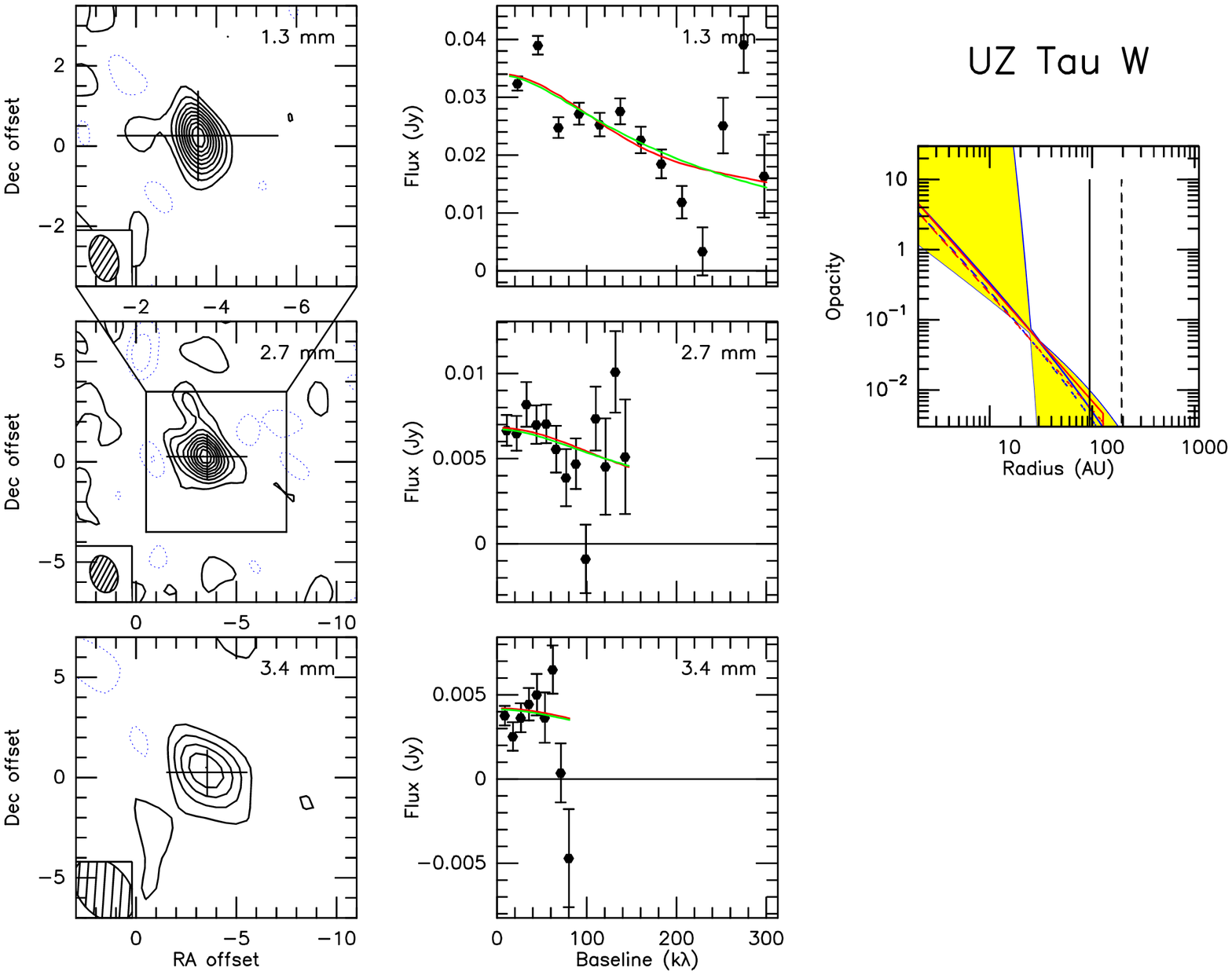}
\caption{As Fig.\ref{fig:alldmtau} but for UZ Tau W. Contour level is 2.6 mJy/beam ($1.9 \sigma$)
at 1.3 mm, 0.6 mJy/beam ($1.5 \sigma$) at 2.8\,mm, and 0.7 mJy/beam ($2 \sigma$) at 3.4\,mm. }
   \label{fig:alluztauw}
\end{figure*}

\end{appendix}

\end{document}